\documentclass[%
 aip,pop,
 amsmath,amssymb,
 preprint,%
]{revtex4-1}

\usepackage{graphicx}
\usepackage{dcolumn}
\usepackage{bm}
\usepackage{color}

\usepackage{hyperref}
\hypersetup{
     colorlinks   = true,
     citecolor    = blue,
     linkcolor     = blue}

\usepackage{amsmath}
\usepackage[a4paper, total={8.5in, 11in},margin=0.8in]{geometry}
\usepackage{graphicx}
\usepackage[makeroom]{cancel}

\renewcommand{\v}[1]{\ensuremath{\mathbf{#1}}} 
\newcommand{\gv}[1]{\ensuremath{\mbox{\boldmath$ #1 $}}} 
\newcommand{\uv}[1]{\ensuremath{\mathbf{\hat{#1}}}} 
\renewcommand{\d}[2]{\frac{d #1}{d #2}} 
\newcommand{\pd}[2]{\frac{\partial #1}{\partial #2}} 
 
\newcommand{\partialx}[1]{\partial_x #1} 
\newcommand{\partialy}[1]{\partial_y #1} 
\newcommand{\grad}[1]{\gv{\nabla} #1} 
\renewcommand{\div}[1]{\gv{\nabla} \cdot #1} 
\newcommand{\curl}[1]{\gv{\nabla} \times #1} 
\newcommand{\delsq}[1]{\nabla^2 #1} 
\newcommand{\xhat}{\uv{x}}
\newcommand{\yhat}{\uv{y}}
\newcommand{\zhat}{\uv{z}}
\newcommand{\gradv}[1]{\gv{\nabla}_v #1} 

\newcommand{\vv}{\v{v}}								
\newcommand{\vperp}{v_{\perp}}						
\newcommand{\vvi}{\v{v}_i}							
\newcommand{\vve}{\v{v}_e}							
\newcommand{\vvExB}{\v{v}_{E}}						
\newcommand{\vvdi}{\v{v}_{di}}						
\newcommand{\vvde}{\v{v}_{de}}						

\newcommand{\kp}{k_\perp}

\newcommand{\pO}{p_0(x)}
\newcommand{\BO}{B_0(x)}
\newcommand{\pper}{\tilde{p}}
\newcommand{\Bper}{\tilde{B}}
\newcommand{\pOp}{p_0'}
\newcommand{\BOp}{B_0'}
\newcommand{\LB}{L_B}
\newcommand{\Ln}{L_n}
\newcommand{\Lp}{L_p}
\newcommand{\omegac}{\Omega}
\newcommand{\omegaci}{\Omega_{ci}}
\newcommand{\omegace}{\Omega_{ce}}
\newcommand{\cs}{C_s}
\newcommand{\vbar}{\bar{v}}
\newcommand{\vxbar}{\bar{v}_x}
\newcommand{\vybar}{\bar{v}_y}
\newcommand{\vzbar}{\bar{v}_z}
\newcommand{\vbarperp}{\bar{v}_\perp}
\newcommand{\Aper}{\tilde{A}}
\newcommand{\vAper}{\v{\tilde{A}}}

\newcommand{\vEper}{\v{\tilde{E}}}
\newcommand{\LF}{L_F}
\newcommand{\LT}{L_T}
\newcommand{\LTe}{L_{Te}}
\newcommand{\LTi}{L_{Ti}}
\newcommand{\LFT}{L_{FT}}
\newcommand{\phiper}{\tilde{\phi}}
\newcommand{\varphiper}{\tilde{\varphi}}
\newcommand{\varphihat}{\hat{\varphi}}
\newcommand{\bper}{\tilde{b}}
\newcommand{\bhat}{\hat{b}}
\newcommand{\fper}{\tilde{f}}
\newcommand{\fhat}{\hat{f}}
\newcommand{\fbarOi}{\bar{f}_{0i}}
\newcommand{\fbarOe}{\bar{f}_{0e}}
\newcommand{\vt}{v_t}
\newcommand{\vti}{v_{ti}}
\newcommand{\vte}{v_{te}}
\newcommand{\omegab}{\omega_b}
\newcommand{\omegabi}{\omega_{bi}}
\newcommand{\omegabe}{\omega_{be}}
\newcommand{\omegabar}{\bar{\omega}}
\newcommand{\omegabari}{\bar{\omega}_i}
\newcommand{\omegabare}{\bar{\omega}_e}
\newcommand{\TiO}{T_{i0}}
\newcommand{\TeO}{T_{e0}}
\newcommand{\IBA}{I_{BA}}
\newcommand{\IphiA}{I_{\phi A}}
\newcommand{\IBQ}{I_{BQ}}
\newcommand{\IphiQ}{I_{\phi Q}}
\newcommand{\taue}{\tau_e}
\newcommand{\taui}{\tau_i}
\newcommand{\etae}{\eta_e}
\newcommand{\etai}{\eta_i}
\newcommand{\betae}{\beta_e}
\newcommand{\betai}{\beta_i}
\newcommand{\LnOLB}{\frac{\Ln}{\LB}}
\newcommand{\LBOLn}{\frac{\LB}{\Ln}}
\newcommand{\de}{d_e}
\newcommand{\di}{d_i}
\newcommand{\tauit}{\tilde{\taui}}
\newcommand{\DeltaGDC}{\Delta_{\text{GDC}}}

\begin{document}


\title{Gyrokinetic theory of slab universal modes and the non-existence of the Gradient Drift Coupling (GDC) instability}

\author{Barrett N. Rogers}
 \email{Barrett.N.Rogers@dartmouth.edu}
 \affiliation{Department of Physics and Astronomy, Dartmouth College\\6127 Wilder Laboratory\\Hanover, New Hampshire 03755-3528, USA}
\author{Ben Zhu}%
 \affiliation{Department of Physics and Astronomy, Dartmouth College\\6127 Wilder Laboratory\\Hanover, New Hampshire 03755-3528, USA}
\author{Manaure Francisquez}%
 \affiliation{Department of Physics and Astronomy, Dartmouth College\\6127 Wilder Laboratory\\Hanover, New Hampshire 03755-3528, USA}

\date{\today}

\begin{abstract}
A gyrokinetic linear stability analysis of a collisionless slab
geometry in the local approximation is presented. We focus on $k_\parallel=0$ universal (or entropy) modes
driven by plasma gradients at small and large plasma $\beta$. 
These are small scale non-MHD instabilities with growth
rates that typically peak near $k_\perp\rho_i\sim 1$ and vanish in the long 
wavelength $k_\perp\to 0$ limit. This work also discusses a mode known as the Gradient Drift
Coupling (GDC) instability previously reported in the gyrokinetic literature, which has
a finite growth rate $\gamma= \sqrt{\beta/[2(1+\beta)]} C_s/|L_p|$ with
$C_s^2=p_0/\rho_0$ for $k_\perp\to 0$ and is universally unstable for $1/L_p\neq 0$. We
show the GDC instability is a spurious, unphysical artifact that erroneously arises due to the failure
to respect the total equilibrium pressure balance $p_0+B_0^2/(8\pi)=\text{constant}$, which
renders the assumption $B_0'=0$ inconsistent if $p_0'\neq 0$.
%
\end{abstract}

\maketitle


\section{Introduction}\label{sec:intro}
We present gyrokinetic linear stability studies of a collisionless slab
geometry in the local approximation to entropy or ``universal" modes with $k_\parallel=0$
at small and large plasma $\beta=8\pi p_0/B_0^2$, $p_0=p_{i0}+p_{e0}$.
The term entropy mode is used in a variety of ways in plasma physics and is 
sometimes associated with the universal mode. The former is not to be confused with the stationary entropy wave eigenmode of ideal MHD~\cite{Goedbloed2004}. The entropy mode has a history that goes back more than five decades~\cite{Kadomtsev1960}. It received this appellation because though the mode perturbs the density and temperature, the plasma pressure remains undisturbed and thus entropy fluctuations ensue ($S=pn^{-\gamma}$ is the specific entropy, $p$ the total plasma kinetic pressure, $n$ the density and $\gamma$ the ratio of specific heats). These are non-MHD instabilities with growth rates that typically peak near either $k_\perp\rho_i\sim 1$, and vanish in the long wavelength $k_\perp\to 0$ limit. They can be unstable even in the absence of finite $k_\parallel$~\cite{Mikhailovskii1971}, magnetic curvature~\cite{Mikhailovskaya1968}, electrical resistivity~\cite{Ware1962},
electron inertia~\cite{KrassVarban1994}, electromagnetic effects~\cite{Simakov2001,Ricci2006}, collisionless damping~\cite{Barnes1966,Coppi1969}, 
wave-particle resonances~\cite{Barnes1966}, magnetic shear~\cite{Simakov2001,Ricci2006} and many other effects.

Past studies of the entropy mode were mostly carried out in the kinetic framework,
using either the Vlasov equation~\cite{Coppi1969}, drift kinetics~\cite{Mikhailovskii1971} or gyrokinetics~\cite{Simakov2001,Ricci2006,Kobayashi2010}. Some fluid estimations also exist~\cite{Mikhailovskaya1968,Hassam1984}; as well as MHD studies, as in Ref.~\cite{Kesner2001NF}, where
entropy modes are also termed drift temperature gradient modes. Aside from the perturbed entropy at constant pressure a characteristic of these modes is that their stability depends on $-d\ln p/d{\ln V}$ ($V$ is the volume per unit flux) and $\eta=d\ln T/d\ln n$. However some of the studies above also explore how collisionality modifies stability regions. The research mentioned thus far has been performed in the electrostatic limit, some for slab geometries as well as curved magnetic fields such as Z-pinches~\cite{Kadomtsev1960,Hassam1984,Ricci2006} and dipoles~\cite{Simakov2001}.

The analysis presented here treats entropy modes in the sense that an MHD-stable
equilibrium with constant total plasma pressure $p+B^2/(8\pi)$ is considered, but
entropy fluctuations are possible. Since the kinetic pressure of the plasma is not
guaranteed to remain constant, this work may not examine entropy modes in the
historical sense of the name. Because we are concerned with instabilities driven by a
density (or temperature) gradient, the term universal instability may be more applicable.
Entropy and universal instabilities are used interchangeably from time to time. The term
universal also goes back to the 1960's~\cite{Galeev1963,Cap1976}, referring to modes
that were excited by the presence of a density gradient. Given that an inhomogeneous density
was a ubiquitous property of (confined) plasmas this phenomenon was deemed universal.
In the later half of the twentieth century it was thought to be the microinstability with the
dominant contribution to transport. The universal instability is one in a class of drift
waves and thus sometimes was referred to as universal drift mode (or wave), or even just a drift mode. Early on it was found that these modes, driven by electron-wave particle resonance coupled with the density gradient, were always unstable in a shearless magnetic field~\cite{Pearlstein1969,Gladd1973}. Weak shear did not appear to stabilize the mode. But in 1978 three papers~\cite{Antonsen1978,Ross1978,Tsang1978} addressing this slab system in the absence of temperature gradients, parallel currents or magnetic curvature, concluded that including the full electron dispersion function absolutely stabilized the mode for any shear strength. The reason was attributed to damping caused by non-resonant electrons. These works, along with an earlier one that considered the stabilizing influence of more complex geometries and temperature gradients~\cite{Krall1965}, disputed the use of the term ``universal" since most properties of real-world plasmas appear to stabilize the mode.

The universal mode appeared to suffer a sudden death after 1978. A short-lived
resurgence occurred when a study of sheaths found that in strongly localized plasmas magnetic shear can in fact have destabilize universal modes~\cite{Marchand1982}. More recently, universal modes were studied in toroidal systems and found to coexist with finite shear and temperature gradients, though considerably stabilized by perpendicular magnetic field perturbations~\cite{Chowdhury2010} (i.e. $\delta A_\parallel$). Moreover, the stability of the slab has been re-examined at the sub-ion Larmor scale and found to be unstable even in the presence of shear~\cite{Landreman2015PRL} due to a relaxation of the assumption that the radial extent (in the direction of the density gradient) of the mode is much larger than the ion Larmor radius ($\rho_i$). This has reinvigorated the exploration of universal modes, with Ref.~\cite{Landreman2015} exploring its role in subcritical turbulence and Ref.~\cite{Helander2015} investigating these modes in general geometries.

The modes to be explored here share similarities with both entropy and universal
instabilities. For that reason, and due to the occasional swapping of these terms, we may
call the modes in this paper by either name. We shall refer to them as universal modes.
Most of the research into both entropy and universal modes has been carried out in the
electrostatic limit, and thus this presents one of the few electromagnetic studies
available. It also offers the first view into the electromagnetic effect of supporting
parallel magnetic field fluctuations ($\delta B_\parallel$). As a start we describe the
system under study in section~\ref{sec:pressure}. In this segment we also comment on
how the spurious gradient-drift coupling (GDC)
instability~\cite{pueschel2015enhanced,pueschel2017basic} arises from a simplified
fluid estimate. Section~\ref{sec:DR} presents the gyrokinetic dispersion relation for
these universal modes, whose derivation is outlined in appendix~\ref{app:kin} and
complemented by the supplemental material. The low $\beta$ limit is examined
in~\ref{sec:lowbeta} and the high $\beta$ case in~\ref{sec:highBeta}. With the
gyrokinetic model of these universal modes in hand we show that one can obtain the GDC
instability via violation of total pressure balance in section~\ref{sec:gdc}. Conclusions
and future directions are summarized in section~\ref{sec:conc}.

\section{Equilibrium pressure balance and the problem with GDC} \label{sec:pressure}
Let us evaluate a shearless slab geometry
in which the magnetic field ($\mathbf{B}=B\mathbf{\hat{z}}$) and its perturbations are 
entirely in the $z$-direction, the equilibrium gradients in the $x$-direction,
and the wavenumber of the perturbations along $y$: $k_\perp=k\hat{\v{y}}$, such that
the perturbations vary with $t$ and $y$ as $\exp(-i\omega t+i k y)$.
We consider low frequency modes in a hydrogenic plasma that maintains the total pressure 
balance condition $p+B^2/(8\pi)=\text{constant}$. Writing $p=p_0(x)+\tilde{p}$, $B=B_0(x)+\tilde{B}$, in the case of the
perturbations this implies 
\begin{equation}
\tilde{p}=-\frac{B_0\tilde{B}}{4\pi},
\label{eq:pbpert}
\end{equation}
while for the equilibium $p_0'=-B_0B_0'/(4\pi)$ (primes denoting $d/dx$), or 
\begin{equation}
\frac{1}{L_B}=\frac{B_0'}{B_0}= - \frac{\beta}{2L_p} \ ,\quad \frac{1}{L_p}=\frac{p_0'}{p_0}\ .
\label{eq:pbeq}
\end{equation}
A key finding of this study is the importance of the total pressure balance relation
of the equilibrium: failure to respect this condition in the stability analysis can lead to unphysical instability.
In particular, if one assumes $p_0'\neq 0$ then one must account for finite ${1}/{L_B}=-(\beta/2)/L_p$ rather than, 
as in some past studies, assume $B'_0=0$.
Consider for example the usual decomposition of the ion and electron fluid drift velocities
into $E\times B$, diamagnetic and polarization drift components (neglecting the latter for electrons) with $\phi=\tilde{\phi}$ the perturbation of the electrostatic potential
\begin{equation}
\v{v}_i=\frac{c}{B}\left( \uv{z}\times \nabla\phi +\frac{1}{ne}\uv{z}\times\nabla p_i+i\frac{\omega}{\Omega_{ci}}\nabla\phi \right)\ ,
\label{eq:vi}
\end{equation}
\begin{equation}
\v{v}_e=\frac{c}{B}\left( \uv{z}\times \nabla\phi -\frac{1}{ne}\uv{z}\times\nabla p_e\right)\ ,
\label{eq:ve}
\end{equation}
and $\Omega_{ci}={eB_0}/(m_i c)$ is the ion gyrofrequency. These lead to the vorticity (or current continuity) equation:
\begin{equation}
\nabla\cdot n\left(\v{v}_i-\v{v}_e\right)\simeq i n  \frac{\omega c}{\Omega_{ci}B}\nabla^2\phi+
\nabla\cdot \frac{c}{eB}\uv{z}\times\nabla p=0 +\dots
\label{eq:vort0}
\end{equation}
Linearizing the last term and assuming $k\gg d/dx$ for the perturbations
one obtains
\begin{equation}
i \omega k^2\phi+
i\frac{k }{cn_0 m_i}\left(p_0'\tilde{B}-B_0'\tilde{p}\right) =0.
\label{eq:vort1}
\end{equation}
Finally, expressing $\tilde{B}$ in terms of $\tilde{p}$ with Eq.~(\ref{eq:pbpert})
and assuming 
\begin{equation}
\dot{p}+\frac{c}{B}\uv{z}\times\nabla\phi \cdot\nabla p=0+\dots\ \ \to\ \ \tilde{p}=-\frac{ckp_0'}{\omega B_0}\phi +\dots
\label{eq:peq}
\end{equation}
yields the dispersion relation
\begin{equation}
\omega^2=-\frac{C_s^2}{L_p}\left(\frac{\beta}{2L_p}+\frac{1}{L_B}\right)\ \ ,\quad C_s^2=\frac{p_0}{n_0m_i}\ \ .
\end{equation}
Noting the equilibrium pressure balance relation Eq.~(\ref{eq:pbeq}): $1/L_B=-\beta/(2L_p)$, the right-hand side
of this equation vanishes. The same cancellation occurs in the gyrokinetic calculations to follow, leaving at low $\beta$
a leading-order ion drift wave solution $\omega=\omega_{di}$, $\omega_{di}=k c p_{i0}'/(n_0eB_0)$
as well as a finite-$\beta$ potentially unstable universal mode solution.
On the other hand, if one inconsistently assumes, as in Ref.~\cite{pueschel2015enhanced}, that $B_0$ is uniform so that $1/L_B=0$ even
though $p_0'\neq 0$,
one recovers a physically spurious instability known in the literature~\cite{pueschel2015enhanced,pueschel2017basic} 
as the Gradient Drift Coupling (GDC)
instability, which at low $\beta$ and $k\to 0$ has the non-vanishing growth rate $\gamma= \sqrt{\beta/2} C_s/|L_p|$.
More generally, the gyrokinetic calculations to follow ({\it e.g.,} section~\ref{sec:gdc}), given the (inconsistent) assumptions $B_0'=0$ and $1/L_p\neq 0$,
obtain {\it unconditional} instability at long wavelength for any $\beta$ with the  $k\to 0$ growth rate 
\begin{equation}\label{eq:gdcgen}
\gamma= \left[\frac{\beta}{2(1+\beta)} \right]^{1/2} \frac{C_s}{|L_p|}\ \ .
\end{equation}
Numerical simulations with this error contain erroneously unstable box-sized modes and
are physically unreliable.

The next section presents the general gyrokinetic dispersion relation with which we
explore the universal mode and derive the GDC instability by neglecting pressure balance.
We have undertaken a more thorough exploration of the universal mode with
$\delta B_\parallel$ using fluid and gyrofluid models, but those studies will be the focus
of a separate publication.

\section{General gyrokinetic dispersion relation} \label{sec:DR}

In the gyrokinetic model the fluctuations are considered to be of the same order as the
ion gyro-radius ($k\sim\rho_i^{-1}$), and one order smaller than the scales of the
equilibrium gradients ($k L\gg 1$, where $L$ is the equilibrium gradient scale length).
For slow frequency modes ($\omega \ll \Omega_i$, where $\Omega_i$ is the ion
gyro-frequency) the fast Larmor motion around the equilbrium magnetic field may be
averaged over to yield a lower-dimensional kinetic
equation~\cite{antonsen1980kinetic,Frieman1982}. This framework has been formally
justified in many parameter regimes and validated in numerous numerical realizations.
For convenience we chose the standard gyrokinetic reduction of the Vlasov 
equation~\cite{antonsen1980kinetic} to derive the general dispersion relation of the
system described in the previous section. The procedure to derive the dispersion relation, 
which is valid for $k\rho_e\ll1$ and any $\beta$, $k\rho_i$,  is described in
appendix~\ref{app:kin}, with the additional aid of the supplemental material. In this
section we also investigate the analytically tractable limits of low and high $\beta$ with
$k\rho_i\ll 1$.

In this work we use the following definitions of the plasma kinetic to magnetic pressure ratio and thermal speed of species $\alpha$:
\begin{equation} \label{eq:defs1}
\beta_\alpha=8\pi p_{\alpha0}/B_0^2\ , \quad v_{t\alpha}^2={2T_{\alpha 0}}/{m_\alpha}\ ,
\end{equation}
while the ion sound gyro-radius, the ratio of density to temperature gradient scale length and the ratio of the temperatures are:
\begin{equation}
\rho_\alpha=v_{t\alpha}/\Omega_{c\alpha}\ , \quad \eta_\alpha=\frac{n_0T_{\alpha 0}'}{n_0'T_{\alpha 0}}\ ,\quad \tau_e=\frac{T_{e0}}{T_{i0}}=\tau_i^{-1}\ .
\end{equation}
For simplicity and physical insight we utilize the following normalization
\begin{equation} \label{eq:defs2}
\omega=\omega_{phys}L_n/v_{ti}\ ,\quad k=k_{phys}\rho_i\ , \quad v=v_{\perp\alpha}/v_{t\alpha}\ .
\end{equation}
The equilibrium condition discussed above can be written in terms of the
parameter $\alpha_0$ as follows:
\begin{equation} \label{eq:defs3}
\frac{L_n}{L_B}= -\frac{\beta}{2}\frac{L_n}{L_p}=-\frac{\beta_i\alpha_0}{2}\ ,\quad
\alpha_0=1+\tau_e+\eta_i+\eta_e\tau_e\ .
\end{equation}
And throughout this paper we employ the following notation for various modified frequencies:
\begin{equation} \label{eq:defs4}
\bar{\omega}_{i}=\omega-(k/2)\left[1+\eta_i\left(v^2-1\right)\right] \ ,\quad \bar{\omega}_e=\omega+(\tau_e k/2)\left[1+\eta_e\left(v^2-1\right)\right]\ ,
\end{equation}
\begin{equation} \label{eq:defs5}
\omega_{bi}=\omega-(k/2)(L_n/L_B)v^2 \ ,\quad \omega_{be}=\omega+(\tau_e k/2)(L_n/L_B)v^2\ .
\end{equation}
By using this notation, the dispersion relation (Eq.~(\ref{eq:dr0})) without finite Larmour radius (FLR) corrections for electrons may be written as
\begin{equation} \label{eq:gendr}
I_{\phi Q} I_{BA}=(2/\beta_i)\left(I_{\phi A}\right)^2\ ,
\end{equation}
where
\begin{equation}
I_{\phi Q}=2\int_0^\infty dv v e^{-v^2}\left[
J_0^2\frac{\bar{\omega}_{i}}{\omega_{bi}}-1+\tau_i \left(\frac{\bar{\omega}_e}{\omega_{be}}-1\right) 
\right]\ ,
\label{eq:IphiQ}
\end{equation}
\begin{equation}
I_{BA}=1+2\int_0^\infty dv v^3 e^{-v^2}\left( J_1^2\frac{2\beta_i \bar{\omega}_i}{k^2\omega_{bi}}+v^2\frac{\beta_e \bar{\omega}_e}{2\omega_{be}}\right)\ ,
\label{eq:IBA}
\end{equation}
\begin{equation}
I_{\phi A}=\beta_i \int_0^\infty dv v^2 e^{-v^2}\left( - 2J_0J_1\frac{ \bar{\omega}_i}{k\omega_{bi}} + v\frac{\bar{\omega}_e}{\omega_{be}}\right)\ ,
\label{eq:IphiA}
\end{equation}
and $J_0(kv)$ and $J_1(kv)$ are Bessel functions of the first kind.

At this stage the pressure balance condition linking $1/L_p$ and $1/L_B$ as
$1/L_B=-\beta/(2L_p)$ has yet to be enforced. The scale length $1/L_B$ arising from the $\nabla B$ drift (see appendix~\ref{app:kin}) 
appears exclusively in the denominators associated with $\omega_{bi,e}$.
As we shall show in the following sections, the {\it unstable} universal modes of interest here (regarding $\eta_i$, $\eta_e$, $T_{i0}/T_{e0}$ as order unity parameters) 
have phase speeds that scale as $(\omega/k)_{phys}\sim \sqrt{\beta}v_d$, 
$v_d\sim cp_0'/(n_0eB_0)$ or in normalized units $u\sim \sqrt{\beta}$.
Thus at low $\beta$, one might expect the corrections due to finite $1/L_B$ in $\omega_{bi,e}$  to be negligible since (for $v\sim 1$) they
are smaller than $\omega$ by a factor
of $\sqrt{\beta}$. But as we show in the next section, this is not the case. Due to cancellations of the leading terms in the dispersion relation 
the $1/L_B$ terms are of the {\it same order} as those arising from $1/L_p$, and are essential to the long-wavelength behavior of the growth rate: discarding them
leads to the spurious GDC instability, while properly retaining them produces universal mode growth rates that vanish for $k\to 0$, as one would
expect of non-MHD instabilities. The same is true at high $\beta\gg 1$ though in that case the $1/L_B$ terms are $\sqrt{\beta}$ {\it larger} than $\omega$
in $\omega_{bi,e}$ and are thus the leading order terms, the neglect of which would be patently incorrect. 

\begin{figure}[h]
\includegraphics[width=0.8\linewidth]{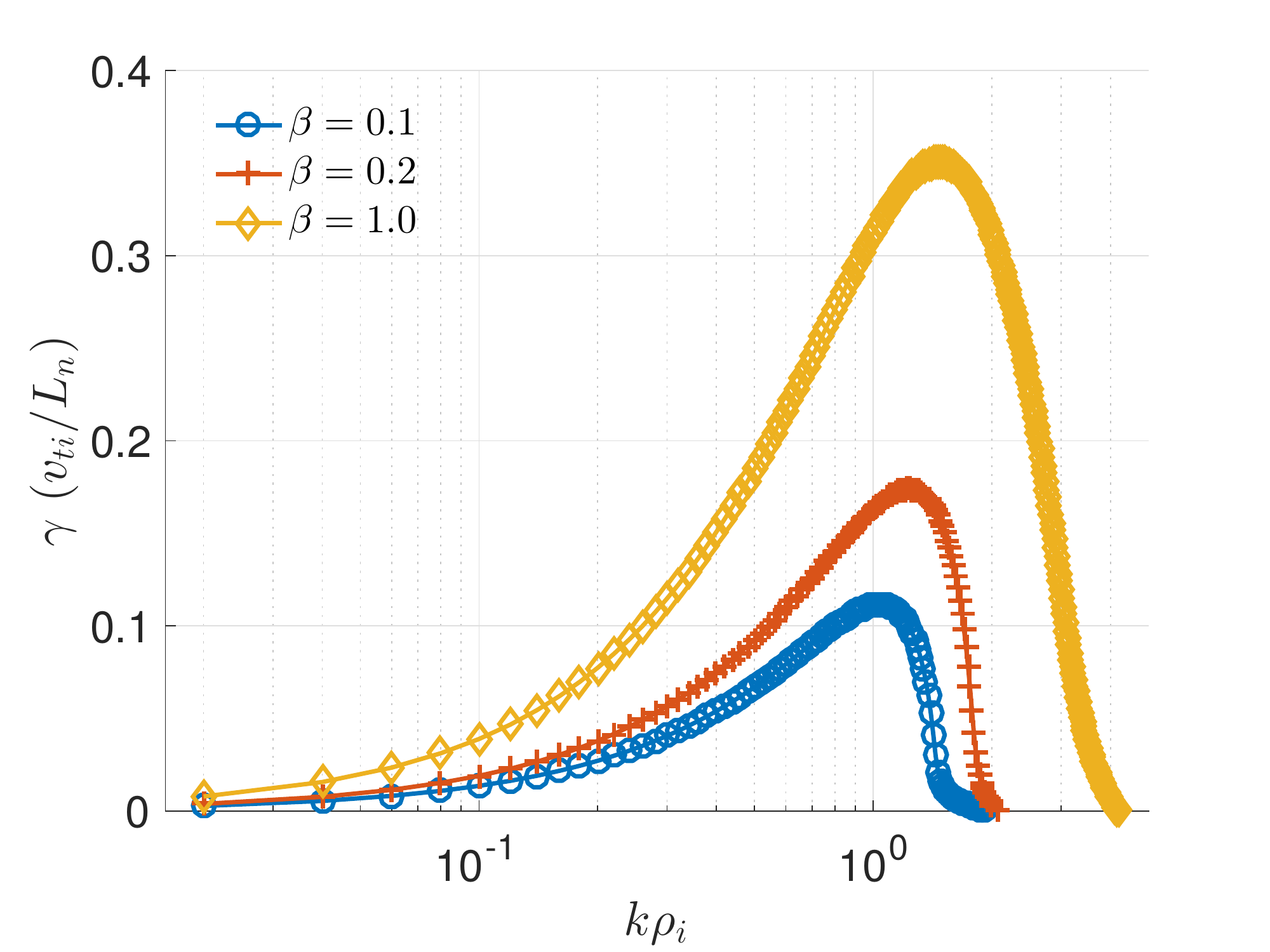}%
\caption{Linear growth rate of the slab universal mode for $\tau_e=1$, $\eta_i=-1$, $\eta_e=2$. \label{fig:entropy}}%
\end{figure}

Without further simplifications Eq.~(\ref{eq:gendr}) is difficult to solve analytically but a numerical solution is feasible as shown in Fig.~\ref{fig:entropy}.
In general, the growth rates of the universal mode (correctly enforcing equilibrium pressure balance, Eq.~(\ref{eq:pbeq})) vanish in the long wavelength $k_\perp\to 0$ limit and peak near $k_\perp\rho_i\sim 1$. We confirmed these properties also in the high and low $\tau_i$ limits (not shown here). Note that the electrostatic study of universal slab modes at sub-Larmor scales~\cite{Landreman2015PRL} showed such modes to peak significantly smaller wavelengths. The stabilization for large $k_\perp$ is due to FLR effects, as gyromotion averages over very small fluctuations. The linear growth rate scaling $\gamma\sim \sqrt{\beta}$ to be shown later leads to stronger
instability at higher $\beta$, as seen in the figure.
The figure also illustrates an unusual property of the universal mode that is further explored in the sections to follow: instability {\it requires}
that either $\eta_i$, $\eta_e$ or both be negative. That is, the density gradient must point in the opposite direction of either the electron or the ion temperature gradient.

\subsection{Low $\beta\ll 1$ limit with $k\rho_i\ll 1$} \label{sec:lowbeta}
Noticing $L_n/L_B\sim O(\beta)$ and expanding $\omega_{bi}$ and $\omega_{be}$ for $\beta\ll 1$ to $O(\beta^2)$ as
\begin{equation}
\frac{1}{\omega_{bi}}=\frac{1}{\omega}\left[1+\frac{k}{2\omega}\frac{L_n}{L_B} v^2+\frac{k^2}{4\omega^2}\left(\frac{L_n}{L_B}\right)^2 v^4+\dots
\right]\ ,
\label{eq:ombi2}
\end{equation}
\begin{equation}
\frac{1}{\omega_{be}}=\frac{1}{\omega}\left[1-\frac{k \tau_e}{2\omega}\frac{L_n}{L_B} v^2+\frac{k^2\tau_e^2}{4\omega^2}\left(\frac{L_n}{L_B}\right)^2 v^4+\dots
\right]\, 
\label{eq:ombe2}
\end{equation}
hence,
\begin{equation}
\begin{split}
\frac{\bar{\omega}_i}{\omega_{bi}}&=1-\frac{k}{2\omega}+\frac{k\eta_i}{2\omega} 
-\frac{k}{2\omega} \left[ \eta_i - \frac{L_n}{L_B}\left( 1-\frac{k}{2\omega}+\frac{k\eta_i}{2\omega} \right) \right]v^2 \\
&
-\frac{k^2}{4\omega^2}\frac{L_n}{L_B}
\left[ \eta_i -\frac{L_n}{L_B} \left( 1-\frac{k}{2\omega}+\frac{k\eta_i}{2\omega} \right) \right] v^4
-\frac{k^3\eta_i}{8\omega^3}\left(\frac{L_n}{L_B}\right)^2 v^6\ ,
\end{split}
\end{equation}
\begin{equation}
\begin{split}
\frac{\bar{\omega}_e}{\omega_{be}}&=1+\frac{\tau_ek}{2\omega}-\frac{\tau_e k \eta_e}{2\omega} 
+\frac{\tau_e k}{2\omega} \left[ \eta_e - \frac{L_n}{L_B}\left( 1+\frac{\tau_e k}{2\omega}-\frac{\tau_e k\eta_e}{2\omega} \right) \right]v^2 \\
&-\frac{\tau_e^2k^2}{4\omega^2}\frac{L_n}{L_B}
\left[ \eta_e -\frac{L_n}{L_B} \left( 1+\frac{\tau_e k}{2\omega}-\frac{\tau_e k\eta_e}{2\omega} \right) \right] v^4
+\frac{\tau_e^3 k^3\eta_e}{8\omega^3}\left(\frac{L_n}{L_B}\right)^2 v^6
\end{split}
\end{equation}
one obtains
\begin{equation}
\begin{split} 
I_{\phi Q} & \simeq   -\frac{k^2}{2} 
\Bigg\{
1-\frac{k}{2\omega}
(1+\eta_i)
+\frac{L_n}{L_B}
\Bigg[
\frac{k}{\omega}-\frac{k^2}{2\omega^2}(1+2\eta_i)+\frac{1}{2\omega^2}(1+\tau_e+\eta_i+\tau_e\eta_e)
\Bigg]
 \\
&  +\frac{1}{\omega^2}
\left( \frac{L_n}{L_B}\right)^2 
\Bigg[
\frac{3}{2}k^2 
\left(
1-\frac{k}{2\omega}(1+3\eta_i)
\right)
-(1+\tau_e)+\frac{k}{2\omega}(1-\tau_e^2+2\eta_i-2\eta_e\tau_e^2)
\Bigg]
\Bigg\}\ ,
\end{split}
\label{eq:IphiQ2}
\end{equation}
\begin{equation}
\begin{split}
I_{BA}= 1+\beta_i
\Bigg\{
1+\tau_e&+\frac{k}{2\omega}
\left[
-(1+2\eta_i)+(1+2\eta_e)\tau_e^2
\right]
\\
&+3\frac{L_n}{L_B}
\left[
\frac{k}{2\omega}(1-\tau_e^2)-\frac{k^2}{4\omega^2}\left(
(1+3\eta_i)+(1+3\eta_e)\tau_e^3\right)
\right]
\Bigg\}\ ,
\end{split}
\label{eq:IBA2}
\end{equation}
\begin{equation}
\begin{split}
\frac{1}{\beta_i} I_{\phi A} & =\frac{3k^2}{8}
\left[
1-\frac{k}{2\omega}(1+2\eta_i)+\frac{L_n}{L_B}
\left( \frac{3k}{2\omega} -\frac{3k^2}{4\omega^2}(1+3\eta_i)\right)
\right] \\
&+\frac{k}{4\omega}(1+\tau_e+\eta_i+\tau_e \eta_e)+\frac{L_n}{L_B}
\left[ -\frac{k}{2\omega}(1+\tau_e)+\frac{k^2}{4\omega^2}\left(1-\tau_e^2+2\eta_i-2\eta_e\tau_e^2\right)\right]\ .
\end{split}
\label{eq:IphiA2}
\end{equation}

First, if one considers the $k\to 0$ limit and (erroneously) assumes 
$\omega\to const$ for $k\to 0$, then to $O(\beta^2)$, Eqs.~(\ref{eq:IphiQ2})-(\ref{eq:IphiA2}) become
\begin{equation}
\begin{split}
I_{\phi Q}I_{BA}&=\frac{-k^2}{2\omega^2}\left\lbrace \left(1+\beta_i+\beta_i\tau_e\right)\omega^2 
+\frac{L_n}{L_B} \left[
\frac{\alpha}{2} -\frac{L_n}{L_B}\left(1+\tau_e\right) 
+\frac{\beta_i\alpha_0}{2}\left(1+\tau_e\right) 
\right] \right\rbrace\ ,\\
\frac{2}{\beta_i} I_{\phi A}^2 &= \frac{\beta_i k^2}{2\omega^2}\left[\frac{\alpha_0^2}{4}-\frac{\alpha_0 L_n}{L_B}\left( 1+\tau_e \right) \right]\ .
\end{split}
\end{equation}
Hence the dispersion relation, Eq.~(\ref{eq:gendr}), gives
\begin{equation}
\label{eq:gdclimit}
\omega^2=-\frac{\beta_i}{4}\alpha_0^2-\frac{L_n}{2L_B}\alpha_0+\frac{\beta\alpha_0}{4}
\left(\beta_i\alpha_0+4\frac{L_n}{L_B}\right)+(1+\tau_e)\left(\frac{L_n}{L_B}\right)^2+O(\beta^3)\ .
\end{equation}
Now, if one disregards the total pressure balancing condition $L_n/L_B=-\beta_i\alpha_0/2$ and takes (for example) $1/L_B=0$,
then the leading order in $\beta$ result is
$\omega^2=-\beta_i\alpha_0^2/4$
which is the GDC mode growth rate discussed earlier at low $\beta$.
On the other hand, upon substitution of $L_n/L_B=-\beta_i\alpha_0/2$, the right-hand side vanishes identically up to $O(\beta^2)$.

Continuing under the assumption of equilibrium pressure balance  $L_n/L_B=-\beta_i\alpha_0/2$,
substituting the expressions for $I_{\phi Q}$, etc into the dispersion relation and 
retaining the leading terms to $O(\beta)$ assuming $\omega\propto k$ yields:
\begin{equation}
\begin{split} 
I_{\phi Q} & \simeq   -\frac{k^2}{2} 
\left[
1-\frac{\alpha_1}{u}
-\frac{\beta_i\alpha_0}{2}
\left(
\frac{2}{u}-\frac{2\alpha_2}{u^2}+\frac{\alpha_0}{2\omega^2}
\right) \right] \ ,
 \\
I_{BA}&= 1+\beta_i\left(1+\tau_e-\frac{\alpha_3}{u}\right)\ ,
\\
I_{\phi A} &=\beta_i \left[\frac{3k^2}{8} \left( 1-\frac{\alpha_2}{u}\right)+\frac{\alpha_0}{2u}\right]\ .
\end{split}
\end{equation}
Here $u$ is the normalized phase velocity
\begin{equation}
\label{eq:phaseu}
u=\frac{2\omega}{k}=\frac{\omega_{phys}}{k_{phys} v_{ni}}\ ,\ \ v_{ni}=\frac{cn_0'T_{i0}}{n_0eB_0}
\end{equation}
and 
\begin{equation}
\label{eq:defsdr2}
\alpha_0  =1+\tau_e+\eta_i+\tau_e\eta_e\ ,\quad \alpha_1=1+\eta_i ,\quad \alpha_2=1+2\eta_i\ , \quad \alpha_3=1-\tau_e^2+2\left(\eta_i-\eta_e\tau_e^2\right).
\end{equation}
Eq.~(\ref{eq:gendr}) in the long-wavelength limit ($k\ll 0$) to $O(\beta)$ then becomes:
\begin{equation}
\label{eq:lowbetaquad}
a_2u^2+a_1u+a_0=0.
\end{equation}
with
\begin{equation}
\label{eq:defsdr3}
a_2=1+\beta_i\left(1+\tau_e\right)\ , \quad 
a_1=-\alpha_1+\beta_i\left[\alpha_0/2-\left(1+\tau_e\right)\alpha_1-\alpha_3\right]\ ,\quad 
a_0=\beta_i\left(\alpha_1\alpha_3-\alpha_0\alpha_2/2\right).
\end{equation}
First, if the finite $\beta$ terms are neglected then $a_0\to 0$, $a_1\to -\alpha_1$, $a_2\to 1$, and one obtains the solutions $u=0$ and
\begin{equation}\label{eq:beta0}
u=1+\eta_i\quad {\rm or} \quad \left(\frac{\omega}{k}\right)_{phys}=v_{di}\ \ ,\ \ v_{di}=\frac{cp'_{i0}}{n_0eB_0}\ ,
\end{equation}
indicating system is always stable to $O(\beta^0)$.
At finite $\beta$, the leading-order universal mode root may be obtained by solving the quadratic dispersion relation Eq.~(\ref{eq:lowbetaquad}), yielding the instability condition 
\begin{equation}
\Delta_a=a_1^2-4a_0a_2<0
\label{eq:inst1}
\end{equation}
and the growth rate
\begin{equation}
\label{eq:lowbetasol}
\gamma=\text{Im}(\omega)=\text{Im}(uk/2)=|\Delta_a|^{1/2}k/(4a_2).
\end{equation}
The scaling $u\sim \sqrt{\beta}$ for instability arises from Eq.~(\ref{eq:inst1}), which shows that instability is only possible for $a_1^2 \sim a_0 a_2\sim \beta$.
As a result all three terms in Eq.~(\ref{eq:lowbetaquad}) are $O(\beta)$ and Eq.~(\ref{eq:lowbetasol}) yields $\gamma/k\sim \sqrt{\beta}$.

\begin{figure}[h]
\includegraphics[width=0.8\linewidth]{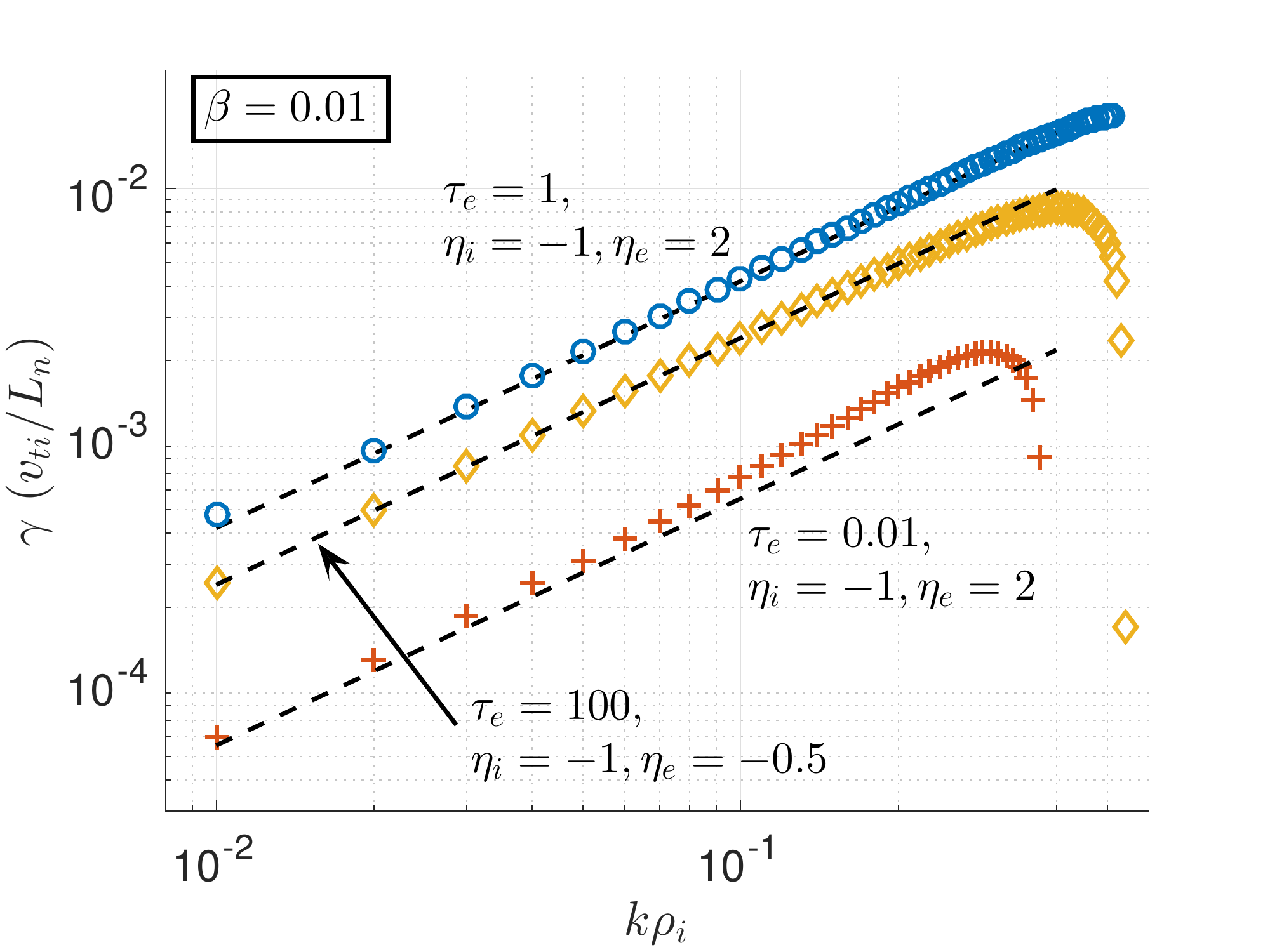}%
\caption{Comparison of numerical (symbols) and analytic (dashed lines) linear growth rates at $\beta=0.01$ for $\tau_e=T_e/T_i=1$ (blue circles), cold electrons ($\tau_e=0.01$, red crosses) and cold ions ($\tau_e=100$, yellow diamonds). \label{fig:lbcomp}}%
\end{figure}

\begin{figure}[h]
\includegraphics[width=0.8\linewidth]{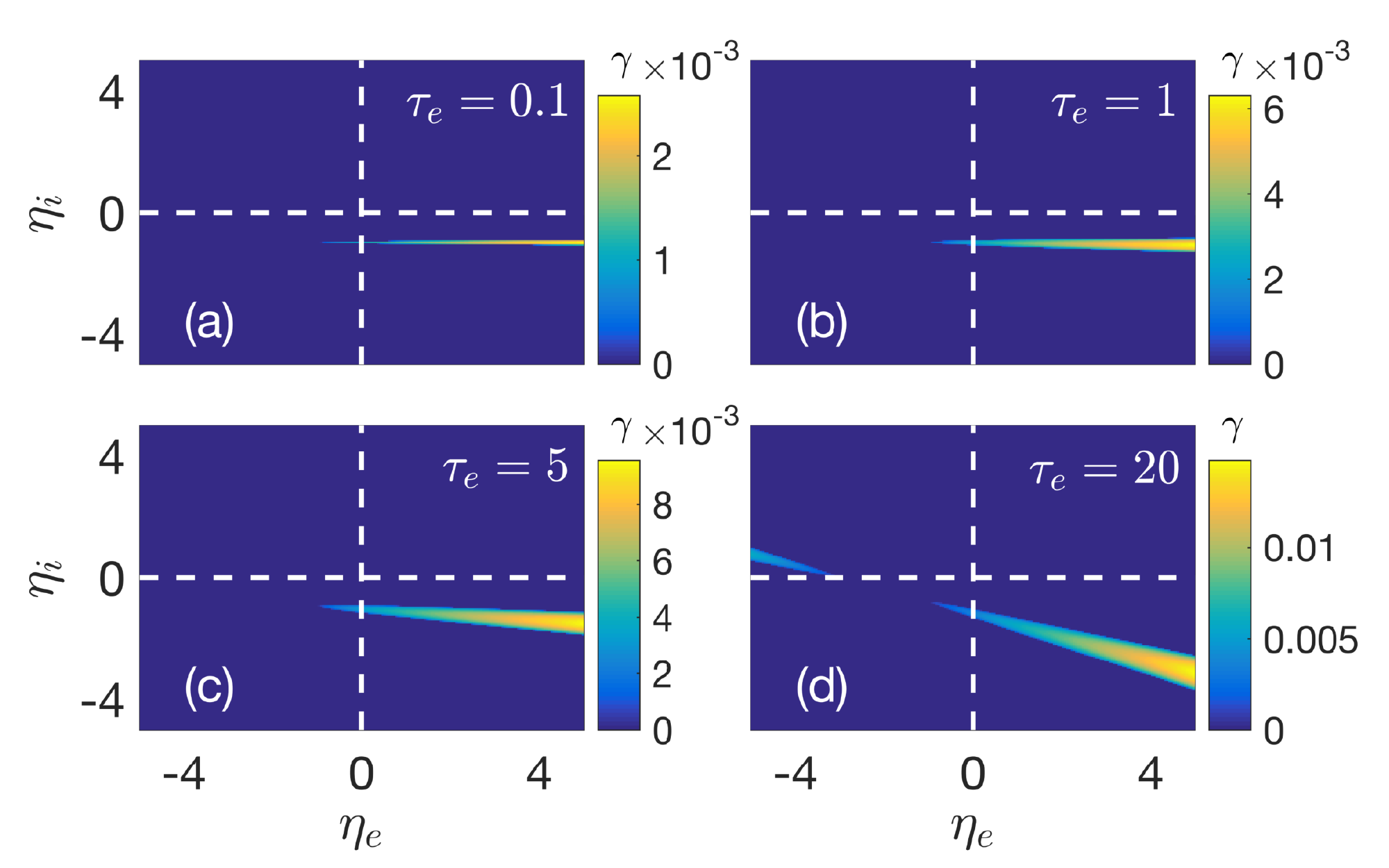}%
\caption{Universal mode linear growth rate as a function of $\eta_i$, $\eta_e$ and $\tau_e$ for $k\rho_i=0.1$,  $\beta=0.01$. \label{fig:lb1}}%
\end{figure}

\begin{figure}[h]
\includegraphics[width=0.8\linewidth]{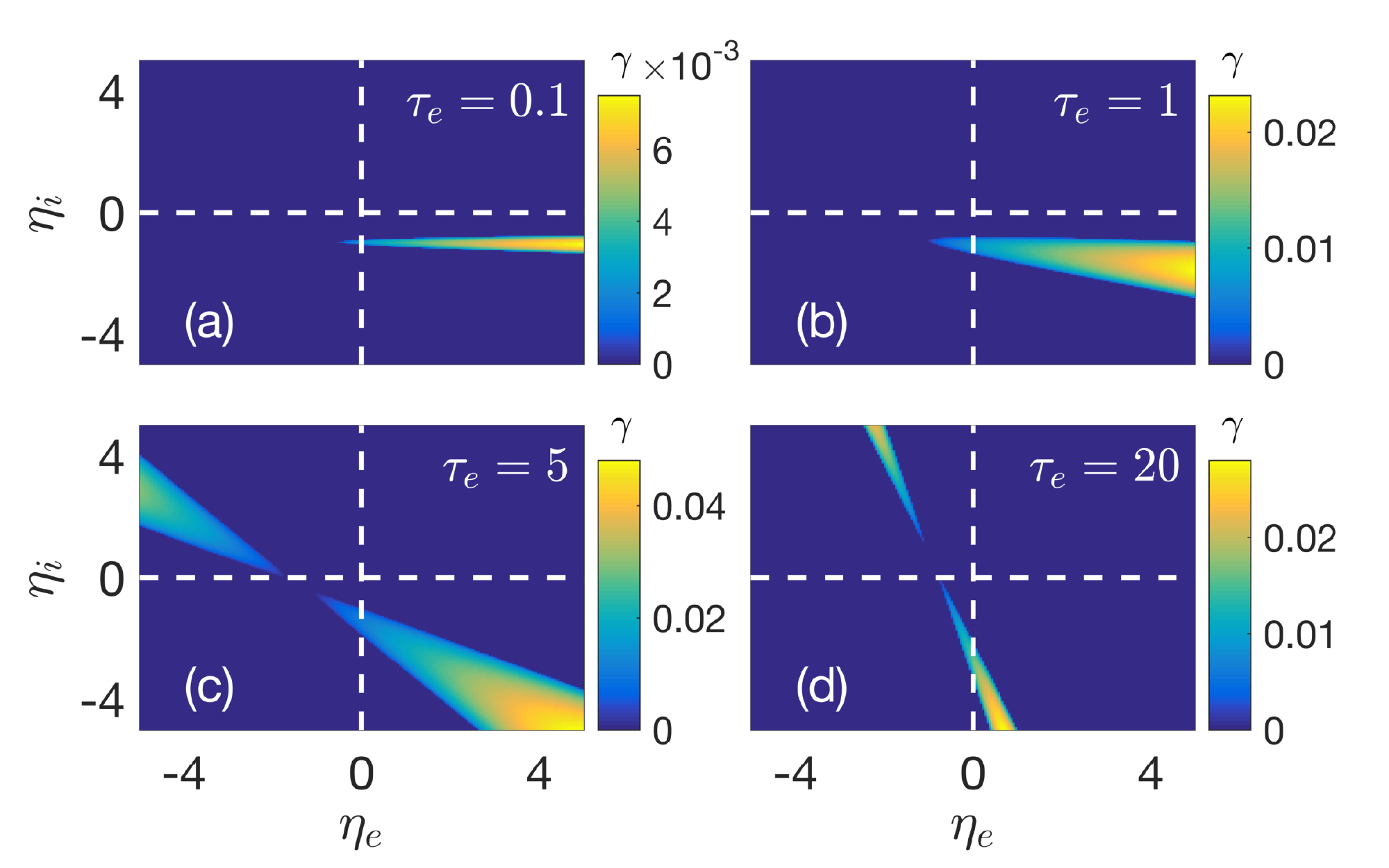}%
\caption{Universal mode linear growth rate versus $\eta_i$, $\eta_e$ and $\tau_e$ for $k\rho_i=0.1$, $\beta=0.1$. \label{fig:lb2}}%
\end{figure}

Fig.~\ref{fig:lbcomp} shows that the analytic expression for the low $\beta$, long wavelength linear growth rate (\ref{eq:lowbetasol}) agrees well with the numerical solution of the full dispersion relation (\ref{eq:gendr}) for both small and large $T_e/ T_i$. The normalized linear growth rate $\gamma$ as a function of $\eta_i$, $\eta_e$ and $\tau_e$ for $k\rho_i=0.1$, $\beta=0.01$ and $\beta=0.1$ are shown in Figs.~\ref{fig:lb1} and~\ref{fig:lb2}. 
These parameter scans, like those at higher $\beta$ described in the next section, indicate the universal mode is always stable for $\eta_i,\eta_e>0$. This condition can also be proved algebraically, stating that this slab universal mode only exists linearly in regions where the density gradient opposes the electron or the ion temperature gradient. However, in other geometries~\cite{Krall1965,Ricci2006} it may be possible to excite a universal or an entropy instability for $\eta_e=\eta_i>0$. It is pertinent to mention that this quadratic approximation becomes less accurate for extreme values of $\tau_e$ if a large $\kp\rho_i$ value is used.


The instability condition (\ref{eq:inst1}) can be further simplified by choosing a particular value of $\tau_e=T_e/T_i$. 
If one assumes $\tau_e=1$, for example, the instability condition becomes:
\begin{equation}
\Delta_a=\alpha_1^2+\frac{\beta}{2}\left[ \alpha_1^2+\alpha_1\left(5+5\eta_e+2\eta_i\right)+2\eta_i\left( 1+\eta_e\right)\right]+O(\beta^2) <0.
\end{equation}
Therefore, this universal mode is unstable only when $\alpha_1\simeq 0$ ($\eta_i\simeq -1$) and $\eta_e>-1$ with, to leading order,
\begin{equation}
\gamma\simeq \sqrt{\left( 1+\eta_e \right)\beta}k/4.
\end{equation}


\subsection{High $\beta\gg 1$ limit with $k\rho_i\ll 1$} \label{sec:highBeta}
Considering the $\beta\gg 1$ limit we expand $\omega_{bi}$ and $\omega_{be}$ as
\begin{equation}
\frac{1}{\omega_{bi}}=-\frac{1}{d_i}-\frac{\omega}{d_i^2}+\dots \ ,\quad d_i=\frac{k}{2}\frac{L_n}{L_B}v^2
\label{eq:ombi3}
\end{equation}
\begin{equation}
\frac{1}{\omega_{be}}=\frac{1}{d_e}-\frac{\omega}{d_e^2}+\dots \ ,\quad d_e=\frac{k\tau_e}{2}\frac{L_n}{L_B}v^2
\label{eq:ombe3}
\end{equation}
hence,
\begin{equation}
\begin{split}
\frac{\bar{\omega}_i}{\omega_{bi}}&=\frac{L_B}{L_n}\left[\eta_i-\left(u-1+\eta_i\right)v^{-2}\right]+\left( \frac{L_B}{L_n} \right)^2 uv^{-2}\left[\eta_i-\left(u-1+\eta_i\right)v^{-2}\right] \ , \\
\frac{\bar{\omega}_e}{\omega_{be}}&=\frac{L_B}{L_n}\frac{1}{\tau_e}\left[\tau_e\eta_e+\left(u+\tau_e-\tau_e\eta_e\right)v^{-2}\right]-\left( \frac{L_B}{L_n} \right)^2 \frac{uv^{-2}}{\tau_e^2}\left[\tau_e\eta_e+\left(u+\tau_e-\tau_e\eta_e\right)v^{-2}\right] \ ,
\end{split}
\end{equation}
which turns the integrals in the dispersion relation into
\begin{equation}
\label{eq:IphiQ3}
I_{\phi Q}=-\tilde{\tau}-\frac{k^2}{\beta_i\alpha_0}\left( u-1 \right)+R_{\phi Q}\ ,
\end{equation}
\begin{equation}
\label{eq:IBA3}
I_{BA}= -\frac{k^2}{2\alpha_0}\left(u-\alpha_2\right) - \frac{2u}{\beta_i\alpha_0^2}\left[\tilde{\tau}u-\frac{k^2}{4}\left(u-\alpha_1\right)\right]\ ,
\end{equation}
\begin{equation}
I_{\phi A}= -\frac{1}{\alpha_0}\left[u\tilde{\tau} -\frac{3}{8}k^2\left(u-\alpha_1\right)\right] \ ,
\label{eq:IphiA3}
\end{equation}
Here, $\tilde{\tau}=1+\tau_i$. In the expression for $I_{\phi Q}$ the residue $R_{\phi Q}$ comes from, to $O\left(\beta^{-1}\right)$, the multiplication of the lowest order term in the expansion of $J_0^2$ and $\bar{\omega}_i/\omega_{bi}$, and the lowest order terms in $\bar{\omega}_e/\omega_{be}$. On can integrate these terms to yield
\begin{equation}
\begin{split}
R_{\phi Q}& \simeq -\frac{2}{\beta_i\alpha_0}\bigg\{ \eta_i+\tau_i \eta_e +\left(u-1+\eta_i\right) \left[\gamma_{EM}+\ln\left( \frac{2u}{\beta_i\alpha_0}\right) \right] \\
&-\tau_i^2\left(u+\tau_e-\tau_e\eta_e\right) \left[\gamma_{EM}+\ln\left( -\frac{2\tau_i u}{\beta_i\alpha_0}\right)\right] \bigg\}\ ,
\end{split}
\end{equation}
where $\gamma_{EM}\simeq0.57721...$ is the Euler-Mascheroni constant.
Substituting Eq. (\ref{eq:IphiQ3})-(\ref{eq:IphiA3}) into the general dispersion relation (\ref{eq:gendr}) yields a transcendental dispersion relation. 
If the small $R_{\phi Q}$ term is neglected\footnote{At first it may appear that $R_{\phi Q}$, being $O\left(\beta^{-1}\right)$, ought to be retained. However, for the parameters we have explored the term $\left(\eta_i+\tau_i\eta_e\right)$ is to lowest order cancelled by the rest of $R_{\phi Q}$, such that neglecting this residue appears justified.}, then to $O(\beta^{-1})$ a simple quadratic dispersion is obtained:
\begin{equation}\label{eq:highbetaquad}
b_2u^2+b_1u+b_0=0
\end{equation}
with
\begin{equation}
b_2=2\ , \quad
b_1=-2\alpha_1+\beta_i\alpha_0\ , \quad
b_0=-\beta_i\alpha_0\alpha_2.
\end{equation}
Similar to the low $\beta$ limit, instability occurs when 
\begin{equation}
\Delta_b=b_1^2-4b_0b_2<0
\label{eq:inst2}
\end{equation}
with 
\begin{equation}
\label{eq:highbetasol}
\gamma=\text{Im}(\omega)=\text{Im}(uk/2)=|\Delta_b|^{1/2}k/8.
\end{equation}
As before, the scaling $u\sim \sqrt{\beta}$ for instability arises from Eq.~(\ref{eq:inst2}), which shows that instability is only possible for $b_1^2 \sim b_0 b_2\sim \beta$.
As a result all three terms in Eq.~(\ref{eq:highbetaquad}) are $O(\beta)$ and Eq.~(\ref{eq:highbetasol}) yields $\gamma/k\sim \sqrt{\beta}$.

\begin{figure}[h]
\includegraphics[width=0.8\linewidth]{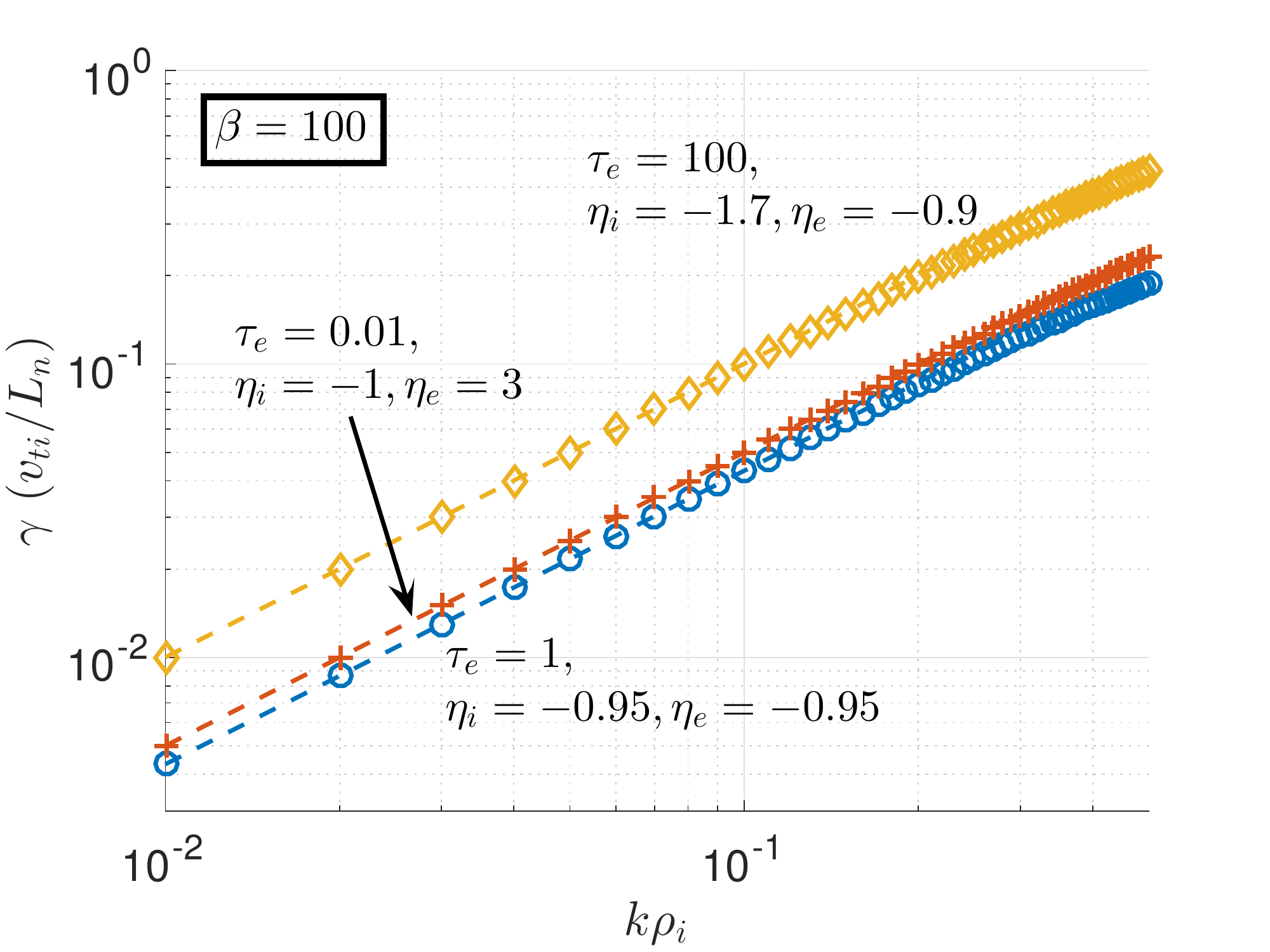}%
\caption{Numerical (symbols) and analytic (dashed lines) linear growth rates at $\beta=100$ for $\tau_e=T_e/T_i=1$ (blue circles), cold electrons ($\tau_e=0.01$, red crosses) and cold ions ($\tau_e=100$, yellow diamonds) \label{fig:hbcomp}}%
\end{figure}

\begin{figure}[h]
\includegraphics[width=0.8\linewidth]{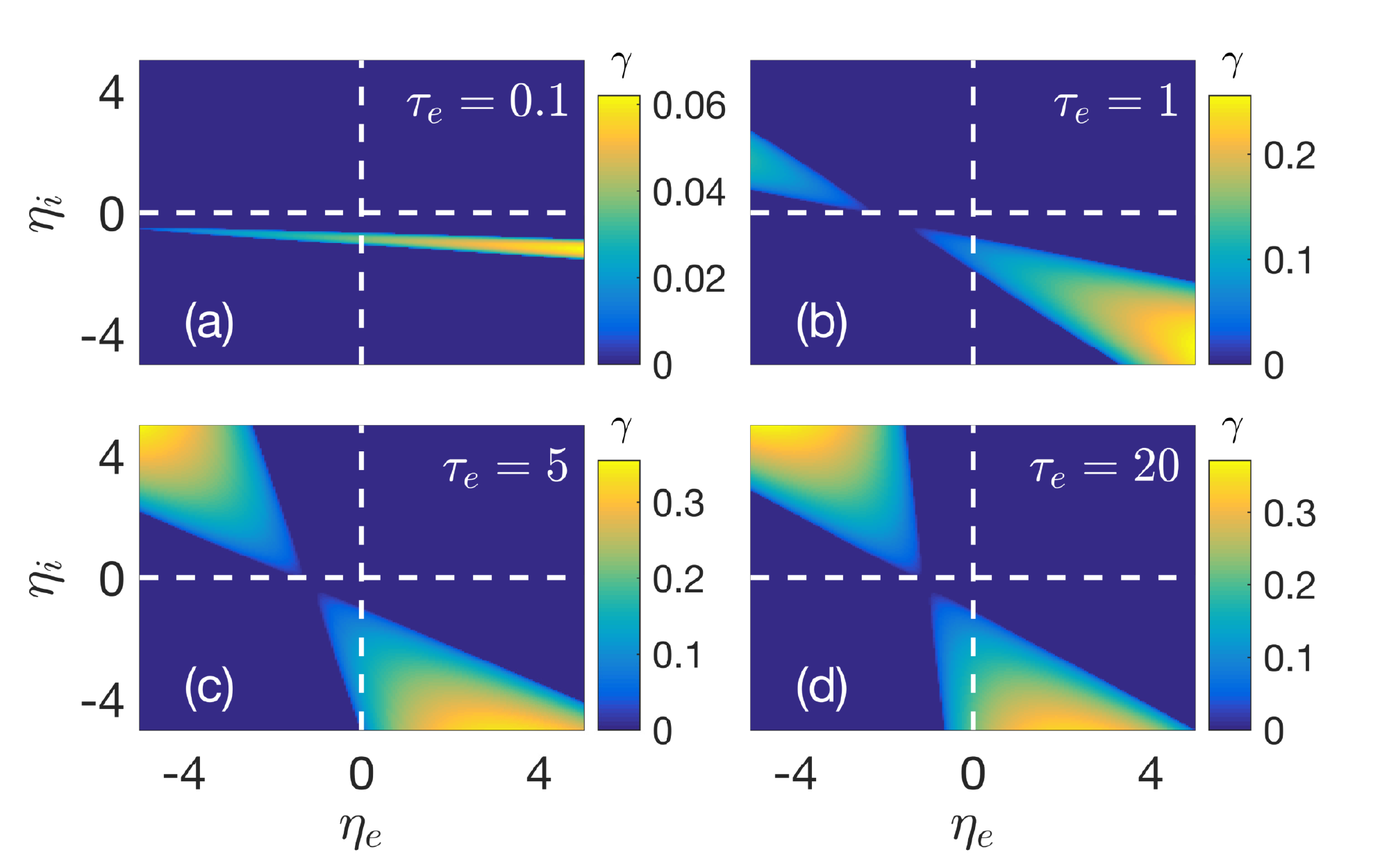}%
\caption{Universal mode linear growth rates versus $\eta_i$, $\eta_e$ and $\tau_e$ for $k\rho_i=0.1$, $\beta=10$. \label{fig:hb1}}%
\end{figure}
\begin{figure}[h]
\includegraphics[width=0.8\linewidth]{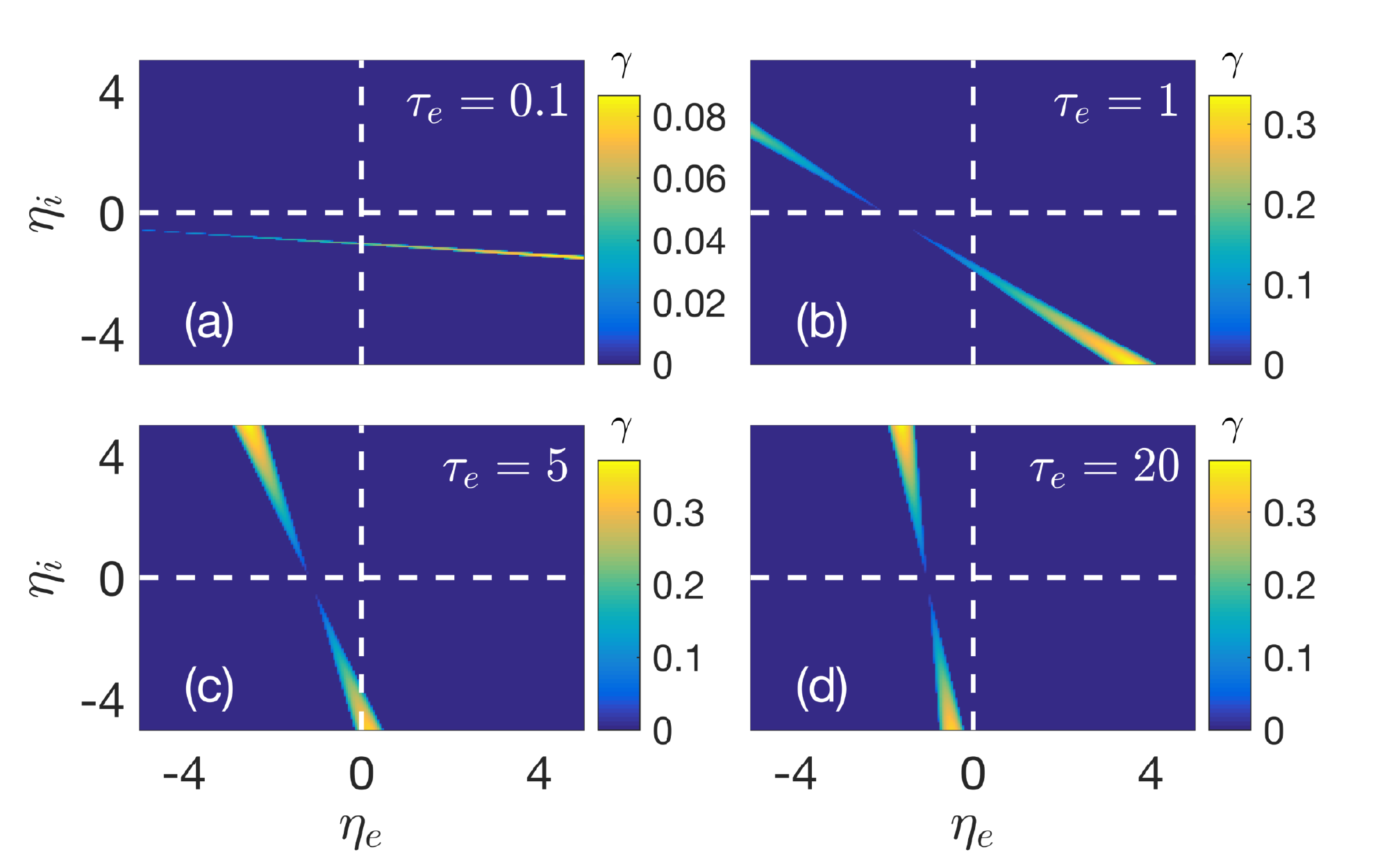}%
\caption{Universal mode linear growth rates versus $\eta_i$, $\eta_e$ and $\tau_e$ for $k\rho_i=0.1$, $\beta=100$. \label{fig:hb2}}%
\end{figure}

Fig.~\ref{fig:hbcomp} shows that Eq.~(\ref{eq:highbetasol}) is in good agreement with the general dispersion relation (\ref{eq:gendr}) in the low $k$ limit for three sets of temperature ratios and $\eta_{i,e}$ values. 
The dependence of the growth rate on the $\eta_i$ and $\eta_e$ ratios is illustrated in Figs.~\ref{fig:hb1} and~\ref{fig:hb2} from the cold electron ($\tau_e\ll 1$) to the cold ion ($\tau_e\gg 1$) limit for $k\rho_i=0.1$, $\beta=10$ and $\beta=100$. 
Again, these results and a simple algebraic proof (not shown here) demonstrate that, in the high $\beta$, long wavelength limit, a necessary condition for this slab universal mode instability is that at least one of 
$\eta_i$ and $\eta_e$ be negative.


\section{The specious origin of the GDC instability}\label{sec:gdc}

\begin{figure}[!htb]
\includegraphics[width=0.6\linewidth]{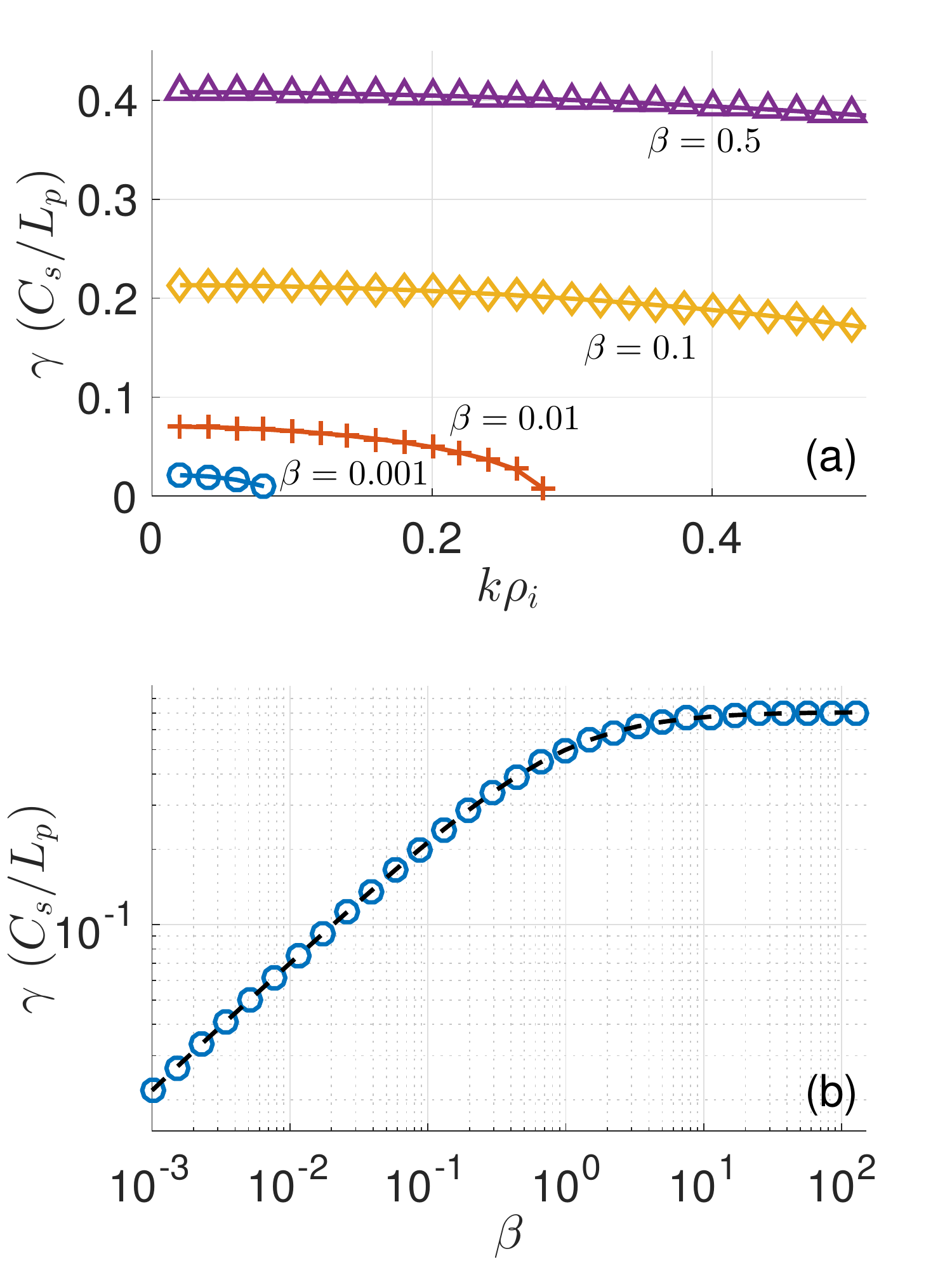}%
\caption{Reproduction of the GDC instabilitiy with $\tau_e=1,~\eta_i=\eta_e=0$ and $B_0'=0$: (a) linear growth rate spectra for different $\beta$; (b) $\beta$ scaling at $k\rho_i=0.02$ from general dispersion relelation Eq.~\ref{eq:dr0} (blue circles) and low $k$ estimation Eq.~\ref{eq:gdclowk} (black dashed line). Similar figures were reported as Figs. 3 and 4 in Ref.~\cite{pueschel2015enhanced}. \label{fig:gdc}}%
\end{figure}

In Ref.~\cite{pueschel2015enhanced} a new gradient-drift coupling (GDC) instability is
discussed as existing in a shearless slab with a density or a temperature gradient, or
both. No parallel variation is allowed, nor are perpendicular magnetic fluctuations
($A_\parallel=0$). These appear as key ingredients for a universal instability. An analysis of turbulence in the Large Plasma Device (LAPD) was found to exhibit signs of GDC
activity. However, the initial work on the GDC instability ignored the equilibrium pressure balance relation (\ref{eq:LB}) by assuming $1/L_B=0$, thus generating this unphysical mode. In the derivation presented here the GDC is obtained by taking $1/L_B=0$ such that $\omega_{bi}=\omega_{be}=\omega$. By doing so we can reproduce the GDC growth rates described in Ref.~\cite{pueschel2015enhanced}, as shown in Fig.~\ref{fig:gdc}.

\begin{figure}[!htb]
\includegraphics[width=0.6\linewidth]{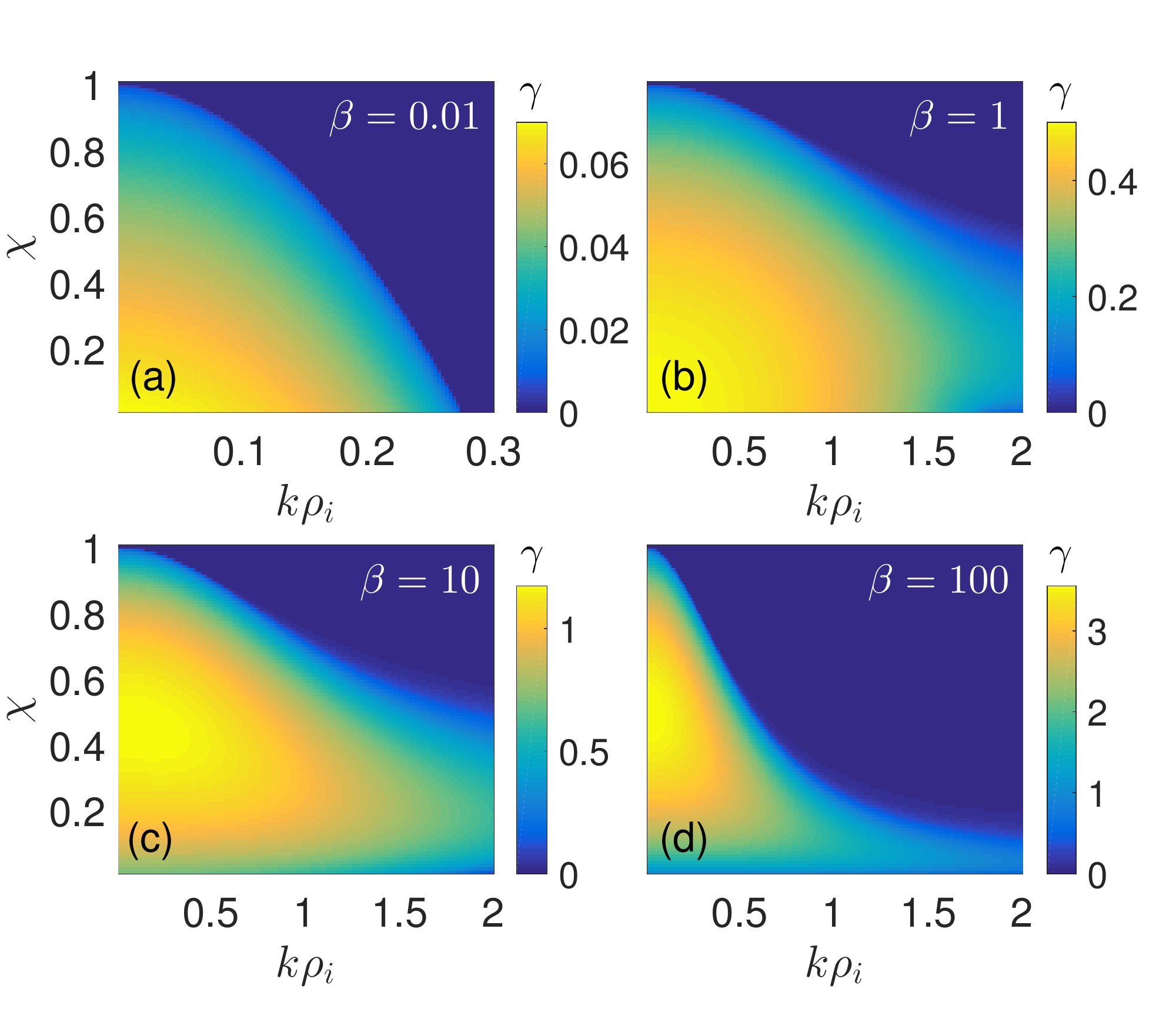}%
\caption{Vanishing of the GDC mode growth rate as $1/L_B$ is increased from zero ($\chi=0$) to its correct value $-\beta/(2L_p)$ ($\chi=1$) at various $\beta$ and $k\rho_i$ for $T_e=T_i$, $\eta_i=\eta_e=0$. \label{fig:gdc_scan}}%
\end{figure}

\begin{figure}[!htb]
\includegraphics[width=0.6\linewidth]{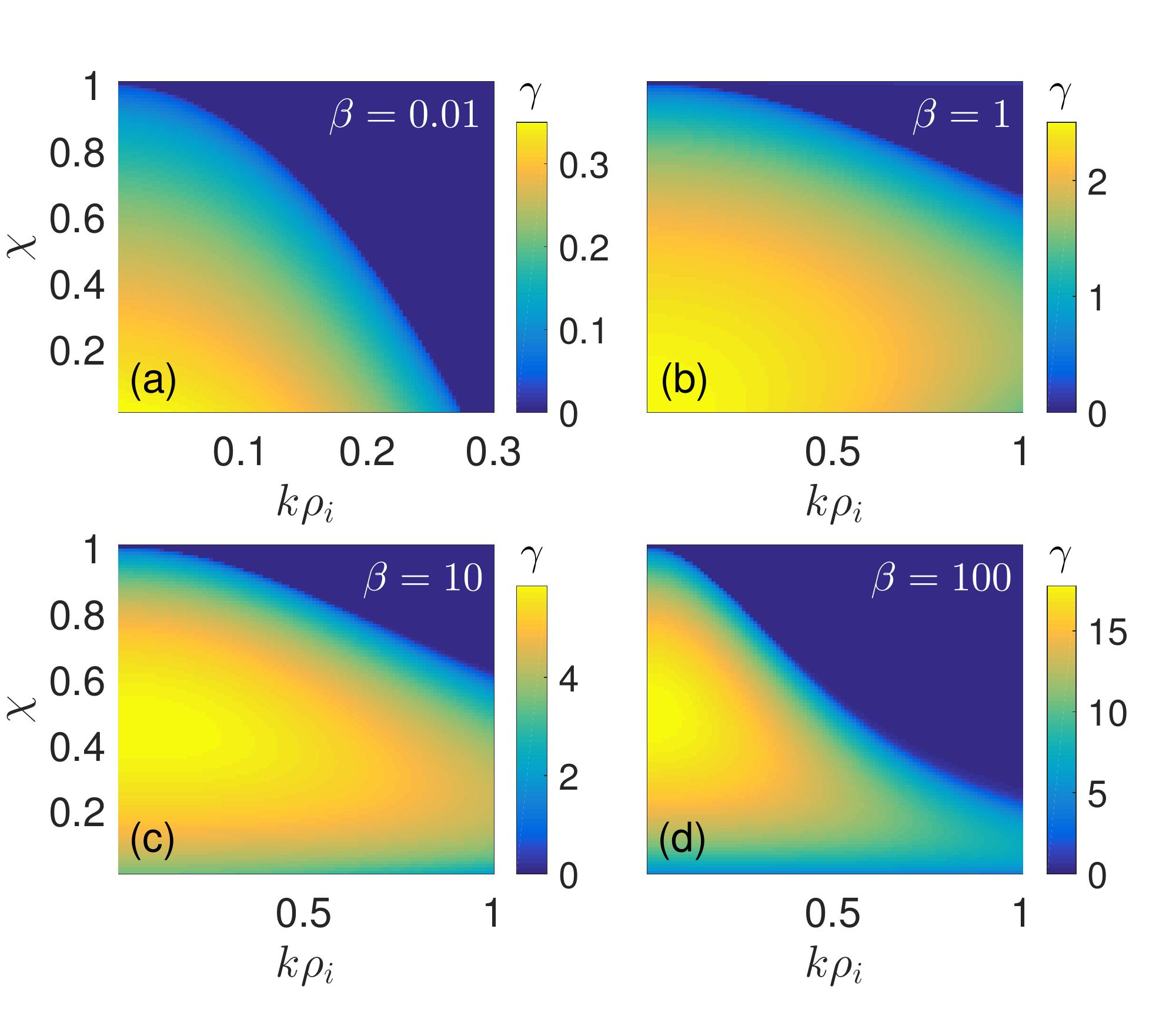}%
\caption{Vanishing of the GDC mode growth rate as $1/L_B$ is increased from zero ($\chi=0$) to its correct value $-\beta/(2L_p)$ ($\chi=1$) at various $\beta$ and $k\rho_i$ for $T_e=T_i$, $\eta_i=\eta_e=4$. \label{fig:gdc_chi}}%
\end{figure}

In order to more gradually demonstrate the effect of violating this pressure balance,
Figs.~\ref{fig:gdc_scan} and \ref{fig:gdc_chi} show numerical solutions of the full
dispersion relation with a factor of $\chi$ inserted in front of $1/L_B=-\beta/(2L_p)$.
This $\chi$ factor is alowed to vary from 0 to 1, illustrating in two representative cases
the shut-off of the GDC mode as one increases $\chi$ from 0 (the GDC limit) to 1 (the
physically correct value). As $\chi$ reaches 1 the growth rate vanishes for all $k_\perp$
and all $\beta$. 

\begin{figure}[!htb]
\includegraphics[width=0.6\linewidth]{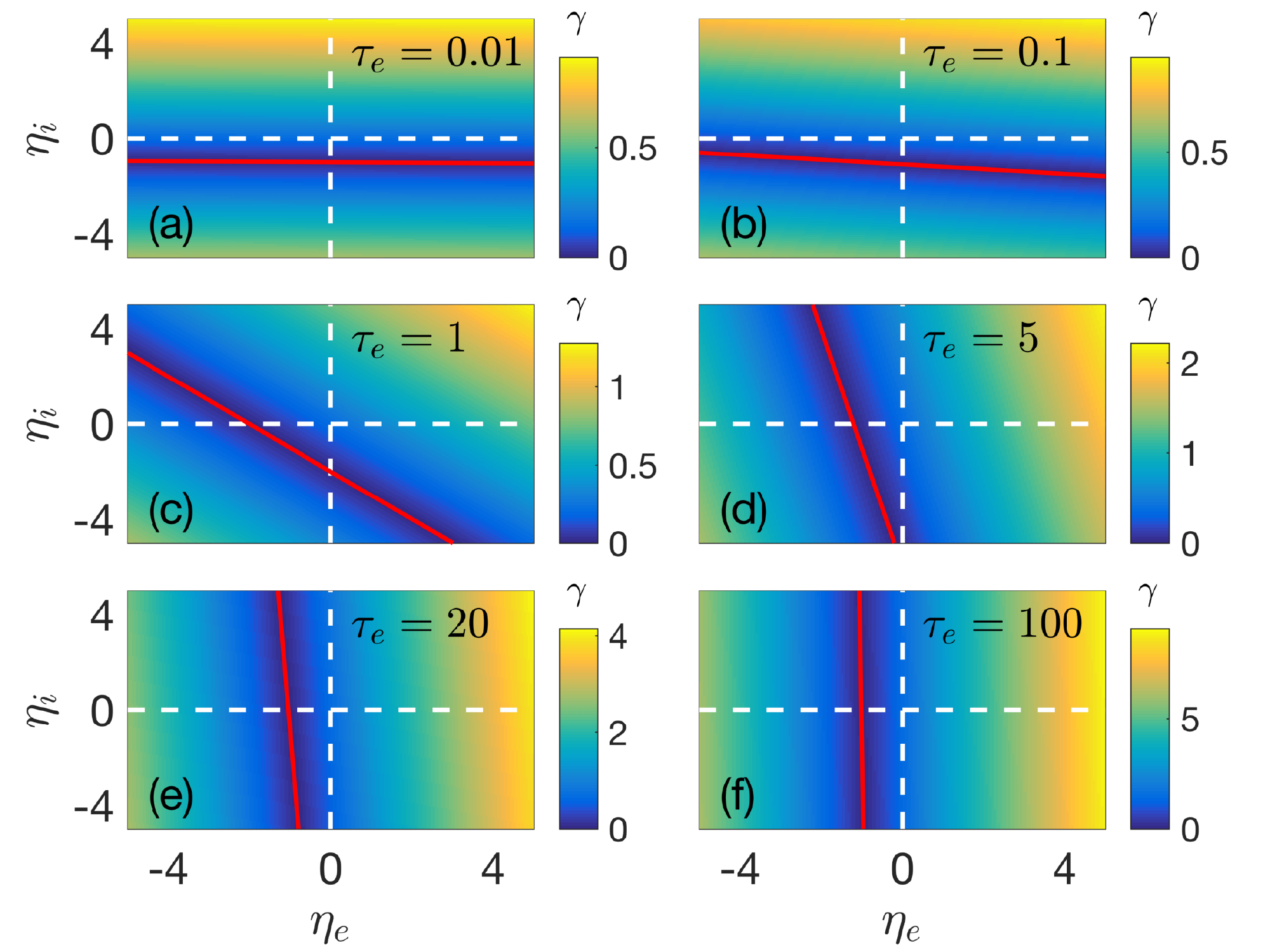}%
\caption{GDC instability linear growth rate $\gamma$ as a function of $\eta_i$ and $\eta_e$ from cold electron ($\tau_e\ll 1$) to cold ion limit $(\tau_e\gg 1$) for $k\rho_i=0.02,~\beta=0.1$. Red line denotes $\alpha_0=0$. \label{fig:gdc_lb}}%
\end{figure}

Finally we consider the analytically tractable limit of $k\rho_i\ll 1$ with the GDC assumption $1/L_B=0$ such that $\omega_{bi}=\omega_{be}=\omega$. With the small-argument expansions of the Bessel functions (\ref{eq:jsmall}), Eqs.~(\ref{eq:IphiQ})-(\ref{eq:IphiA}) become
\begin{equation}\label{eq:gdcI}
\begin{split}
I_{\phi Q} &=  -\frac{k^2}{2} \left(1-\frac{\alpha_1}{u}\right)\ ,
\\
I_{BA} &= 1+\beta_i\left(1+\tau_e-\frac{\alpha_3}{u}\right)\ ,
\\
I_{\phi A} &= \beta_i \left[\frac{3k^2}{8} \left( 1-\frac{\alpha_2} {u}\right)+\frac{\alpha_0}{2u}\right]\ ,
\end{split}
\end{equation}
where $u,~\alpha_{0,1,2,3}$ are defined in (\ref{eq:phaseu}) and (\ref{eq:defsdr2}) respectively.
Substituting Eqs.~(\ref{eq:gdcI}) into Eq.~(\ref{eq:dr0}) yields the dispersion relation of the GDC instability at low $k$ for arbitrary $\beta$:
\begin{equation}
c_2u^2+c_1u+c_0=0
\end{equation}
where
\begin{equation}
\begin{split}
c_2&=\left(1+\beta \right)k^2+\frac{9}{16}\beta_ik^4\ , \\
c_1&=\left[ \frac{3}{2}\beta_i\alpha_0 - \alpha_1\left(1+\beta\right) -\beta_i \alpha_3\right]k^2 -\frac{9}{8}\beta_i\alpha_2k^4\ , \\
c_0&=\beta_i\alpha_0^2+\left( \alpha_1\alpha_3-\frac{3}{2}\alpha_0\alpha_2\right)\beta_ik^2+\frac{9}{16}\beta_i\alpha_2^2k^4\ .
\end{split}
\end{equation}
The instability condition is $\Delta_{GDC}=c_1^2-4c_2c_0< 0$ with 
\begin{equation}\label{eq:gdclowk}
\gamma=\text{Im}(\omega)=\text{Im}(uk/2)=|\Delta_{GDC}|^{1/2}k/(4c_2),
\end{equation}
where in this case the discriminant is
\begin{equation}
\begin{split}
\Delta_{GDC}&=-4\beta_i\left( 1+\beta\right)\alpha_0^2k^2 
\\
&\quad+\left\lbrace \left[\left(1+\beta\right)\alpha_1-\beta_i\alpha_3\right]^2+\beta_i\alpha_0\left[3\left(1+\beta\right)\left(2\alpha_2-\alpha_1\right)-3\beta_i\alpha_3\right]\right\rbrace k^4+O(k^6)\ .
\end{split}
\end{equation}
To $O(k^2)$, $\Delta_{GDC}\simeq -4\beta_i\left( 1+\beta\right)\alpha_0^2k^2 < 0$ if $\alpha_0$ is finite. This means in the long wavelength limit the GDC instability is unconditionally unstable 
(Fig.~\ref{fig:gdc_lb}, and the structure of the high $\beta$ case is identical albeit with different magnitudes)
unless $\alpha_0=0$ in which case the $O(k^4)$ stability condition becomes $\Delta_{GDC}=\left[\left(1+\beta\right)\alpha_1-\beta_i\alpha_3\right]^2k^4\geq 0$. Physically, $\alpha_0=0$ implies an isobaric plasma, i.e., $1/L_p=0$. More intuitively speaking, the phenomenological explanation offered in Ref.~\cite{pueschel2015enhanced} (see paragraph below figure 2 therein) disregards the necessary equilibrium gradient in $B_0$ responsible for a $\nabla B$ drift that helps break the positive feedback loop behind the instability.

\section{Conclusion}\label{sec:conc}
We have explored the gyrokinetic stability of electromagnetic ($\delta B_\parallel$) ``universal" or entropy modes driven by density and temperature gradients in slab geometry. As is typical of such modes, the instability conditions sometimes exhibit a complicated dependence on the density and temperature gradients. We have shown that the failure to respect the total
pressure/force balance of the equilibrium in low frequency theories that order-out compressional magnetosonic waves 
can lead to fictitious instability, such as the gradient drift coupling (GDC) mode in the gyrokinetic literature.
Fortunately for studies which violate equilibrium force balance, this fictitious mode can sometimes be stabilized by magnetic shear~\cite{pueschel2015enhanced}.

There is no mathematical or physical relationship between the GDC and universal modes described here. The former is universally and spuriously unstable at arbitrarily long
wavelengths in all parameter regimes with $1/L_p\neq 0$ due to order-unity mathematical errors. This will produce unphysical behavior
in nonlinear turbulence simulations, for example, which tend to be dominated by the longest-wavelength unstable modes in the system. In contrast, the universal
modes are small-scale instabilities with growth rates that vanish for $k\to 0$ and are unstable only in constrained parameter regions 
($i.e.$ negative $\eta_i$ and/or $\eta_e$). We conclude that when the underlying mathematical errors are corrected the GDC mode does
not continue to exist in the theory in any recognizable form.

These universal instabilities are interesting and potentially existent in a variety of space
and laboratory plasmas. The non-linear saturation mechanism of these modes is another
unresolved issue of interest that ought to be explored in the future, just as is quantifying
the particle and energy transport that these modes are responsible for. Such studies will
be carried out in the future. In the meantime we have also pursued the analysis of this
electromagnetic universal mode system with fluid and gyrofluid models, results that are
reported in a separate publication.

\begin{acknowledgments}
This work is supported by DOE-SC-0010508. This research used resources of the Discovery cluster supported by the Research Computing Group at Dartmouth College.
\end{acknowledgments}

\appendix

\section{Derivation of the gyrokinetic dispersion relation}\label{app:kin}
In this section we derive the gyrokinetic dispersion relation discussed
in the previous sections, starting from the Vlasov equation for species $\alpha=i,e$ with $q_i=e=-q_e$. 
Our derivation, presented here for the convenience of the reader, is equivalent to the linear stability calculations presented
in Ref.~\cite{antonsen1980kinetic}, for example, and numerous other sources.

We assume a magnetic field purely in the $\hat{z}$ direction and an equilibrium state that depends only on $x$, such that
\begin{equation} \label{eq:b}
\v{B}=\left(B_0(x)+\tilde{B}\right)\v{\hat{z}}
\end{equation}
where the perturbed magnetic field $\tilde{B}\propto e^{-i\omega t+iky}$. The electric current is related to the magnetic field by Ampere's law
\begin{equation}\label{eq:amp0}
\v{J}=\frac{c}{4\pi}\nabla\times \v{B}\to \v{J}_{0}=J_0\v{\hat{y}} \text{ with } J_{0}=-\frac{c}{4\pi}B_0'
\end{equation}
where we denote $x$-derivatives by primes: $d(\dots)/dx=(\dots)'$.
This current must be carried by a drift velocity consistent with the force balance relation
\begin{equation}\label{eq:driftv}
\begin{split}
n q_\alpha \left(\v{E}+\v{v}_\alpha \times \v{B}/c\right)=\nabla p_\alpha \to  
\v{v}_\alpha=\left(v_E+v_{d\alpha}\right)\v{\hat{y}} \\ 
\text{ with } \  v_E=c\phi_0'/B_0\ , v_{d\alpha}=c p_{\alpha 0}'/(n_0 q_\alpha B_0)
\end{split}
\end{equation}
where $\phi_0$ is a possible equilibrium electric potential (soon to be dropped) and thus
\begin{equation}\label{eq:jcrossb}
\frac{1}{c}\v{J}_0\times \v{B}_0=\frac{1}{c}J_{0}B_0\v{\hat{x}}= -\frac{B_0' B_0}{4\pi}\v{\hat{x}}
=\frac{n_0 e}{c} (v_{di}-v_{de})\v{\hat{y}}\times \v{B}_0=p_0' \v{\hat{x}},
\end{equation}
or,
\begin{equation}\label{eq:jcrossb2}
p_0'+\frac{B_0B_0'}{4\pi}=0
\end{equation}
which is essentially the total pressure balance relation:
\begin{equation}\label{eq:pbal}
p_{0}+B_0^2/(8\pi)=const \text{ with } p_0=p_{i0}+p_{e0}.
\end{equation}
Eq.~(\ref{eq:jcrossb2}) also implies that
\begin{equation}\label{eq:LB}
L_B^{-1}=-\beta L_p^{-1}/2
\end{equation}
where $L_B^{-1}=B_0'/B_0$, $L_p^{-1}=p_0'/p_0$ and $\beta=8\pi p_0/B_0^2$.

Denoting
\begin{equation}\label{eq:norms1}
\bar{v}=\frac{v}{v_{t\alpha}} \ ,\quad v_{t\alpha}^2=\frac{2T_\alpha}{m_\alpha}\ ,
\end{equation}
\begin{equation}\label{eq:norms2}
\bar{v}_{d\alpha}=\frac{v_{d\alpha}}{v_{t\alpha}}=\frac{v_{t\alpha}}{2L_{p\alpha}\Omega_{\alpha}}\ ,\quad
\Omega_\alpha=\frac{q_\alpha B_0}{m_\alpha c}\ ,\quad
\frac{1}{L_{p\alpha}}=\frac{p_\alpha'}{p_\alpha}
\end{equation}
we proceed with a solution to the linearized Vlasov equation for species 
$\alpha=i,e$ as an expansion in $\epsilon\sim \rho_\alpha/L$,
$\rho_\alpha=v_{t\alpha}/\Omega_{\alpha}$ where $L$ is a typical equilibrium scale length such as $L_n$, $L_p$, etc.
Suppressing the species label $\alpha$ and omitting for simplicity the collision operator and 
an equilibrium electric field, $\v{E}=\v{\tilde E}$:
\begin{equation}\label{eq:vlasov1}
\left\{\partial_t+\v{v}\cdot\nabla+\left[\frac{q}{m}{\tilde{\v{E}}}
+\Omega\left(1+\frac{\tilde{B}}{B_0}\right)
\v{v}\times\v{\hat{z}}\right]\cdot \nabla_v\right\}\left(F_{eq}+ \tilde{f}\right)=0
\end{equation}
where the equilibrium distribution function to $O(\epsilon)$: $F_{eq}=F_0+F_1+\dots$  obeys
\begin{equation}\label{eq:vlasoveq}
\left(\v{v}\cdot\nabla +\Omega\v{v}\times\v{\hat{z}}\cdot \nabla_v\right)\left(F_0+F_1+\dots\right)=0 \ .
\end{equation}
Noting that
\begin{equation}\label{eq:dxi2}
\v{v}=v_x\hat{\v{x}}+v_y\hat{\v{y}}+v_z\hat{\v{z}}\ , \ v_x=v_\perp\cos\xi\ , \ v_y=v_\perp\sin\xi\ ,
\end{equation}
\begin{equation}\label{eq:dxi}
\v{v}\times\v{\hat{z}}\cdot \nabla_v=v_y\frac{\partial}{\partial v_x}-v_x\frac{\partial}{\partial v_y}=-\frac{\partial}{\partial\xi}
\end{equation}
where $\xi$ is the gyro-angle and choosing $F_0$ to be independent of $\xi$ as
\begin{equation}\label{eq:F0}
F_0=\left(\frac{n_0}{\pi^{3/2}v_t^3}\right)e^{-\bar{v}^2}\ \to\  \partial_\xi F_0=0\ ,
\end{equation}
Eq.~(\ref{eq:vlasoveq}) to first order in $\epsilon$ reduces to
\begin{equation}\label{eq:F1}
\Omega \partial_\xi F_1=\v{v}\cdot\nabla F_0 \ \to\ F_1=\frac{v_y}{\Omega L_F}F_0\ ,\quad 
\frac{1}{L_F}=\frac{1}{L_n}+\frac{1}{L_T}\left(\bar{v}^2-\frac{3}{2}\right)\ .
\end{equation}
The perturbation $\tilde{f}$ which is proportional to $\exp(iky-i\omega t)$ obeys
\begin{equation}\label{eq:f}
\left(-i\omega+\v{v}\cdot\nabla+\Omega\v{v}\times\v{\hat{z}}\cdot \nabla_v\right)\tilde{f}+
\left(\frac{q}{m}{\tilde{\v{E}}}
+\Omega\frac{\tilde{B}}{B_0}
\v{v}\times\v{\hat{z}}\right)\cdot \nabla_v\left(F_0+ F_1\right)=0
\end{equation}
where to the required $O(\epsilon)$ with $1/L_{FT}=1/L_F-1/L_T$,
\begin{equation}\label{eq:vecpot}
\tilde{\v{E}}=-\nabla\phi - \frac{1}{c}\partial_t\v{A}=-\frac{m}{q}\left(i k \frac{q\tilde{\phi}}{m}\hat{\v{y}}+
\frac{\omega\Omega}{k}\frac{\tilde{B}}{B_0}\hat{\v{x}}\right)\ ,
\end{equation}
\begin{equation}\label{eq:eeq1}
-i k \frac{q\tilde{\phi}}{m}
\frac{\partial}{\partial {v_y}}
\left(F_0+F_1\right)=i k \tilde{\varphi}\left[
v_y+\frac{v_y^2}{\Omega L_{FT}}-\frac{v_t^2}{2\Omega L_F}
\right]\ ,\quad \tilde{\varphi}=\frac{q\tilde{\phi}}{T}F_0\ ,
\end{equation}
\begin{equation}\label{eq:eeq2}
-\frac{\omega\Omega}{k}\frac{\tilde{B}}{B_0}\frac{\partial}{\partial{v_x}}\left(F_0+F_1\right)=
\frac{2\omega \bar{v}_x\Omega}{k v_t}\tilde{b}+O(\epsilon^2),\quad \tilde{b}=\frac{\tilde{B}}{B_0}F_0\ .
\end{equation}
Defining
\begin{equation}\label{eq:hats}
\tilde{\varphi}=\hat{\varphi}e^{iky-i\omega t}\ ,\  \ \tilde{b}=\hat{b}e^{iky-i\omega t}\ ,\  \ \tilde{f}=\hat{f}e^{iky-i\omega t}
\end{equation}
and expanding $\hat{f}=\hat{f}_0+\hat{f}_1+\dots$, 
Eq.~(\ref{eq:f}) to first order in $\epsilon$ may be written
\begin{equation}\label{eq:f2}
\begin{split}
\left(-\frac{i \omega}{\Omega}+\frac{v_x\partial_x}{\Omega}+\frac{ik v_y}{\Omega}-\frac{\partial}{\partial \xi}\right)\left(\hat{f}_0+ \hat{f}_1\right)+
\frac{ik\hat{\varphi}}{\Omega}\left[
v_y+\frac{v_y^2}{\Omega L_{FT}}-\frac{v_t^2}{2\Omega L_F}\right] \\
+\hat{b}\bar{v}_x\left(\frac{2\omega }{k v_t}-\frac{v_t}{\Omega L_F}\right)=0\ .
\end{split}
\end{equation}
To leading order this becomes
\begin{equation}\label{eq:lead}
\left(\frac{ik v_y}{\Omega}-\frac{\partial}{\partial \xi}\right)\hat{f}_0+\frac{ikv_y}{\Omega}\hat{\varphi}=0\ \ \to \ \ 
\hat{f}_0=-\hat{\varphi}+g e^{-ikv_x/\Omega}
\end{equation}
where $g$ is independent of $\xi$ and remains to be determined. At first order
\begin{equation}\label{eq:first}
\left(-\frac{i \omega}{\Omega}+\frac{v_x\partial_x}{\Omega}\right)\hat{f}_0
+\left(\frac{ik v_y}{\Omega}-\frac{\partial}{\partial \xi}\right)\hat{f}_1+
\frac{ik\hat{\varphi}}{\Omega}\left[
\frac{v_y^2}{\Omega L_{FT}}-\frac{v_t^2}{2\Omega L_F}\right] 
+\hat{b}\bar{v}_x\left(\frac{2\omega}{k v_t}-\frac{v_t}{\Omega L_F}\right)=0\ .
\end{equation}
Multiplying by $e^{ikv_x/\Omega}$, noting
\begin{equation}\label{eq:note}
e^{ikv_x/\Omega} \left(\frac{ik v_y}{\Omega}-\frac{\partial}{\partial \xi}\right)\hat{f}_1=-
\frac{\partial}{\partial\xi}\left(e^{ikv_x/\Omega}\hat{f}_1\right)
\end{equation}
and taking the gyro-angle average, we obtain
\begin{equation}\label{eq:gyro1}
\begin{split}
\int_{-\pi}^\pi\frac{d\xi}{2\pi}e^{ikv_x/\Omega}\Bigg\{
-\frac{i \omega}{\Omega}\left(-\hat{\varphi}+g e^{-ikv_x/\Omega}\right)
+\frac{v_x}{\Omega}\partial_x\left(-\hat{\varphi}+g e^{-ikv_x/\Omega}\right) +\\
\frac{ik\hat{\varphi}}{\Omega}
\left[\frac{v_y^2}{\Omega L_{FT}}-\frac{v_t^2}{2\Omega L_F}\right] 
+\hat{b}\bar{v}_x\left(\frac{2\omega }{k v_t}-\frac{v_t}{\Omega L_F}\right)\Bigg\}=0\ .
\end{split}
\end{equation}
Collecting terms, taking the various derivatives noting the $\partial_x g$ term vanishes upon integration and
defining 
\begin{equation}\label{eq:ombs}
\omega_{b}=\omega -\frac{k v_\perp^2}{2\Omega L_B}\ ,\quad \bar{\omega}=\omega -\frac{k v_{t}^2}{2\Omega L_F}\ ,
\end{equation}
\begin{equation}\label{eq:gyro2}
{i\omega_b}g=\int_{-\pi}^\pi\frac{d\xi}{2\pi}e^{ikv_x/\Omega}\left\lbrace
{\hat{\varphi}}\left[ i\bar{\omega} -\frac{v_x}{L_{FT}}+
\frac{ik}{\Omega}\frac{v_y^2}{L_{FT}}\right] 
+\hat{b}\bar{\omega}\frac{2\Omega \bar{v}_x}{k v_t}\right\rbrace\ .
\end{equation}
The two terms proportional to $1/L_{FT}$ cancel noting the identity (integrating $v_y\partial_\xi \exp(ikv_x/\Omega)$ by parts)
\begin{equation}\label{eq:id}
\int_{-\pi}^\pi\frac{d\xi}{2\pi}e^{ikv_x/\Omega}\frac{ik v_y^2}{\Omega}=
\int_{-\pi}^\pi\frac{d\xi}{2\pi}e^{ikv_x/\Omega}v_x\ ,
\end{equation}
leaving
\begin{equation}\label{eq:geq}
g=\int_{-\pi}^\pi\frac{d\xi}{2\pi}e^{ikv_x/\Omega}\frac{\bar{\omega}}{\omega_b}
\left(\hat{\varphi}-\frac{2i\Omega \bar{v}_x}{k v_t} \hat{b}\right)\ .
\end{equation}
Noting the identities with $v_x=v_\perp\cos\xi$,
\begin{equation}\label{eq:js}
J_0(k v_\perp/\Omega)=\int_{-\pi}^\pi\frac{d\xi}{2\pi}e^{\pm ikv_x/\Omega}\ , \quad iJ_1(k v_\perp/\Omega)=\int_{-\pi}^\pi\frac{d\xi}{2\pi}e^{ikv_x/\Omega}\cos\xi\,
\end{equation}
with the small-agument expansions
\begin{equation}\label{eq:jsmall}
J_0(b)\simeq1-b^2/4+O(b^4)\ , \quad J_1(b)\simeq b/2-b^3/16+O(b^5)\ ,
\end{equation}
one obtains 
\begin{equation}\label{eq:geq2}
g=\frac{\bar{\omega}}{\omega_b}
\left(J_0\hat{\varphi}+\frac{2\Omega \bar{v}_\perp }{k v_t} J_1\hat{b}\right).
\end{equation}
Expanding the exponents and Bessel functions for the electrons to the required order assuming $k\rho_e\ll1$ yields
\begin{equation}\label{eq:fs}
\begin{split}
\bar{f}_{0i}&=\frac{\tilde{f}_{0i}}{F_{0i}}=\left(e^{-ikv_x/\Omega_i}J_0\frac{\bar{\omega}_i}{\omega_{bi}}-1\right)\frac{e\tilde{\phi}}{T_{i0}}+
e^{-ikv_x/\Omega_i}J_1\frac{\bar{\omega}_i}{\omega_{bi}}\frac{2\bar{v}_\perp}{k\rho_i}\frac{\tilde{B}}{B_0}\ , \\
\bar{f}_{0e}&=\frac{\tilde{f}_{0e}}{F_{0e}}=-\left[\left(1-\frac{ikv_x}{\Omega_e}\right)\frac{\bar{\omega}_e}{\omega_{be}}-1\right]\frac{e\tilde{\phi}}{T_{e0}}+
\left(1-\frac{ikv_x}{\Omega_e}\right)\frac{\bar{\omega}_e}{\omega_{be}}\bar{v}_\perp^2\frac{\tilde{B}}{B_0}
\end{split}
\end{equation}
where the terms proportional to $v_x$ in $\bar{f}_{0e}$ will contribute only in Ampere's law due to an additional multiplicative factor of $v_x$. 

\subsection{Ampere's law} \label{sec:Amperes}
According to Ampere's law, the perturbed magnetic field follows
\begin{equation}\label{eq:amp1}
\begin{split}
\frac{\tilde{B}}{B_0}=\frac{(\nabla\times \tilde{\v{B}})_x}{ikB_0} =& -\frac{4\pi i}{ckB_0}\sum_{i,e}\int d^3vqv_x\tilde{f}_0\\
=& -\frac{4\pi i e n_0}{ck B_0\pi^{3/2}}\int d^3\bar{v} e^{-\bar{v}^2}\bar{v}_x\left(v_{ti}\bar{f}_{0i}-v_{te}\bar{f}_{0e}\right)\ .
\end{split}
\end{equation}
Defining
\begin{equation}\label{eq:f1def}
f_{1\alpha}=\frac{i\Omega_\alpha}{kv_x}\left(e^{-ikv_x/\Omega_\alpha}-1+\frac{ikv_x}{\Omega_\alpha}\right)\ \to\ 
e^{-ikv_x/\Omega_\alpha}=1-\frac{ikv_x}{\Omega_\alpha}\left(1+f_{1\alpha}\right)
\end{equation}
where for electrons $f_{1e}=-{ikv_x}/({2\Omega_e})+\dots$ may be neglected, substituting Eq.~(\ref{eq:fs}) and collecting real terms this may be written as
\begin{equation}\label{eq:aamp0}
I_{BA}\frac{\tilde{B}}{B_0}=I_{\phi A}\frac{e\tilde{\phi}}{T_{i0}}
\end{equation}
where
\begin{equation}\label{eq:IBA0}
\begin{split}
I_{BA} & =1+\int \frac{d^3\bar{v}}{\pi^{3/2}}e^{-\bar{v}^2}\bar{v}_x^2\bar{v}_\perp^2
\left[\beta_i \frac{2J_1}{k \rho_i\bar{v}_\perp} (1+f_{1i})\frac{\bar{\omega}_i}{\omega_{bi}}+\beta_e \frac{\bar{\omega}_e}{\omega_{be}} \right], \\
I_{\phi A} & = \beta_i \int \frac{d^3\bar{v}}{\pi^{3/2}}e^{-\bar{v}^2}\bar{v}_x^2
\left[-J_0(1+f_{1i})\frac{\bar{\omega}_i}{\omega_{bi}}+\frac{\bar{\omega}_e}{\omega_{be}}\right]\ .
\end{split}
\end{equation}
Transforming to plane-polar perpendicular velocity components with 
\begin{equation}\label{eq:polar}
\int \frac{d^3\bar{v}}{\pi^{3/2}}=\frac{1}{\pi^{3/2}} \int_{-\infty}^{\infty}d\bar{v}_z \int_{-\infty}^{\infty}d\bar{v}_y  \int_{-\infty}^{\infty}d\bar{v}_x=
2\int _{-\infty}^{\infty}\frac{d\bar{v}_z}{\sqrt{\pi}}\int_0^{\infty}d\bar{v}_\perp \bar{v}_\perp \int_{-\pi}^{\pi} \frac{d\xi}{2\pi}\ \ ,
\end{equation}
using the identity 
\begin{equation}\label{eq:f2i}
\frac{\bar{v}_\perp}{k\rho_i}J_1\left(k\rho_i \bar{v}_\perp\right)=\int_{-\pi}^{\pi} \frac{d\xi}{2\pi}\bar{v}_x^2\left(1+f_{1i}\right)\ ,
\end{equation}
carrying out the $v_z$ integrals with $\bar{v}^2=\bar{v}^2_z+ \bar{v}^2_\perp$ and noting the generalized Gaussian integral
\begin{equation}\label{eq:itable}
\begin{split}
\int _{-\infty}^{\infty}\frac{dx}{\sqrt{\pi}}e^{-x^2}\left(1+a x^2+bx^4+cx^6\right)&=1+\frac{1}{2}a+ \frac{3}{4}b+\frac{15}{8}c\ , \\
\int _{0}^{\infty}dx \ e^{-x^2}\left(x+a x^3+bx^5+cx^7+dx^9\right)&=\frac{1}{2}+\frac{1}{2}a+ b+ 3c +12d\ ,
\end{split}
\end{equation}
Eq.~(\ref{eq:IBA0}) becomes
\begin{equation}
\begin{split}
I_{BA}&=1+2\int_0^\infty d\bar{v}_{\perp} \bar{v}_{\perp}^3 e^{-\bar{v}_{\perp}^2}\left( J_1^2\frac{2\beta_i \bar{\omega}_i}{k^2\rho_i^2\omega_{bi}}+\bar{v}_{\perp}^2\frac{\beta_e \bar{\omega}_e}{2\omega_{be}}\right)\ , \\
I_{\phi A}&=\beta_i \int_0^\infty d\bar{v}_{\perp} \bar{v}_{\perp}^2 e^{-\bar{v}_{\perp}^2}\left( - 2J_0J_1\frac{ \bar{\omega}_i}{k\rho_i \omega_{bi}} + \bar{v}_{\perp}\frac{\bar{\omega}_e}{\omega_{be}}\right)\ .
\end{split}
\end{equation}
Note that the $L_F$ expression (\ref{eq:F1}) embedded in $\bar{\omega}$ is also modified after the $v_z$ integration:
\begin{equation}
\frac{1}{L_F}=\frac{1}{L_n}+\frac{1}{L_T}\left(\bar{v}_\perp^2-1\right).
\end{equation}
Normalizing frequencies to $v_{ti}/L_n$ and lengths to $\rho_i$ (so for example $\omega=\omega_{phys} L_n/v_{ti}$ and $k=k_{phys}\rho_i$) the expressions
for $I_{BA}$, $I_{\phi A}$ become those given earlier in the text with the simplified notation for the integration variable: $\bar{v}_\perp\to v$.

\subsection{Quasineutrality condition $n_i=n_e$} \label{sec:quasi}
The quasineutrality condition to the required order may be written as
\begin{equation}\label{eq:qn1}
\frac{1}{n_0}\int d^3{v}\left(\tilde{f}_{0i}-\tilde{f}_{0e}\right)=\int \frac{d^3\bar{v}}{\pi^{3/2}}e^{-\bar{v}^2}\left(\bar{f}_{0i}-\bar{f}_{0e}\right)=0\ .
\end{equation}
Therefore,
\begin{equation}\label{eq:qn2}
\begin{split}
\int \frac{d^3\bar{v}}{\pi^{3/2}}e^{-\bar{v}^2}\Bigg[ \left(e^{-ikv_x/\Omega_i}J_0\frac{\bar{\omega}_i}{\omega_{bi}}-1\right)\frac{e\tilde{\phi}}{T_{i0}}+
e^{-ikv_x/\Omega_i}\frac{\bar{\omega}_i}{\omega_{bi}}\frac{2J_1}{k\rho_i}\bar{v}_\perp\frac{\tilde{B}}{B_0}\\
+\left(\frac{\bar{\omega}_e}{\omega_{be}}-1\right)\frac{e\tilde{\phi}}{T_{e0}}-
\frac{\bar{\omega}_e}{\omega_{be}}\bar{v}_\perp^2\frac{\tilde{B}}{B_0}
\Bigg]=0\ .
\end{split}
\end{equation}
Collecting terms this becomes
\begin{equation}\label{eq:qn3}
I_{\phi Q}\frac{e\tilde{\phi}}{T_{i0}}=I_{BQ}\frac{\tilde{B}}{B_0}
\end{equation}
with
\begin{equation}\label{eq:iphiq0}
\begin{split}
I_{\phi Q}&=\int \frac{d^3\bar{v}}{\pi^{3/2}}e^{-\bar{v}^2}\left[
e^{-ikv_x/\Omega_i}J_0\frac{\bar{\omega}_i}{\omega_{bi}}-1+\left(\frac{\bar{\omega}_e}{\omega_{be}}-1\right)\frac{T_{i0}}{T_{e0}}\right]\ , \\
I_{B Q}&=\int \frac{d^3\bar{v}}{\pi^{3/2}}e^{-\bar{v}^2}\bar{v}_\perp^2\left(
-e^{-ikv_x/\Omega_i}\frac{2J_1}{k\rho_i\bar{v}_\perp}\frac{\bar{\omega}_i}{\omega_{bi}}+\frac{\bar{\omega}_e}{\omega_{be}}\right)\ .
\end{split}
\end{equation}
Following the same steps as before one obtains
\begin{equation}
\begin{split}
I_{\phi Q}&=2\int_0^\infty d\bar{v}_{\perp} \bar{v}_{\perp} e^{-\bar{v}_{\perp}^2}\left[
J_0^2\frac{\bar{\omega}_{i}}{\omega_{bi}}-1+\left(\frac{\bar{\omega}_e}{\omega_{be}}-1\right)\frac{T_{i0}}{T_{e0}}
\right]\ ,\\
I_{BQ}&=2\int_0^\infty d\bar{v}_{\perp} \bar{v}_{\perp}^2 e^{-\bar{v}_{\perp}^2}\left( - 2J_0J_1\frac{ \bar{\omega}_i}{k\rho_i \omega_{bi}} + \bar{v}_{\perp}\frac{\bar{\omega}_e}{\omega_{be}}\right)=2I_{\phi A}/\beta_i
\end{split}
\end{equation}
with the aid of Eq.~(\ref{eq:js}), implying with Eq.~(\ref{eq:aamp0}) the dispersion relation:
\begin{equation}\label{eq:dr0}
I_{\phi Q} I_{BA}=I_{BQ}I_{\phi A}=(2/\beta_i)\left(I_{\phi A}\right)^2\ .
\end{equation}

\bibliography{rogers18gdc}

\pagebreak
\newgeometry{margin=0.5in}
\begin{center}
{\large Universal/Entropy/GDC supplement} \\
{\small Barrett N. Rogers, Ben Zhu, Manaure Francisquez} \\
{\footnotesize 6127 Wilder Laboratory, Dartmouth College, Hanover, NH, 03755 USA}
\end{center}

This supplement fills some gaps between the equations in the main manuscript. We begin by adding clarification to the equations in section II, followed by the derivation of the dispersion relation in appendix A. Then we treat the special cases discussed in sections IIIA, IIIB and IV. The equation numbers used here are consistent with those used in the main manuscript.

\section*{Equilibrium pressure balance and the problem with GDC}

From the equilibrium condition:
\begin{align*}
p + \frac{1}{8\pi}B^2 &= \text{const} \\
\pO + \pper + \frac{1}{8\pi}\left(\BO+\Bper\right)^2 &= \pO + \frac{1}{8\pi}\BO^2 \\
\pper + \frac{1}{8\pi}\left(2\BO\Bper+\Bper^2\right) &= 0 \\
\pper + \frac{1}{8\pi}\left(2\BO\Bper+\cancelto{\text{\scriptsize small}}{\Bper^2}\right) &= 0
\end{align*}
We obtain the relationship between the pressure and magnetic field fluctuations:
\begin{equation}
\begin{aligned}
\pper &= - \frac{1}{4\pi}\BO\Bper.
\end{aligned}
\end{equation}

Since $\pOp=-B_0\BOp/(4\pi)$, the gradient scale lengths are related by:

\begin{equation}
\begin{aligned}
\frac{\BOp}{B_0} &=\frac{1}{\LB} = -\frac{8\pi\pOp}{2B_0^2} = -\frac{8\pi p_0}{B_0^2} \frac{\pOp}{2p_0} = -\frac{\beta}{2\Lp}.
\end{aligned}
\end{equation}

Using fluid velocities
\begin{equation}
\begin{aligned}
\vvi &= \vvExB+\vvdi+\v{v}_{\text{pol}}=\frac{c}{B}\zhat\times\grad{\phi}+\frac{c}{enB}\zhat\times \grad{p_i}+\frac{1}{\omegaci}\zhat\times\d{\vvExB}{t} \\
&= \frac{c}{B}\left(\zhat\times\grad{\phi}+\frac{1}{en}\zhat\times \grad{p_i}-\frac{i\omega}{\omegaci}\zhat\times\zhat\times\grad{\phi}\right) \\
&= \frac{c}{B}\left(\zhat\times\grad{\phi}+\frac{1}{en}\zhat\times \grad{p_i}+\frac{i\omega}{\omegaci}\grad{\phi}\right),
\end{aligned}
\end{equation}
\begin{equation}
\vve = \vvExB+\vvde= \frac{c}{B}\left(\zhat\times\grad{\phi}-\frac{1}{en}\zhat\times \grad{p_e}\right),
\end{equation}
current continuity reads:
\begin{equation}
\div{n\left(\vvi-\vve\right)} = i n \frac{\omega c}{\omegaci B_0}\delsq{\phi}+\div{\frac{c}{eB}\zhat\times\grad{p}} = 0,
\end{equation}
Linearize the right side assuming $k\gg d/dx$ for perturbations:
\begin{equation}
\begin{aligned}
-i\frac{c\omega k^2}{B_0} \phi + \frac{B}{n_0m_i}\left[\grad{\frac{1}{B}}\cdot\zhat\times\grad{p}+\frac{1}{B}\div{\zhat\times\grad{p}}\right] &= 0	\\
-i\frac{c\omega k^2}{B_0}\phi + \frac{B}{n_0m_i}\left[\frac{\partialx B}{B^2}\partialy p-\frac{\partialy B}{B^2}\partialx p+\frac{1}{B}\left(-\partialx\partialy p +\partialy\partialx p\right)\right] &= 0	\\
-i\frac{c\omega k^2}{B_0}\phi + \frac{ikB}{n_0m_i}\left(\frac{\BOp}{B_0^2}\pper-\frac{\Bper}{B_0^2}\pOp\right) &= 0	\\
-i\omega k^2\phi + \frac{ik}{cn_0m_i}\left(\BOp\pper-\Bper\pOp\right) &= 0	\\
i\omega k^2\phi + \frac{ik}{cn_0m_i}\left(\Bper\pOp-\BOp\pper\right) &= 0	.
\end{aligned}
\end{equation}

Now using $\Bper=-4\pi\pper/B_0$ from equation 1 and
\begin{equation}
\begin{aligned}
\dot{p}+\frac{c}{B}\zhat\times\grad{\phi}\cdot\grad{p}&=0\\
-i\omega\pper+\frac{c}{B}\zhat\times ik\phi\yhat\cdot\grad{p}&=0 \\
-i\omega\pper-\frac{c}{B}ik\phi p_0'&=0
\end{aligned}
\longrightarrow\pper = -\frac{ck\pOp}{\omega B_0}\phi
\end{equation}
from the fluid pressure equation, equation 5 yields the dispersion relation:
\begin{equation}
\begin{aligned}
i\omega k^2 &+ \frac{ik}{cn_0m_i}\left(\frac{4\pi}{B_0}\pOp+\BOp\right)\frac{ck\pOp}{\omega B_0} = 0 \\
\omega^2 &= - \frac{1}{cn_0m_i}\left(\frac{4\pi}{B_0}\pOp+\BOp\right)\frac{c\pOp}{B_0} \\
\omega^2 &= - \frac{1}{n_0m_i}\left(\frac{4\pi}{B_0}\pOp+\BOp\right)\frac{B_0\beta}{8\pi\Lp} \\
\omega^2 &= - \frac{p_0}{n_0m_i\Lp}\left(\frac{4\pi}{B_0}\pOp+\BOp\right)\frac{B_0\beta}{8\pi p_0} \\
\omega^2 &= -\frac{\cs^2}{\Lp}\left(\frac{\beta}{2\Lp}+\frac{1}{\LB}\right)
\end{aligned}
\end{equation}

\appendix
\section*{Appendix A}
\renewcommand{\theequation}{A\arabic{equation}}

Since
\begin{equation} \tag{A12}
v_x=\vperp\cos\xi, \quad v_y=\vperp\sin\xi, 
\end{equation}
the following operator can be written as:
\begin{equation} \tag{A13}
\begin{aligned}
\vv\times\zhat\cdot\grad{_v} &= v_y\pd{}{v_x}-v_x\pd{}{v_y} = \vperp\sin\xi\pd{\xi}{v_x}\pd{}{\xi}-\vperp\cos\xi\pd{\xi}{v_y}\pd{}{\xi} \\
&= -\vperp\sin\xi\frac{v_y}{v_x^2+v_y^2}\pd{}{\xi}-\vperp\cos\xi\frac{v_x}{v_x^2+v_y^2}\pd{}{\xi} \\
&= -\vperp\sin\xi\frac{\vperp\sin\xi}{\vperp^2}\pd{}{\xi}-\vperp\cos\xi\frac{\vperp\cos\xi}{\vperp^2}\pd{}{\xi} \\
&= -\pd{}{\xi}
\end{aligned}
\end{equation}

The lowest order equilibrium Maxwellian distribution function is
\begin{equation} \tag{A14}
F_0 = \frac{n_0}{\pi^{3/2}\vt^3}e^{-\bar{v}^2} = \frac{n_0}{\pi^{3/2}\vt^3}e^{-\frac{v^2}{\vt^2}},
\end{equation}
such that a spatial derivative of it gives:
\begin{equation} \tag{A15}
\begin{aligned}
\partial_i{F_0} &= \left(\frac{1}{\pi^{3/2}\vt^3}\partial_in_0 + \frac{n_0}{\pi^{3/2}}\partial_i\vt^{-3} - \frac{n_0}{\pi^{3/2}\vt^3}\partial_i\frac{v^2}{\vt^2}\right)e^{-\bar{v}^2} \\
&= \frac{1}{\pi^{3/2}}\left(\frac{1}{\vt^3}\partial_in_0 - n_03\vt^{-4}\partial_i\vt + \frac{2v^2}{\vt^3}\frac{n_0}{\vt^3}\partial_i\vt\right)e^{-\bar{v}^2} \\
&= \frac{n_0}{\pi^{3/2}\vt^3}\left(\frac{\partial_in_0}{n_0} -\frac{3}{\vt}\partial_i\vt + \frac{2\bar{v}^2}{\vt}\partial_i\vt\right)e^{-\bar{v}^2} \\
&= \left[\frac{1}{L_n} +\left(\frac{m}{2T}\right)^{1/2}\left(\frac{2}{mT}\right)^{1/2}\partial_i T\left(\bar{v}^2-\frac{3}{2}\right)\right]F_0 \\
&= \left[\frac{1}{L_n}+\frac{1}{L_T}\left(\vbar^2-\frac{3}{2}\right)\right]F_0=\frac{1}{\LF}F_0
\end{aligned}
\end{equation}

The total magnetic field can be written as:
\begin{equation*}
\begin{aligned}
\v{B} &= \BO\left(1+\frac{\Bper}{B_0}\right)\zhat = \curl{\v{A}} = \curl{\left(\v{A}_0+\vAper\right)} \\
\Bper\zhat &= \curl{\vAper} = -ik\Aper_x\zhat \Longrightarrow \vAper = -\frac{\Bper}{ik}\xhat.
\end{aligned}
\end{equation*}
The fluctuating component of the fields are
\begin{equation} \tag{A17}
\begin{aligned}
\vEper = -\grad{\phi}-\frac{1}{c}\partial_t\v{A} &= -\frac{m}{q} ik\frac{q\phi}{m}\yhat +\frac{i\omega}{c}\v{\Aper} \\
&=  -\frac{m}{q} \left(ik\frac{q\phi}{m}\yhat +\frac{\omega \omegac}{k}\frac{\Bper}{B_0}\xhat\right)
\end{aligned}
\end{equation}

In the gyrokinetic equation one has the velocity-space convection term
\begin{equation*}
\begin{aligned}
\left(\frac{q}{m}\vEper+\omegac\frac{\Bper}{B_0}\vv\times\zhat\right)\cdot\gradv{\left(F_0+F_1\right)} &= \left(-ik\frac{q\phi}{m}\yhat -\frac{\omega \omegac}{k}\frac{\Bper}{B_0}\xhat+\omegac\frac{\Bper}{B_0}\vv\times\zhat\right)\cdot\gradv{\left(F_0+F_1\right)}.
\end{aligned}
\end{equation*}
The first term on the right side is
\begin{equation} \tag{A18}
\begin{aligned}
-ik\frac{q\phi}{m}\yhat\cdot\gradv{\left(F_0+F_1\right)} &= -ik\frac{q\phi}{m}\pd{}{v_y}{\left(1+\frac{v_y}{\omegac \LF}\right)F_0} \\
&= -ik\frac{q\phi}{m}\left[\frac{1}{\omegac \LF}F_0+\frac{v_y}{\omegac}F_0\pd{}{v_y}\frac{1}{\LF}+\left(1+\frac{v_y}{\omegac \LF}\right)\pd{F_0}{v_y}\right] \\
&= -ik\frac{q\phi}{m}\left[\frac{1}{\omegac \LF}+\frac{v_y}{\omegac}\pd{}{v_y}\frac{\vbar^2}{\LT}+\left(1+\frac{v_y}{\omegac \LF}\right)\pd{}{v_y}\left(-\frac{v^2}{\vt^2}\right)\right]F_0 \\
&= -ik\frac{q\phi}{m}\left[\frac{1}{\omegac \LF}+\frac{v_y}{\omegac}\frac{2v_y}{\vt^2\LT}+\left(1+\frac{v_y}{\omegac \LF}\right)\left(-\frac{2v_y}{\vt^2}\right)\right]F_0 \\
&= -ik\frac{T}{m}\left(\frac{1}{\omegac \LF}+\frac{2}{\omegac\LT}\frac{v_y^2}{\vt^2}-\frac{2v_y}{\vt^2}-\frac{2v_y^2}{\omegac \LF \vt^2}\right)\varphiper \\
&= i\frac{2kT}{m\vt^2}\left(v_y+\frac{v_y^2}{\omegac \LF}-\frac{v_y^2}{\omegac\LT}-\frac{\vt^2}{2\omegac \LF}\right)\varphiper \\
&= i\frac{2kT}{m\vt^2}\left[v_y+\left(\frac{1}{\LF}-\frac{1}{\LT}\right)\frac{v_y^2}{\omegac}-\frac{\vt^2}{2\omegac \LF}\right]\varphiper \\
&= ik\left(v_y+\frac{v_y^2}{\omegac \LFT}-\frac{\vt^2}{2\omegac \LF}\right)\varphiper,
\end{aligned}
\end{equation}
while the second term is written as
\begin{equation} \tag{A19}
\begin{aligned}
-\frac{\omega \omegac}{k}\frac{\Bper}{B_0}\xhat\cdot\gradv{\left(F_0+F_1\right)} &= -\frac{\omega \omegac}{k}\frac{\Bper}{B_0}\pd{}{v_x}{\left(1+\frac{v_y}{\omegac \LF}\right)F_0} \\
&= -\frac{\omega \omegac}{k}\frac{\Bper}{B_0}\pd{}{v_x}{\left(1+\frac{v_y}{\omegac \LF}\right)F_0} \\
&= -\frac{\omega \omegac}{k}\frac{\Bper}{B_0}{\left(1+\frac{v_y}{\omegac \LF}\right)\pd{F_0}{v_x}} \\
&= -\frac{\omega \omegac}{k}\frac{\Bper}{B_0}{\left(1+\frac{v_y}{\omegac \LF}\right)\pd{}{v_x}\left(-\frac{v^2}{\vt^2}\right)F_0} \\
&= -\frac{\omega \omegac}{k}\frac{\Bper}{B_0}{\left(1+\frac{v_y}{\omegac \LF}\right)\left(-\frac{2v_x}{\vt^2}\right)F_0} \\
&= \frac{\omega \omegac}{k}\frac{\Bper}{B_0}\frac{2\vxbar}{\vt}F_0 + \frac{\omega}{k}\frac{\Bper}{B_0}\frac{2}{\LF}\vxbar\vybar F_0 \\
&= \frac{2\omega \omegac}{k\vt}\vxbar\bper + \frac{\omega}{k}\frac{\Bper}{B_0}\frac{2}{\LF}\vxbar\vybar F_0.
\end{aligned}
\end{equation}
We also need
\begin{equation*}
\begin{aligned}
\omegac\frac{\Bper}{B_0}\vv\times\zhat\cdot\gradv{F_1} &= -\omegac\frac{\Bper}{B_0}\pd{}{\xi}\left(\frac{v_y}{\omegac \LF}F_0\right) = -\frac{v_x}{\LF}\bper = -\frac{\vt}{\LF}\vxbar\bper 
\end{aligned}
\end{equation*}

Now define
\begin{equation} \tag{A20}
\varphiper = \varphihat e^{iky-i\omega t}, \quad
\bper = \bhat e^{iky-i\omega t}, \quad
\fper = \fhat e^{iky-i\omega t},
\end{equation}
and expanding $\fhat = \fhat_0+\fhat_1+\dots$ we obtain the equation
\begin{equation} \tag{A22a}
\begin{aligned}
\left(\frac{ikv_y}{\omegac}-\pd{}{\xi}\right)\fhat_0 +\frac{ikv_y}{\omegac}\varphihat=0.
\end{aligned}
\end{equation}

The homogeneous solution obeys 
\begin{equation*}
\begin{aligned}
\pd{}{\xi}\fhat_{0,H}=\frac{ik\vperp}{\omegac}\fhat_{0,H}\sin\xi,
\end{aligned}
\end{equation*}
and thus this solution is:
\begin{align*}
\fhat_{0,H} = ge^{-\frac{ik\vperp}{\omegac}\cos\xi} = ge^{-\frac{ikv_x}{\omegac}} \qquad g\neq g(\xi),
\end{align*}
while the inhomogeneous equation is
\begin{equation*}
\begin{aligned}
\pd{}{\xi}\fhat_0-\frac{ikv_y}{\omegac}\fhat_0 =\frac{ikv_y}{\omegac}\varphihat.
\end{aligned}
\end{equation*}
By inspection we see the solution is 
\begin{equation} \tag{A22b}
\begin{aligned}
\fhat_0 = ge^{-\frac{ikv_x}{\omegac}} - \varphihat
\end{aligned}
\end{equation}

Now we gyro-average the equation:
\begin{equation} \tag{A25}
\begin{aligned}
\int_{-\pi}^\pi\frac{d\xi}{2\pi}e^{\frac{ikv_x}{\omegac}}\left[ -\frac{i\omega}{\omegac}\left(ge^{-\frac{ikv_x}{\omegac}} - \varphihat\right)+\frac{v_x}{\omegac}\partialx\left(ge^{-\frac{ikv_x}{\omegac}} - \varphihat\right) \right. \\
\left. +\frac{ik}{\omegac}\varphihat\left(\frac{v_y^2}{\omegac \LFT}-\frac{\vt^2}{2\omegac \LF}\right)+\bhat\vxbar\left(\frac{2\omega}{k\vt}-\frac{\vt}{\omegac\LF}\right)\right]=0 \\
\int_{-\pi}^\pi\frac{d\xi}{2\pi}e^{\frac{ikv_x}{\omegac}}\left\lbrace -\frac{i\omega}{\omegac}\left(ge^{-\frac{ikv_x}{\omegac}} - \varphihat\right)+\frac{v_x}{\omegac}e^{-\frac{ikv_x}{\omegac}}\left[g\partialx\left(-\frac{ikv_x}{\omegac}\right) +\partialx g\right]- \frac{v_x}{\omegac}\partialx\varphihat \right. \\
\left. +\frac{ik}{\omegac}\varphihat\left(\frac{v_y^2}{\omegac \LFT}-\frac{\vt^2}{2\omegac \LF}\right)+\bhat\vxbar\left(\frac{2\omega}{k\vt}-\frac{\vt}{\omegac\LF}\right)\right\rbrace=0 \\
\int_{-\pi}^\pi\frac{d\xi}{2\pi}e^{\frac{ikv_x}{\omegac}}\left\lbrace -\frac{i\omega}{\omegac}ge^{-\frac{ikv_x}{\omegac}}+\frac{ik}{\omegac^2}\frac{\BOp}{B_0}gv_x^2e^{-\frac{ikv_x}{\omegac}} - \frac{v_x}{\omegac}\partialx\varphihat \right. \\
\left. +\left[\frac{ik}{\omegac}\left(\frac{v_y^2}{\omegac \LFT}-\frac{\vt^2}{2\omegac \LF}\right)+\frac{i\omega}{\omegac}\right]\varphihat+\bhat\vxbar\left(\frac{2\omega}{k\vt}-\frac{\vt}{\omegac\LF}\right)\right\rbrace=0 \\
ig\int\frac{d\xi}{2\pi} \left(-\frac{\omega}{\omegac}+\frac{k\vperp^2}{\omegac^2\LB}\sin^2\xi \right)= \\
\int_{-\pi}^\pi\frac{d\xi}{2\pi}e^{\frac{ikv_x}{\omegac}}\left\lbrace \frac{v_x}{\omegac}\partialx\varphihat -\left[\frac{ik}{\omegac}\left(\frac{v_y^2}{\omegac \LFT}-\frac{\vt^2}{2\omegac \LF}\right)+\frac{i\omega}{\omegac}\right]\varphihat-\bhat\vxbar\left(\frac{2\omega}{k\vt}-\frac{\vt}{\omegac\LF}\right)\right\rbrace \\
-\frac{i}{\omegac} g \left(\omega-\frac{k\vperp^2}{2\omegac\LB} \right)= \\
\int_{-\pi}^\pi\frac{d\xi}{2\pi}e^{\frac{ikv_x}{\omegac}}\left\lbrace \frac{v_x}{\omegac}\partialx\varphihat -\frac{i}{\omegac}\left[\frac{kv_y^2}{\omegac \LFT}+\omega-\frac{k\vt^2}{2\omegac \LF}\right]\varphihat-\bhat\vxbar\left(\frac{2\omega}{k\vt}-\frac{\vt}{\omegac\LF}\right)\right\rbrace \\
-\frac{i}{\omegac}\omegab g=
\int_{-\pi}^\pi\frac{d\xi}{2\pi}e^{\frac{ikv_x}{\omegac}}\left\lbrace \frac{v_x}{\omegac}\partialx\varphihat -\frac{i}{\omegac}\left[\frac{kv_y^2}{\omegac \LFT}+\omegabar\right]\varphihat-\bhat\vxbar\left(\frac{2\omega}{k\vt}-\frac{\vt}{\omegac\LF}\right)\right\rbrace
\end{aligned}
\end{equation}

and $\partialx\varphihat=\varphihat/\LFT$, so

\begin{equation*}
\begin{aligned}
-\frac{i}{\omegac}\omegab g=
\int_{-\pi}^\pi\frac{d\xi}{2\pi}e^{\frac{ikv_x}{\omegac}}\left\lbrace\left[ \frac{v_x}{\omegac\LFT} -\frac{i}{\omegac}\omegabar-\frac{ikv_y^2}{\omegac^2 \LFT}\right]\varphihat-\bhat\vxbar\left(\frac{2\omega}{k\vt}-\frac{\vt}{\omegac\LF}\right)\right\rbrace \\
i\omegab g=
\int_{-\pi}^\pi\frac{d\xi}{2\pi}e^{\frac{ikv_x}{\omegac}}\left\lbrace\left[i\omegabar - \frac{v_x}{\LFT} + \frac{ikv_y^2}{\omegac \LFT}\right]\varphihat+\bhat\vxbar\frac{2\omegac}{k\vt}\left(\omega-\frac{k\vt^2}{2\omegac\LF}\right)\right\rbrace \\
\end{aligned}
\end{equation*}
\begin{equation} \tag{A27}
\begin{aligned}
i\omegab g=
\int_{-\pi}^\pi\frac{d\xi}{2\pi}e^{\frac{ikv_x}{\omegac}}\left\lbrace\left[i\omegabar - \frac{v_x}{\LFT} + \frac{ikv_y^2}{\omegac \LFT}\right]\varphihat+\bhat\vxbar\frac{2\omegac\omegabar}{k\vt}\right\rbrace
\end{aligned}
\end{equation}

Use the identity
\begin{equation} \tag{A28}
\begin{aligned}
\int^\pi_{-\pi}d\xi\frac{ikv_y^2}{\omegac}e^{\frac{ikv_x}{\omegac}} &= -\int^\pi_{-\pi}d\xi v_y \pd{}{\xi}e^{\frac{ikv_x}{\omegac}} = -\int^\pi_{-\pi}d\xi \left(\pd{}{\xi}v_y e^{\frac{ikv_x}{\omegac}} -e^{\frac{ikv_x}{\omegac}}\pd{}{\xi}v_y \right) \\
&= \int^\pi_{-\pi}d\xi \left(v_x e^{\frac{ikv_x}{\omegac}} - \pd{}{\xi}v_y e^{\frac{ikv_x}{\omegac}} \right) = \int^\pi_{-\pi}d\xi v_x e^{\frac{ikv_x}{\omegac}}
\end{aligned}
\end{equation}
one sees that the two terms proportional to $\LFT^{-1}$ cancel, leaving

\begin{equation} \tag{A29}
\begin{aligned}
g=
\int_{-\pi}^\pi\frac{d\xi}{2\pi}e^{\frac{ikv_x}{\omegac}}\frac{\omegabar}{\omegab}\left(\varphihat-\bhat\vxbar\frac{2i\omegac}{k\vt}\right)
\end{aligned}
\end{equation}

The following small argument expansions are used for the electrons:
\begin{equation} \tag{A33}
\begin{aligned}
e^{-\frac{ikv_x}{\omegac}}J_0\left(\frac{k\vperp}{\omegac}\right) &\approx \left(1-\frac{ikv_x}{\omegac}\right)\left[1-\frac{1}{4}\left(\frac{k\vperp}{\omegac}\right)^2\right] \\
&\approx 1-\frac{ikv_x}{\omegac} \\
e^{-\frac{ikv_x}{\omegac}}J_1\left(\frac{k\vperp}{\omegac}\right) &\approx \left(1-\frac{ikv_x}{\omegac}\right)\left[\frac{k\vperp}{2\omegac}-\frac{1}{16}\left(\frac{k\vperp}{\omegac}\right)^3\right] \\
&\approx \left(1-\frac{ikv_x}{\omegac}\right)\frac{k\vperp}{2\omegac}
\end{aligned}
\end{equation}

\subsection*{Ampere's law}

\begin{equation} \tag{A34}
\begin{aligned} 
\frac{\Bper}{B_0} &=-\frac{4\pi ien_0}{ckB_0\pi^{3/2}}\int d^3\vbar e^{-\vbar^2}\vbar_x\left(\vti\fbarOi-\vte\fbarOe\right) \\
&=-\frac{4\pi ien_0}{ckB_0\pi^{3/2}}\int d^3\vbar e^{-\vbar^2}\vbar_x\left\lbrace\vti\left[\left(e^{-\frac{ikv_x}{\omegaci}}J_0\frac{\omegabari}{\omegabi}-1\right)\frac{e\phiper}{\TiO}+e^{-\frac{ikv_x}{\omegaci}}J_1\frac{\omegabari}{\omegabi}\frac{2\vbarperp}{k\rho_i}\frac{\Bper}{B_0}\right] \right. \\
&\left. \quad-\vte\left[-\left[\left(1-\frac{ikv_x}{\omegace}\right)\frac{\omegabare}{\omegabe}-1\right]\frac{e\phiper}{\TeO}+\left(1-\frac{ikv_x}{\omegace}\right)\frac{\omegabare}{\omegabe}\vbarperp^2\frac{\Bper}{B_0}\right]\right\rbrace
\end{aligned}
\end{equation}
Move terms proportional to $\Bper/B_0$ to the left side:
\begin{equation*}
\begin{aligned}
\frac{\Bper}{B_0} \left\lbrace 1+\frac{4\pi ien_0}{ckB_0\pi^{3/2}}\int d^3\vbar e^{-\vbar^2}\vbar_x\left[\vti e^{-\frac{ikv_x}{\omegaci}}J_1\frac{\omegabari}{\omegabi}\frac{2\vbarperp}{k\rho_i}-\vte\left(1-\frac{ikv_x}{\omegace}\right)\frac{\omegabare}{\omegabe}\vbarperp^2\right] \right\rbrace \\
=-\frac{4\pi ien_0}{ckB_0\pi^{3/2}}\int d^3\vbar e^{-\vbar^2}\vbar_x\left\lbrace\vti\left(e^{-\frac{ikv_x}{\omegaci}}J_0\frac{\omegabari}{\omegabi}-1\right)+\vte\left[\left(1-\frac{ikv_x}{\omegace}\right)\frac{\omegabare}{\omegabe}-1\right]\frac{\TiO}{\TeO}\right\rbrace\frac{e\phiper}{\TiO} \\
\frac{\Bper}{B_0} \left\lbrace 1+\frac{4\pi ien_0}{ckB_0\pi^{3/2}}\int d^3\vbar e^{-\vbar^2}\vbar_x\vbarperp^2\left[\vti \left[1-\frac{ikv_x}{\omegaci}\left(1+f_{1i}\right)\right]\frac{\omegabari}{\omegabi}\frac{2J_1}{k\rho_i\vbarperp}-\vte\left(1-\frac{ikv_x}{\omegace}\right)\frac{\omegabare}{\omegabe}\right] \right\rbrace \\
=-\frac{4\pi ien_0}{ckB_0\pi^{3/2}}\int d^3\vbar e^{-\vbar^2}\vbar_x\left\lbrace\vti\left[\left[1-\frac{ikv_x}{\omegaci}\left(1+f_{1i}\right)\right]J_0\frac{\omegabari}{\omegabi}-1\right] \right. \\
\left.+\vte\left[\left(1-\frac{ikv_x}{\omegace}\right)\frac{\omegabare}{\omegabe}-1\right]\frac{\TiO}{\TeO}\right\rbrace\frac{e\phiper}{\TiO} \\
\end{aligned}
\end{equation*}
Examine the left hand side only:
\begin{equation*}
\begin{aligned}
\text{LHS}\frac{B_0}{\Bper}&=1+\frac{1}{\pi^{3/2}}\int d^3\vbar e^{-\vbar^2}\vbar_x\vbarperp^2\left\lbrace\left[\frac{4\pi ien_0}{ckB_0}\vti+\vxbar\beta_i\left(1+f_{1i}\right)\right]\frac{\omegabari}{\omegabi}\frac{2J_1}{k\rho_i\vbarperp} \right.\\
&\left.\qquad\qquad\qquad\qquad\qquad\qquad\quad-\vte\frac{4\pi ien_0}{ckB_0}\left(1-\frac{ikv_x}{\omegace}\right)\frac{\omegabare}{\omegabe}\right\rbrace \\
&=1+\frac{1}{\pi^{3/2}}\int d^3\vbar e^{-\vbar^2}\vbar_x\vbarperp^2\left\lbrace\left[\frac{4\pi ien_0}{ckB_0}\vti+\vxbar\beta_i\left(1+f_{1i}\right)\right]\frac{\omegabari}{\omegabi}\frac{2J_1}{k\rho_i\vbarperp} \right.\\
&\left.\qquad\qquad\qquad\qquad\qquad\qquad\quad+\left(\betae\vxbar-\vte\frac{4\pi ien_0}{ckB_0}\right)\frac{\omegabare}{\omegabe}\right\rbrace \\
&=1+\frac{1}{\pi^{3/2}}\int d^3\vbar e^{-\vbar^2}\vxbar^2\vbarperp^2\left[\beta_i\left(1+f_{1i}\right)\frac{\omegabari}{\omegabi}\frac{2J_1}{k\rho_i\vbarperp}+\betae\frac{\omegabare}{\omegabe}\right]
\end{aligned}
\end{equation*}

\begin{equation} \tag{A37a}
\IBA=1+\frac{1}{\pi^{3/2}}\int d^3\vbar e^{-\vbar^2}\vxbar^2\vbarperp^2\left[\beta_i\left(1+f_{1i}\right)\frac{\omegabari}{\omegabi}\frac{2J_1}{k\rho_i\vbarperp}+\betae\frac{\omegabare}{\omegabe}\right]
\end{equation}

Now reorganize the right hand side:
\begin{equation*}
\begin{aligned}
\text{RHS}&=-\frac{4\pi ien_0}{ckB_0\pi^{3/2}}\int d^3\vbar e^{-\vbar^2}\vbar_x\left\lbrace\vti\left[\left[1-\frac{ikv_x}{\omegaci}\left(1+f_{1i}\right)\right]J_0\frac{\omegabari}{\omegabi}-1\right]
 \right.\\
 &\left.\qquad\qquad\qquad\qquad\qquad\qquad\quad+\vte\left[\left(1-\frac{ikv_x}{\omegace}\right)\frac{\omegabare}{\omegabe}-1\right]\frac{\TiO}{\TeO}\right\rbrace\frac{e\phiper}{\TiO} \\
&=-\frac{4\pi ien_0}{ckB_0\pi^{3/2}}\int d^3\vbar e^{-\vbar^2}\vbar_x\left\lbrace-\vti\frac{ikv_x}{\omegaci}\left(1+f_{1i}\right)J_0\frac{\omegabari}{\omegabi}
-\vte\left(\frac{ikv_x}{\omegace}\right)\frac{\omegabare}{\omegabe}\frac{\TiO}{\TeO}\right\rbrace\frac{e\phiper}{\TiO} \\
&=-\frac{1}{\pi^{3/2}}\int d^3\vbar e^{-\vbar^2}\vxbar^2\left[-\frac{4\pi ien_0}{ckB_0}\vti^2\frac{ik}{\omegaci}\left(1+f_{1i}\right)J_0\frac{\omegabari}{\omegabi}
-\frac{4\pi ien_0}{ckB_0}\vte^2\left(\frac{ik}{\omegace}\right)\frac{\omegabare}{\omegabe}\frac{\TiO}{\TeO}\right]\frac{e\phiper}{\TiO} \\
&=-\frac{1}{\pi^{3/2}}\int d^3\vbar e^{-\vbar^2}\vxbar^2\left[\beta_i\left(1+f_{1i}\right)J_0\frac{\omegabari}{\omegabi}
-\betae\frac{\omegabare}{\omegabe}\frac{\TiO}{\TeO}\right]\frac{e\phiper}{\TiO} \\
&=-\frac{1}{\pi^{3/2}}\beta_i\int d^3\vbar e^{-\vbar^2}\vxbar^2\left[\left(1+f_{1i}\right)J_0\frac{\omegabari}{\omegabi}
-\frac{\omegabare}{\omegabe}\right]\frac{e\phiper}{\TiO}
\end{aligned}
\end{equation*}

\begin{equation} \tag{A37b}
\IphiA=\frac{1}{\pi^{3/2}}\beta_i\int d^3\vbar e^{-\vbar^2}\vxbar^2\left[-\left(1+f_{1i}\right)J_0\frac{\omegabari}{\omegabi}
+\frac{\omegabare}{\omegabe}\right]
\end{equation}

For the transformation:
\begin{equation*}
\vperp = \sqrt{v_x^2+v_y^2} \qquad \xi=\arctan\left(\frac{v_y}{v_x}\right)
\end{equation*}
the Jacobian matrix is
\begin{equation*}
\begin{bmatrix}
\pd{\vperp}{v_x} & \pd{\vperp}{v_y} \\
\pd{\xi}{v_x} & \pd{\xi}{v_y}
\end{bmatrix} = \begin{bmatrix}
\frac{v_x}{\vperp} & \frac{v_y}{\vperp} \\
-\frac{v_y}{\vperp} & \frac{v_x}{\vperp}
\end{bmatrix} = \frac{1}{\vperp}.
\end{equation*}

\begin{align*}
ik\rho_i\vxbar\left(1+f_{1i}\right) = 1-e^{-\frac{ikv_x}{\omegaci}} = \frac{e^{\frac{ikv_x}{\omegaci}}-1}{e^{\frac{ikv_x}{\omegaci}}} \\
e^{\frac{ikv_x}{\omegaci}}ik\rho_i\vxbar\left(1+f_{1i}\right) +1= e^{\frac{ikv_x}{\omegaci}}
\end{align*}

\begin{equation} \tag{A39}
\begin{aligned}
\frac{\vbarperp}{k\rho_i}J_1\left(k\rho_i\vbarperp\right) &= \int^\pi_{-\pi}\frac{d\xi}{2\pi}\vxbar^2\left(1+f_{1i}\right) \\
&= \int^\pi_{-\pi}\frac{d\xi}{2\pi}\vxbar^2 \frac{i\omegaci}{kv_x}\left(e^{-ikv_x/\omegaci}-1\right) \\
&= \frac{i}{k\rho_i}\int^\pi_{-\pi}\frac{d\xi}{2\pi}\vxbar\left(e^{-ikv_x/\omegaci}-1\right) = \frac{i\vbarperp}{k\rho_i}\int^\pi_{-\pi}\frac{d\xi}{2\pi}e^{-ikv_x/\omegaci}\cos\xi \\
iJ_1\left(k\rho_i\vbarperp\right) &= -\int^\pi_{-\pi}\frac{d\xi}{2\pi}e^{-ikv_x/\omegaci}\cos\xi = - iJ_1\left(-k\rho_i\vbarperp\right) = iJ_1\left(k\rho_i\vbarperp\right)
\end{aligned}
\end{equation}

The $\IBA$ integral is
\begin{equation} \tag{A41a}
\begin{aligned}
\IBA &= 1+\frac{1}{\pi^{3/2}}\int d^3\vbar e^{-\vbar^2}\vxbar^2\vbarperp^2\left[\beta_i\left(1+f_{1i}\right)\frac{\omegabari}{\omegabi}\frac{2J_1}{k\rho_i\vbarperp}+\betae\frac{\omegabare}{\omegabe}\right] \\
&= 1+2\int \frac{d\vzbar}{\pi^{1/2}}\int d\vbarperp e^{-\vbar^2}\vbarperp^3\int \frac{d\xi}{2\pi}\vxbar^2\left[\beta_i\left(1+f_{1i}\right)\frac{\omegabari}{\omegabi}\frac{2J_1}{k\rho_i\vbarperp}+\betae\frac{\omegabare}{\omegabe}\right] \\
&= 1+2\int d\vbarperp e^{-\vbarperp^2}\vbarperp^3\left[\beta_i\frac{\omegabari}{\omegabi}\frac{2J_1^2}{\left(k\rho_i\right)^2}+\vbarperp^2\frac{\betae\omegabare}{2\omegabe}\right]
\end{aligned}
\end{equation}
and the $\IphiA$ integral becomes
\begin{equation} \tag{A41b}
\begin{aligned}
\IphiA &= \frac{1}{\pi^{3/2}}\beta_i\int d^3\vbar e^{-\vbar^2}\vxbar^2\left[-\left(1+f_{1i}\right)J_0\frac{\omegabari}{\omegabi}
+\frac{\omegabare}{\omegabe}\right] \\
&= 2\beta_i\int \frac{d\vzbar}{\pi^{1/2}}\int d\vbarperp \vbarperp e^{-\vbar^2}\int\frac{d\xi}{2\pi} \vxbar^2\left[-\left(1+f_{1i}\right)J_0\frac{\omegabari}{\omegabi}
+\frac{\omegabare}{\omegabe}\right] \\
&= \beta_i\int d\vbarperp \vbarperp^2 e^{-\vbarperp^2} \left[-2J_0J_1\frac{\omegabari}{k\rho_i\omegabi}
+\vbarperp\frac{\omegabare}{\omegabe}\right]
\end{aligned}
\end{equation} 

\subsection*{Quasineutrality condition}

\begin{equation*}
\begin{aligned}
&\int\frac{d^3\vbar}{\pi^{3/2}}e^{-\vbar^2} \left(\fbarOi-\fbarOe\right) = 0 \\
&\int\frac{d^3\vbar}{\pi^{3/2}}e^{-\vbar^2} \left[\left(e^{-\frac{ikv_x}{\omegaci}}J_0\frac{\omegabari}{\omegabi}-1\right)\frac{e\phiper}{\TiO}+e^{-\frac{ikv_x}{\omegaci}}J_1\frac{\omegabari}{\omegabi}\frac{2\vbarperp}{k\rho_i}\frac{\Bper}{B_0} + \right.\\
&\left.\qquad\qquad\qquad\quad \left[\left(1-\frac{ikv_x}{\omegace}\right)\frac{\omegabare}{\omegabe}-1\right]\frac{e\phiper}{\TeO}-\left(1-\frac{ikv_x}{\omegace}\right)\frac{\omegabare}{\omegabe}\vbarperp^2\frac{\Bper}{B_0}\right] = 0
\end{aligned}
\end{equation*}
\begin{align} \tag{A44}
\int\frac{d^3\vbar}{\pi^{3/2}}e^{-\vbar^2}\left[\left(e^{-\frac{ikv_x}{\omegaci}}J_0\frac{\omegabari}{\omegabi}-1\right)\frac{e\phiper}{\TiO}+e^{-\frac{ikv_x}{\omegaci}}J_1\frac{\omegabari}{\omegabi}\frac{2\vbarperp}{k\rho_i}\frac{\Bper}{B_0}+
\left(\frac{\omegabare}{\omegabe}-1\right)\frac{e\phiper}{\TeO}-\frac{\omegabare}{\omegabe}\vbarperp^2\frac{\Bper}{B_0}\right] &= 0 
\end{align}
Collecting the terms proportional to $\Bper/B_0$ on the right side, we obtain:
\begin{equation}\tag{A45}
\begin{aligned}
&\int\frac{d^3\vbar}{\pi^{3/2}}e^{-\vbar^2}\left[\left(e^{-\frac{ikv_x}{\omegaci}}J_0\frac{\omegabari}{\omegabi}-1\right)+
\left(\frac{\omegabare}{\omegabe}-1\right)\frac{\TiO}{\TeO}\right]\frac{e\phiper}{\TiO} \\
&\qquad= \int\frac{d^3\vbar}{\pi^{3/2}}e^{-\vbar^2}\left(\frac{\omegabare}{\omegabe}\vbarperp^2-e^{-\frac{ikv_x}{\omegaci}}J_1\frac{\omegabari}{\omegabi}\frac{2\vbarperp}{k\rho_i}\right)\frac{\Bper}{B_0}
\end{aligned}
\end{equation}

The integral on the left side is:

\begin{equation} \tag{A47a}
\begin{aligned}
\IphiQ &= \int\frac{d^3\vbar}{\pi^{3/2}}e^{-\vbar^2}\left[\left(e^{-\frac{ikv_x}{\omegaci}}J_0\frac{\omegabari}{\omegabi}-1\right)+
\left(\frac{\omegabare}{\omegabe}-1\right)\frac{\TiO}{\TeO}\right]\\
&= 2\int\frac{d\vzbar}{\pi^{1/2}}e^{-\vzbar^2}\int d\vbarperp e^{-\vbarperp^2}\vbarperp\int\frac{d\xi}{2\pi}\left\lbrace e^{-\frac{ikv_x}{\omegaci}}J_0\frac{\omegabari}{\omegabi}-1+
\left(\frac{\omegabare}{\omegabe}-1\right)\frac{\TiO}{\TeO}\right\rbrace\\
&= 2\int d\vbarperp e^{-\vbarperp^2}\vbarperp\left[ J_0^2\frac{\omegabari}{\omegabi}-1+
\left(\frac{\omegabare}{\omegabe}-1\right)\frac{\TiO}{\TeO}\right]
\end{aligned}
\end{equation}

while the integral on the right side can be written as:
\begin{equation} \tag{A47b}
\begin{aligned}
\IBQ &= \int\frac{d^3\vbar}{\pi^{3/2}}e^{-\vbar^2}\left(\frac{\omegabare}{\omegabe}\vbarperp^2-e^{-\frac{ikv_x}{\omegaci}}J_1\frac{\omegabari}{\omegabi}\frac{2\vbarperp}{k\rho_i}\right) \\
&= 2\int\frac{d\vzbar}{\pi^{1/2}}e^{-\vzbar^2}\int d\vbarperp e^{-\vbarperp^2}\vbarperp\int\frac{d\xi}{2\pi}\left(\frac{\omegabare}{\omegabe}\vbarperp^2-e^{-\frac{ikv_x}{\omegaci}}J_1\frac{\omegabari}{\omegabi}\frac{2\vbarperp}{k\rho_i}\right) \\
&= 2\int d\vbarperp e^{-\vbarperp^2}\vbarperp^2\left(\frac{\omegabare}{\omegabe}\vbarperp-J_0J_1\frac{2\omegabari}{k\rho_i\omegabi}\right)
\end{aligned}
\end{equation}

\section*{Low $\beta$ limit with $k\rho_i\ll 1$}

\begin{equation} \tag{20}
\frac{1}{\omegabi}=\frac{1}{\omega - \frac{k \Ln v^2}{2\LB}}=\frac{1}{\omega}\left[1+\frac{k}{2\omega}\frac{\Ln}{\LB}v^2+\frac{k^2}{4\omega^2}\left(\frac{\Ln}{\LB}\right)^2v^4+\frac{k^3}{8\omega^3}\left(\frac{\Ln}{\LB}\right)^3v^6+\dots\right]
\end{equation}

\begin{equation} \tag{21}
\begin{aligned}
\frac{1}{\omegabe} &= \frac{1}{\omega - \frac{k \Ln \vperp^2}{2\vti\omegace\LB}} = \frac{1}{\omega+\frac{k \Ln\taue\vti v^2}{2\omegaci\LB}} = \frac{1}{\omega+\frac{\taue k}{2}\frac{\Ln}{\LB}v^2} \\
&= \frac{1}{\omega}\left[1-\frac{\taue k}{2\omega}\frac{\Ln}{\LB}v^2+\frac{\taue^2 k^2}{4\omega^2}\left(\frac{\Ln}{\LB}\right)^2v^4-\frac{\taue^3k^3}{8\omega^3}\left(\frac{\Ln}{\LB}\right)^3v^6+\dots\right]
\end{aligned}
\end{equation}

\begin{equation}\tag{22}
\begin{aligned}
\frac{\omegabari}{\omegabi} &= \left[1-\frac{k}{2\omega}\left(1-\etai\right)-\frac{k}{2\omega}\etai v^2\right]\left[1+\frac{k}{2\omega}\frac{\Ln}{\LB}v^2+\frac{k^2}{4\omega^2}\left(\frac{\Ln}{\LB}\right)^2v^4 \right.\\
&\left.\qquad\qquad\qquad\qquad\qquad\qquad\qquad+\frac{k^3}{8\omega^3}\left(\frac{\Ln}{\LB}\right)^3v^6+\frac{k^4}{16\omega^4}\left(\frac{\Ln}{\LB}\right)^4v^8\dots\right] \\
&= 1-\frac{k}{2\omega}\left(1-\etai\right)-\frac{k}{2\omega}\left\lbrace\etai-\frac{\Ln}{\LB}\left[1-\frac{k}{2\omega}\left(1-\etai\right)\right]\right\rbrace v^2 \\
&\quad-\frac{k^2}{4\omega^2}\frac{\Ln}{\LB}\left\lbrace\etai-\frac{\Ln}{\LB}\left[1-\frac{k}{2\omega}\left(1-\etai\right)\right]\right\rbrace v^4-\frac{k^3}{8\omega^3}\left(\frac{\Ln}{\LB}\right)^2\left\lbrace\etai-\frac{\Ln}{\LB}\left[1-\frac{k}{2\omega}\left(1-\etai\right)\right]\right\rbrace v^6 \\
&\quad-\frac{k^4}{16\omega^4}\left(\frac{\Ln}{\LB}\right)^3\left\lbrace\etai-\frac{\Ln}{\LB}\left[1-\frac{k}{2\omega}\left(1-\etai\right)\right]\right\rbrace v^8
\end{aligned}
\end{equation}

\begin{equation} \tag{23}
\begin{aligned}
\frac{\omegabare}{\omegabe} &= \left[1+\frac{\taue k}{2\omega}\left(1-\etae\right)+\frac{\taue k}{2\omega}\etae v^2\right]\left[1-\frac{\taue k}{2\omega}\frac{\Ln}{\LB}v^2+\frac{\taue^2 k^2}{4\omega^2}\left(\frac{\Ln}{\LB}\right)^2v^4 \right.\\
&\left.\qquad\qquad\qquad\qquad\qquad\qquad\qquad\quad-\frac{\taue^3k^3}{8\omega^3}\left(\frac{\Ln}{\LB}\right)^3v^6+\frac{\taue^4k^4}{16\omega^4}\left(\frac{\Ln}{\LB}\right)^4v^8+\dots\right] \\
&= 1+\frac{\taue k}{2\omega}\left(1-\etae\right)+\frac{\taue k}{2\omega}\left\lbrace\etae-\frac{\Ln}{\LB}\left[1+\frac{\taue k}{2\omega}\left(1-\etae\right)\right]\right\rbrace v^2 \\
&\quad-\frac{\taue^2k^2}{4\omega^2}\frac{\Ln}{\LB}\left\lbrace\etae-\frac{\Ln}{\LB}\left[1+\frac{\taue k}{2\omega}\left(1-\etae\right)\right]\right\rbrace v^4+\frac{\taue^3k^3}{8\omega^3}\left(\frac{\Ln}{\LB}\right)^2\left\lbrace\etae-\frac{\Ln}{\LB}\left[1+\frac{\taue k}{2\omega}\left(1-\etae\right)\right]\right\rbrace v^6 \\
&\quad-\frac{\taue^4k^4}{16\omega^4}\left(\frac{\Ln}{\LB}\right)^3\left\lbrace\etae-\frac{\Ln}{\LB}\left[1+\frac{\taue k}{2\omega}\left(1-\etae\right)\right]\right\rbrace v^8
\end{aligned}
\end{equation}

We now calculate the $\IphiQ$ integral:
\begin{equation*}
\begin{aligned}
\IphiQ &= 2\int dv e^{-v^2}v\left[ J_0^2\frac{\omegabari}{\omegabi}-1+
\taui\left(\frac{\omegabare}{\omegabe}-1\right)\right] \\
&\simeq 2\int dv e^{-v^2}v\left[ \left(1-\frac{k^2}{4}v^2+\frac{k^4}{64}v^4\right)^2\frac{\omegabari}{\omegabi}-1+
\taui\left(\frac{\omegabare}{\omegabe}-1\right)\right] \\
&\simeq 2\int dv e^{-v^2}v\left[ \left(1-\frac{k^2}{2}v^2+\frac{3k^4}{32}v^4-\frac{k^6}{128}v^6+\frac{k^8}{4096}v^8\right)\frac{\omegabari}{\omegabi}-1+
\taui\left(\frac{\omegabare}{\omegabe}-1\right)\right] \\
&\simeq 2\int dv e^{-v^2}v\left[ \left(1-\frac{k^2}{2}v^2+\frac{3k^4}{32}v^4-\frac{k^6}{128}v^6+\frac{k^8}{4096}v^8\right)\left(1-\frac{k}{2\omega}\left(1-\etai\right) \right.\right. \\
&\left.\left.\quad-\frac{k}{2\omega}\left\lbrace\etai-\frac{\Ln}{\LB}\left[1-\frac{k}{2\omega}\left(1-\etai\right)\right]\right\rbrace v^2-\frac{k^2}{4\omega^2}\frac{\Ln}{\LB}\left\lbrace\etai-\frac{\Ln}{\LB}\left[1-\frac{k}{2\omega}\left(1-\etai\right)\right]\right\rbrace v^4 \right.\right. \\
&\left.\left.\quad-\frac{k^3}{8\omega^3}\left(\frac{\Ln}{\LB}\right)^2\left\lbrace\etai-\frac{\Ln}{\LB}\left[1-\frac{k}{2\omega}\left(1-\etai\right)\right]\right\rbrace v^6 \right.\right. \\
&\left.\left.\quad -\frac{k^4}{16\omega^4}\left(\frac{\Ln}{\LB}\right)^3\left\lbrace\etai-\frac{\Ln}{\LB}\left[1-\frac{k}{2\omega}\left(1-\etai\right)\right]\right\rbrace v^8\right)-1+\taui\left(\frac{\omegabare}{\omegabe}-1\right)\right]
\end{aligned}
\end{equation*}
\begin{equation*}
\begin{aligned}
\IphiQ &\simeq 2\int dv e^{-v^2}v\left[ 1-\frac{k}{2\omega}\left(1-\etai\right)-\left(\frac{k^2}{2}\left[1-\frac{k}{2\omega}\left(1-\etai\right)\right]+\frac{k}{2\omega}\left\lbrace\etai-\frac{\Ln}{\LB}\left[1-\frac{k}{2\omega}\left(1-\etai\right)\right]\right\rbrace\right)v^2 \right. \\
&\left.\quad+\left(\frac{3k^4}{32}\left[1-\frac{k}{2\omega}\left(1-\etai\right)\right]+\left(\frac{k^2}{2}\frac{k}{2\omega}-\frac{k^2}{4\omega^2}\frac{\Ln}{\LB}\right)\left\lbrace\etai-\frac{\Ln}{\LB}\left[1-\frac{k}{2\omega}\left(1-\etai\right)\right]\right\rbrace\right)v^4 \right. \\
&\left.\quad-\left(\frac{k^6}{128}\left[1-\frac{k}{2\omega}\left(1-\etai\right)\right]+\left[\frac{3k^4}{32}\frac{k}{2\omega}-\frac{k^2}{2}\frac{k^2}{4\omega^2}\frac{\Ln}{\LB}+\frac{k^3}{8\omega^3}\left(\frac{\Ln}{\LB}\right)^2\right]\left\lbrace\etai-\frac{\Ln}{\LB}\left[1-\frac{k}{2\omega}\left(1-\etai\right)\right]\right\rbrace\right) v^6 \right. \\
&\left.\quad +\left(\frac{k^8}{4096}\left[1-\frac{k}{2\omega}\left(1-\etai\right)\right]+\left[\frac{k^6}{128}\frac{k}{2\omega}-\frac{3k^4}{32}\frac{k^2}{4\omega^2}\frac{\Ln}{\LB}+\frac{k^2}{2}\frac{k^3}{8\omega^3}\left(\frac{\Ln}{\LB}\right)^2-\frac{k^4}{16\omega^4}\left(\frac{\Ln}{\LB}\right)^3\right] \right.\right. \\
&\left.\left.\qquad\left\lbrace\etai-\frac{\Ln}{\LB}\left[1-\frac{k}{2\omega}\left(1-\etai\right)\right]\right\rbrace\right)v^8 -1+\taui\left(\frac{\omegabare}{\omegabe}-1\right)\right] \\
&\simeq 1-\frac{k}{2\omega}\left(1-\etai\right)-\frac{k^2}{2}\left[1-\frac{k}{2\omega}\left(1-\etai\right)\right]-\frac{k}{2\omega}\left\lbrace\etai-\frac{\Ln}{\LB}\left[1-\frac{k}{2\omega}\left(1-\etai\right)\right]\right\rbrace \\
&\quad+2\left(\frac{3k^4}{32}\left[1-\frac{k}{2\omega}\left(1-\etai\right)\right]+\left(\frac{k^2}{2}\frac{k}{2\omega}-\frac{k^2}{4\omega^2}\frac{\Ln}{\LB}\right)\left\lbrace\etai-\frac{\Ln}{\LB}\left[1-\frac{k}{2\omega}\left(1-\etai\right)\right]\right\rbrace\right) \\
&\quad-6\left(\frac{k^6}{128}\left[1-\frac{k}{2\omega}\left(1-\etai\right)\right]+\left[\frac{3k^4}{32}\frac{k}{2\omega}-\frac{k^2}{2}\frac{k^2}{4\omega^2}\frac{\Ln}{\LB}+\frac{k^3}{8\omega^3}\left(\frac{\Ln}{\LB}\right)^2\right]\left\lbrace\etai-\frac{\Ln}{\LB}\left[1-\frac{k}{2\omega}\left(1-\etai\right)\right]\right\rbrace\right) \\
&\quad +24\left(\frac{k^8}{4096}\left[1-\frac{k}{2\omega}\left(1-\etai\right)\right]+\left[\frac{k^6}{128}\frac{k}{2\omega}-\frac{3k^4}{32}\frac{k^2}{4\omega^2}\frac{\Ln}{\LB}+\frac{k^2}{2}\frac{k^3}{8\omega^3}\left(\frac{\Ln}{\LB}\right)^2-\frac{k^4}{16\omega^4}\left(\frac{\Ln}{\LB}\right)^3\right] \right. \\
&\left.\qquad\left\lbrace\etai-\frac{\Ln}{\LB}\left[1-\frac{k}{2\omega}\left(1-\etai\right)\right]\right\rbrace\right) -\left(1+\taui\right)+2\taui\int dv e^{-v^2}v\frac{\omegabare}{\omegabe}
\end{aligned}
\end{equation*}

\begin{equation*}
\begin{aligned}
\IphiQ &\simeq -\frac{k}{2\omega}\left(1-\etai\right)-\frac{k^2}{2}\left[1-\frac{k}{2\omega}\left(1-\etai\right)\right]-\frac{k}{2\omega}\left\lbrace\etai-\frac{\Ln}{\LB}\left[1-\frac{k}{2\omega}\left(1-\etai\right)\right]\right\rbrace \\
&\quad+2\left(\frac{3k^4}{32}\left[1-\frac{k}{2\omega}\left(1-\etai\right)\right]+\left(\frac{k^2}{2}\frac{k}{2\omega}-\frac{k^2}{4\omega^2}\frac{\Ln}{\LB}\right)\left\lbrace\etai-\frac{\Ln}{\LB}\left[1-\frac{k}{2\omega}\left(1-\etai\right)\right]\right\rbrace\right) \\
&\quad-6\left(\frac{k^6}{128}\left[1-\frac{k}{2\omega}\left(1-\etai\right)\right]+\left[\frac{3k^4}{32}\frac{k}{2\omega}-\frac{k^2}{2}\frac{k^2}{4\omega^2}\frac{\Ln}{\LB}+\frac{k^3}{8\omega^3}\left(\frac{\Ln}{\LB}\right)^2\right]\left\lbrace\etai-\frac{\Ln}{\LB}\left[1-\frac{k}{2\omega}\left(1-\etai\right)\right]\right\rbrace\right) \\
&\quad +24\left(\frac{k^8}{4096}\left[1-\frac{k}{2\omega}\left(1-\etai\right)\right]+\left[\frac{k^6}{128}\frac{k}{2\omega}-\frac{3k^4}{32}\frac{k^2}{4\omega^2}\frac{\Ln}{\LB}+\frac{k^2}{2}\frac{k^3}{8\omega^3}\left(\frac{\Ln}{\LB}\right)^2-\frac{k^4}{16\omega^4}\left(\frac{\Ln}{\LB}\right)^3\right] \right. \\
&\left.\qquad\left\lbrace\etai-\frac{\Ln}{\LB}\left[1-\frac{k}{2\omega}\left(1-\etai\right)\right]\right\rbrace\right)
+\frac{k}{2\omega}\left(1-\etae\right)+ \\
&\quad\frac{k}{2\omega}\left\lbrace\etae-\frac{\Ln}{\LB}\left[1+\frac{\taue k}{2\omega}\left(1-\etae\right)\right]\right\rbrace -\frac{\taue k^2}{2\omega^2}\frac{\Ln}{\LB}\left\lbrace\etae-\frac{\Ln}{\LB}\left[1+\frac{\taue k}{2\omega}\left(1-\etae\right)\right]\right\rbrace \\
&\quad+\frac{3\taue^2k^3}{4\omega^3}\left(\frac{\Ln}{\LB}\right)^2\left\lbrace\etae-\frac{\Ln}{\LB}\left[1+\frac{\taue k}{2\omega}\left(1-\etae\right)\right]\right\rbrace-\frac{3\taue^3k^4}{2\omega^4}\left(\frac{\Ln}{\LB}\right)^3\left\lbrace\etae-\frac{\Ln}{\LB}\left[1+\frac{\taue k}{2\omega}\left(1-\etae\right)\right]\right\rbrace
\end{aligned}
\end{equation*}
To second order in $\beta$ and fifth order in $k$ this is
\begin{equation*}
\begin{aligned}
\IphiQ &\simeq -\frac{k^2}{2}\left[1-\frac{k}{2\omega}\left(1-\etai\right)\right]-\frac{k}{2\omega}\frac{\Ln}{\LB}\frac{k}{2\omega}\left(1-\etai\right)+\frac{3k^4}{16}\left[1-\frac{k}{2\omega}\left(1-\etai\right)\right] \\
&\quad+\left(\frac{k^3}{2\omega}-\frac{k^2}{2\omega^2}\frac{\Ln}{\LB}\right)\left\lbrace\etai-\frac{\Ln}{\LB}\left[1-\frac{k}{2\omega}\left(1-\etai\right)\right]\right\rbrace-6\frac{3k^4}{32}\frac{k}{2\omega}\left(\etai-\frac{\Ln}{\LB}\right) \\
&\quad+6\frac{k^2}{2}\frac{k^2}{4\omega^2}\frac{\Ln}{\LB}\left\lbrace\etai-\frac{\Ln}{\LB}\left[1-\frac{k}{2\omega}\left(1-\etai\right)\right]\right\rbrace-6\frac{k^3}{8\omega^3}\left(\frac{\Ln}{\LB}\right)^2\etai+24\frac{k^2}{2}\frac{k^3}{8\omega^3}\left(\frac{\Ln}{\LB}\right)^2\etai \\
&\quad-\frac{k}{2\omega}\frac{\Ln}{\LB}\frac{\taue k}{2\omega}\left(1-\etae\right)-\frac{\taue k^2}{2\omega^2}\frac{\Ln}{\LB}\left\lbrace\etae-\frac{\Ln}{\LB}\left[1+\frac{\taue k}{2\omega}\left(1-\etae\right)\right]\right\rbrace+\frac{3\taue^2k^3}{4\omega^3}\left(\frac{\Ln}{\LB}\right)^2\etae \\
&\simeq -\frac{k^2}{2}\left\lbrace 1-\frac{k}{2\omega}\left(1-\etai\right)+\frac{\Ln}{\LB}\frac{1}{2\omega^2}\left(1-\etai\right)-\frac{3k^2}{8}\left[1-\frac{k}{2\omega}\left(1-\etai\right)\right] \right.\\
&\left.\quad-\left(\frac{k}{\omega}-\frac{1}{\omega^2}\frac{\Ln}{\LB}\right)\left\lbrace\etai-\frac{\Ln}{\LB}\left[1-\frac{k}{2\omega}\left(1-\etai\right)\right]\right\rbrace+6\frac{3k^3}{32\omega}\left(\etai-\frac{\Ln}{\LB}\right) \right.\\
&\left.\quad-6\frac{k^2}{4\omega^2}\frac{\Ln}{\LB}\left\lbrace\etai-\frac{\Ln}{\LB}\left[1-\frac{k}{2\omega}\left(1-\etai\right)\right]\right\rbrace+6\frac{k}{4\omega^3}\left(\frac{\Ln}{\LB}\right)^2\etai-24\frac{k^3}{8\omega^3}\left(\frac{\Ln}{\LB}\right)^2\etai \right.\\
&\left.\quad+\frac{\Ln}{\LB}\frac{\taue}{2\omega^2}\left(1-\etae\right)+\frac{\taue}{\omega^2}\frac{\Ln}{\LB}\left\lbrace\etae-\frac{\Ln}{\LB}\left[1+\frac{\taue k}{2\omega}\left(1-\etae\right)\right]\right\rbrace-\frac{3\taue^2k}{2\omega^3}\left(\frac{\Ln}{\LB}\right)^2\etae\right\rbrace \\
&\simeq -\frac{k^2}{2}\left\lbrace 1-\frac{k}{2\omega}\left(1+\etai\right)-\frac{3k^2}{8} \left[1-\frac{k}{2\omega}\left(1-\etai\right)\right]+\frac{9k^3}{16\omega}\etai\right. \\
&\left.\quad+\frac{\Ln}{\LB}\left[\frac{1}{2\omega^2}\left(1-\etai\right)+\frac{k}{\omega}\left[1-\frac{k}{2\omega}\left(1-\etai\right)\right]+\frac{\etai}{\omega^2}-\frac{9k^3}{16\omega}-\frac{3k^2}{2\omega^2}\etai +\frac{\taue}{2\omega^2}\left(1-\etae\right)+\frac{\taue}{\omega^2}\etae\right] \right. \\
&\left.\quad+\frac{1}{\omega^2}\left(\frac{\Ln}{\LB}\right)^2\left[-1+\frac{k}{2\omega}\left(1-\etai\right)+\frac{3k^2}{2}\left[1-\frac{k}{2\omega}\left(1-\etai\right)\right]+\frac{3k}{2\omega}\etai-\frac{6k^3}{2\omega}\etai \right.\right. \\
&\left.\left.\quad -\taue\left[1+\frac{\taue k}{2\omega}\left(1-\etae\right)\right]-\frac{3\taue^2k}{2\omega}\etae\right]\right\rbrace \\
&\simeq -\frac{k^2}{2}\left\lbrace 1-\frac{k}{2\omega}\left(1+\etai\right)-\frac{3k^2}{8}+\frac{3k^3}{16\omega}\left(1-\etai\right)+\frac{9k^3}{16\omega}\etai \right. \\
&\left.\quad+\frac{\Ln}{\LB}\left[\frac{1}{2\omega^2}\left(1+\etai\right)+\frac{k}{\omega}\left[1-\frac{k}{2\omega}\left(1+2\etai\right)\right]-\frac{9k^3}{16\omega}+\frac{\taue}{2\omega^2}\left(1+\etae\right)\right] \right. \\
&\left.\quad+\frac{1}{\omega^2}\left(\frac{\Ln}{\LB}\right)^2\left[-\left(1+\taue\right)+\frac{k}{2\omega}\left(1+2\etai\right)+\frac{3k^2}{2}-\frac{3k^3}{4\omega}\left(1-\etai\right)-\frac{6k^3}{2\omega}\etai-\frac{\taue^2 k}{2\omega}\left(1+2\etae\right)\right]\right\rbrace
\end{aligned}
\end{equation*}
\begin{equation}\tag{24}
\begin{aligned}
\IphiQ &\simeq -\frac{k^2}{2}\left\lbrace 1-\frac{k}{2\omega}\left(1+\etai\right)-\frac{3k^2}{8}+\frac{3k^3}{16\omega}\left(1+2\etai\right) \right. \\
&\left.\quad+\frac{\Ln}{\LB}\left[\frac{1}{2\omega^2}\left(1+\etai+\taue+\taue\etae\right)+\frac{k}{\omega}-\frac{k^2}{2\omega^2}\left(1+2\etai\right)-\frac{9k^3}{16\omega}\right] \right. \\
&\left.\quad+\frac{1}{\omega^2}\left(\frac{\Ln}{\LB}\right)^2\left[-\left(1+\taue\right)+\frac{k}{2\omega}\left(1+2\etai-\taue^2-2\taue^2\etae\right)+\frac{3k^2}{2}-\frac{3k^3}{4\omega}\left(1+3\etai\right)\right]\right\rbrace
\end{aligned}
\end{equation}

In the following I will at first keep terms of up to order 1 in $\beta$:
\begin{align*}
\IBA &= 1+2\int dv e^{-v^2}v^3\left(J_1^2\frac{2\betai\omegabari}{k^2\omegabi}+v^2\frac{\betae\omegabare}{2\omegabe}\right) \\
&= 1+2\int dv e^{-v^2}v^3\left[\left(\frac{k}{2}v-\frac{k^3}{16}v^3+\frac{k^5}{384}v^5\right)^2\frac{2\betai\omegabari}{k^2\omegabi}+v^2\frac{\betae\omegabare}{2\omegabe}\right] \\
&= 1+2\int dv e^{-v^2}v^3\left[\frac{2\betai}{k^2}\left(\frac{k^2}{4}v^2-\frac{k^4}{16}v^4+\frac{k^6}{256}v^6+\frac{k^6}{768}v^6\right)\left(1-\frac{k}{2\omega}\left(1-\etai\right) \right.\right.\\
&\left.\left.\quad-\frac{k}{2\omega}\left\lbrace\etai-\frac{\Ln}{\LB}\left[1-\frac{k}{2\omega}\left(1-\etai\right)\right]\right\rbrace v^2 
-\frac{k^2}{4\omega^2}\frac{\Ln}{\LB}\etai v^4\right) \right. \\
&\left.\quad+v^2\frac{\betae}{2}\left(1+\frac{\taue k}{2\omega}\left(1-\etae\right)+\frac{\taue k}{2\omega}\left\lbrace\etae-\frac{\Ln}{\LB}\left[1+\frac{\taue k}{2\omega}\left(1-\etae\right)\right]\right\rbrace v^2-\frac{\taue^2k^2}{4\omega^2}\frac{\Ln}{\LB}\etae v^4\right)\right] \\
&= 1+2\int dv e^{-v^2}v^3\left[\frac{2\betai}{k^2}\left(\frac{k^2}{4}\left[1-\frac{k}{2\omega}\left(1-\etai\right)\right]v^2 \right.\right. \\
&\left.\left.-\left(\frac{k^4}{16}\left[1-\frac{k}{2\omega}\left(1-\etai\right)\right]+\frac{k^2}{4}\frac{k}{2\omega}\left\lbrace\etai-\frac{\Ln}{\LB}\left[1-\frac{k}{2\omega}\left(1-\etai\right)\right]\right\rbrace\right)v^4 \right.\right. \\
&\left.\left.\quad-\left(\frac{k^2}{4}\frac{k^2}{4\omega^2}\frac{\Ln}{\LB}\etai-\frac{k^4}{16}\frac{k}{2\omega}\left\lbrace\etai-\frac{\Ln}{\LB}\left[1-\frac{k}{2\omega}\left(1-\etai\right)\right]\right\rbrace-\frac{k^6}{192}\left[1-\frac{k}{2\omega}\left(1-\etai\right)\right]\right)v^6\right) \right.\\
&\left.\quad+v^2\frac{\betae}{2}\left(1+\frac{\taue k}{2\omega}\left(1-\etae\right)+\frac{\taue k}{2\omega}\left\lbrace\etae-\frac{\Ln}{\LB}\left[1+\frac{\taue k}{2\omega}\left(1-\etae\right)\right]\right\rbrace v^2-\frac{\taue^2k^2}{4\omega^2}\frac{\Ln}{\LB}\etae v^4\right)\right] \\
&= 1+2\int dv e^{-v^2}v^3\left(\left\lbrace\frac{2\betai}{k^2}\frac{k^2}{4}\left[1-\frac{k}{2\omega}\left(1-\etai\right)\right]+\frac{\betae}{2}\left[1+\frac{\taue k}{2\omega}\left(1-\etae\right)\right]\right\rbrace v^2 \right. \\
&\left.\quad+\left[-\frac{2\betai}{k^2}\left(\frac{k^4}{16}\left[1-\frac{k}{2\omega}\left(1-\etai\right)\right]+\frac{k^2}{4}\frac{k}{2\omega}\left\lbrace\etai-\frac{\Ln}{\LB}\left[1-\frac{k}{2\omega}\left(1-\etai\right)\right]\right\rbrace\right) \right.\right. \\
&\left.\left.\qquad+\frac{\betae}{2}\frac{\taue k}{2\omega}\left\lbrace\etae-\frac{\Ln}{\LB}\left[1+\frac{\taue k}{2\omega}\left(1-\etae\right)\right]\right\rbrace\right]v^4-\left[\frac{2\betai}{k^2}\left(\frac{k^2}{4}\frac{k^2}{4\omega^2}\frac{\Ln}{\LB}\etai \right.\right.\right. \\
&\left.\left.\left.\quad-\frac{k^4}{16}\frac{k}{2\omega}\left\lbrace\etai-\frac{\Ln}{\LB}\left[1-\frac{k}{2\omega}\left(1-\etai\right)\right]\right\rbrace-\frac{k^6}{192}\left[1-\frac{k}{2\omega}\left(1-\etai\right)\right]\right)+\frac{\betae}{2}\frac{\taue^2k^2}{4\omega^2}\frac{\Ln}{\LB}\etae\right]v^6\right) \\
&= 1+\frac{2\betai}{k^2}\frac{k^2}{2}\left[1-\frac{k}{2\omega}\left(1-\etai\right)\right]+\betae\left[1+\frac{\taue k}{2\omega}\left(1-\etae\right)\right] \\
&\quad-\frac{12\betai}{k^2}\left(\frac{k^4}{16}\left[1-\frac{k}{2\omega}\left(1-\etai\right)\right]+\frac{k^2}{4}\frac{k}{2\omega}\left\lbrace\etai-\frac{\Ln}{\LB}\left[1-\frac{k}{2\omega}\left(1-\etai\right)\right]\right\rbrace\right) \\
&\qquad+3\betae\frac{\taue k}{2\omega}\left\lbrace\etae-\frac{\Ln}{\LB}\left[1+\frac{\taue k}{2\omega}\left(1-\etae\right)\right]\right\rbrace-24\frac{2\betai}{k^2}\left(\frac{k^2}{4}\frac{k^2}{4\omega^2}\frac{\Ln}{\LB}\etai \right. \\
&\left.\quad-\frac{k^4}{16}\frac{k}{2\omega}\left\lbrace\etai-\frac{\Ln}{\LB}\left[1-\frac{k}{2\omega}\left(1-\etai\right)\right]\right\rbrace-\frac{k^6}{192}\left[1-\frac{k}{2\omega}\left(1-\etai\right)\right]\right)-3\betae\frac{\taue^2k^2}{\omega^2}\frac{\Ln}{\LB}\etae \\
&= 1+\frac{2\betai}{k^2}\frac{k^2}{2}\left[1-\frac{k}{2\omega}\left(1-\etai\right)\right]+\betae\left[1+\frac{\taue k}{2\omega}\left(1-\etae\right)\right]-\frac{12\betai}{k^2}\left\lbrace\frac{k^4}{16}\left[1-\frac{k}{2\omega}\left(1-\etai\right)\right]+\frac{k^2}{4}\frac{k}{2\omega}\etai\right\rbrace \\
&\quad+3\betae\frac{\taue k}{2\omega}\etae+24\frac{2\betai}{k^2}\left\lbrace\frac{k^4}{16}\frac{k}{2\omega}\etai+\frac{k^6}{192}\left[1-\frac{k}{2\omega}\left(1-\etai\right)\right]\right\rbrace+3\frac{\Ln}{\LB}\left(\betai\frac{k}{2\omega}\left[1-\frac{k}{2\omega}\left(1-\etai\right)\right] \right.\\
&\left.\quad-\betae\frac{\taue k}{2\omega}\left[1+\frac{\taue k}{2\omega}\left(1-\etae\right)\right]-8\frac{2\betai}{k^2}\left\lbrace\frac{k^2}{4}\frac{k^2}{4\omega^2}\etai+\frac{k^4}{16}\frac{k}{2\omega}\left[1-\frac{k}{2\omega}\left(1-\etai\right)\right]\right\rbrace-\betae\frac{\taue^2k^2}{\omega^2}\etae\right) \\
&= 1+\betai\left[1-\frac{k}{2\omega}\left(1-\etai\right)\right]+\betae\left[1+\frac{\taue k}{2\omega}\left(1-\etae\right)\right]-3\betai\left\lbrace\frac{k^2}{4}\left[1-\frac{k}{2\omega}\left(1-\etai\right)\right]+\frac{k}{2\omega}\etai\right\rbrace \\
&\quad+3\betae\frac{\taue k}{2\omega}\etae+3\betai\left\lbrace k^2\frac{k}{2\omega}\etai+\frac{k^4}{12}\left[1-\frac{k}{2\omega}\left(1-\etai\right)\right]\right\rbrace+3\frac{\Ln}{\LB}\left(\betai\frac{k}{2\omega}\left[1-\frac{k}{2\omega}\left(1-\etai\right)\right] \right.\\
&\left.\quad-\betae\frac{\taue k}{2\omega}\left[1+\frac{\taue k}{2\omega}\left(1-\etae\right)\right]-\betai\left\lbrace\frac{k^2}{\omega^2}\etai+k^2\frac{k}{2\omega}\left[1-\frac{k}{2\omega}\left(1-\etai\right)\right]\right\rbrace-\betae\frac{\taue^2k^2}{\omega^2}\etae\right)
\end{align*}
\begin{align*}
\IBA &= 1+\betai\left(1-\frac{k}{2\omega}\left(1-\etai\right)+\taue\left[1+\frac{\taue k}{2\omega}\left(1-\etae\right)\right]-3\left\lbrace\frac{k^2}{4}\left[1-\frac{k}{2\omega}\left(1-\etai\right)\right]+\frac{k}{2\omega}\etai\right\rbrace+3\frac{\taue^2 k}{2\omega}\etae \right. \\
&\left.\quad+3\left\lbrace k^2\frac{k}{2\omega}\etai+\frac{k^4}{12}\left[1-\frac{k}{2\omega}\left(1-\etai\right)\right]\right\rbrace+3\frac{\Ln}{\LB}\left\lbrace\frac{k}{2\omega}\left[1-\frac{k}{2\omega}\left(1-\etai\right)\right] -\frac{\taue^2 k}{2\omega}\left[1+\frac{\taue k}{2\omega}\left(1-\etae\right)\right] \right.\right.\\
&\left.\left.\quad-\frac{k^2}{\omega^2}\etai-k^2\frac{k}{2\omega}\left[1-\frac{k}{2\omega}\left(1-\etai\right)\right]-\frac{\taue^3k^2}{\omega^2}\etae\right\rbrace\right) \\
&= 1+\betai\left(1+\taue+\frac{k}{2\omega}\left[-\left(1-\etai\right)+\taue^2\left(1-\etae\right)-3\etai+3\taue^2\etae\right]-3\frac{k^2}{4}+3\frac{k^3}{2\omega}\left[\frac{1}{4}\left(1-\etai\right)+\etai\right] \right. \\
&\left.\quad+3\frac{k^4}{12}\left[1-\frac{k}{2\omega}\left(1-\etai\right)\right]+3\frac{\Ln}{\LB}\left\lbrace\frac{k}{2\omega}\left(1-\taue^2\right)-\left(\frac{k}{2\omega}\right)^2\left[\left(1-\etai\right)+\taue^3 \left(1-\etae\right)+4\etai+4\taue^3\etae\right] \right.\right.\\
&\left.\left.\quad-k^2\frac{k}{2\omega}\left[1-\frac{k}{2\omega}\left(1-\etai\right)\right]\right\rbrace\right)
\end{align*}

\begin{equation}\tag{25}
\begin{aligned}
\IBA &= 1+\betai\left(1+\taue+\frac{k}{2\omega}\left[-\left(1+2\etai\right)+\taue^2\left(1+2\etae\right)\right]-3\frac{k^2}{4}\left\lbrace1-\frac{k}{2\omega}\left(1+3\etai\right) \right.\right. \\
&\left.\left.\quad-\frac{k^2}{3}\left[1-\frac{k}{2\omega}\left(1-\etai\right)\right]\right\rbrace+3\frac{\Ln}{\LB}\left\lbrace\frac{k}{2\omega}\left(1-\taue^2\right)-\left(\frac{k}{2\omega}\right)^2\left[\left(1+3\etai\right)+\taue^3 \left(1+3\etae\right)\right] \right.\right.\\
&\left.\left.\quad-k^2\frac{k}{2\omega}\left[1-\frac{k}{2\omega}\left(1-\etai\right)\right]\right\rbrace\right)
\end{aligned}
\end{equation}

Now the $\IphiA$ integral is:
\begin{align*}
\frac{1}{\betai}\IphiA &= \int dv e^{-v^2}v^2\left(-2J_0J_1\frac{\omegabari}{k\omegabi}+v\frac{\omegabare}{\omegabe}\right) \\
&= \int dv e^{-v^2}v^2\left[-2\left(1-\frac{k^2}{4}v^2+\frac{k^4}{64}v^4-\frac{k^6}{2304}v^6\right)\left(\frac{k}{2}v-\frac{k^3}{16}v^3+\frac{k^5}{384}v^5-\frac{k^7}{18432}v^7\right)\frac{\omegabari}{k\omegabi}+v\frac{\omegabare}{\omegabe}\right] \\
&= \int dv e^{-v^2}v^2\left[-\frac{2}{k}\left(\frac{k}{2}v-\frac{3k^3}{16}v^3+\frac{5k^5}{192}v^5-\frac{35k^7}{18432}v^7\right)\left(1-\frac{k}{2\omega}\left(1-\etai\right) \right.\right.\\
&\left.\left.\quad-\frac{k}{2\omega}\left\lbrace\etai-\frac{\Ln}{\LB}\left[1-\frac{k}{2\omega}\left(1-\etai\right)\right]\right\rbrace v^2-\frac{k^2}{4\omega^2}\frac{\Ln}{\LB}\etai v^4\right)+v\left(1+\frac{\taue k}{2\omega}\left(1-\etae\right) \right.\right.\\
&\left.\left.\quad+\frac{\taue k}{2\omega}\left\lbrace\etae-\frac{\Ln}{\LB}\left[1+\frac{\taue k}{2\omega}\left(1-\etae\right)\right]\right\rbrace v^2-\frac{\taue^2k^2}{4\omega^2}\frac{\Ln}{\LB}\etae v^4\right)\right] \\
&= \int dv e^{-v^2}v^3\left[-1+\frac{k}{2\omega}\left(1-\etai\right)+1+\frac{\taue k}{2\omega}\left(1-\etae\right)+\left(\frac{3k^2}{8}\left[1-\frac{k}{2\omega}\left(1-\etai\right)\right] \right.\right.\\
&\left.\left.\quad+\frac{k}{2\omega}\left\lbrace\etai-\frac{\Ln}{\LB}\left[1-\frac{k}{2\omega}\left(1-\etai\right)\right]\right\rbrace+\frac{\taue k}{2\omega}\left\lbrace\etae-\frac{\Ln}{\LB}\left[1+\frac{\taue k}{2\omega}\left(1-\etae\right)\right]\right\rbrace\right)v^2 \right.\\
&\left.\quad-\left(\frac{5k^4}{96}\left[1-\frac{k}{2\omega}\left(1-\etai\right)\right]+\frac{3k^2}{8}\frac{k}{2\omega}\left\lbrace\etai-\frac{\Ln}{\LB}\left[1-\frac{k}{2\omega}\left(1-\etai\right)\right]\right\rbrace-\frac{k^2}{4\omega^2}\frac{\Ln}{\LB}\etai+\frac{\taue^2k^2}{4\omega^2}\frac{\Ln}{\LB}\etae\right)v^4 \right.\\
&\left.\quad+\left(\frac{35k^6}{9216}\left[1-\frac{k}{2\omega}\left(1-\etai\right)\right]+\frac{5k^4}{96}\frac{k}{2\omega}\left\lbrace\etai-\frac{\Ln}{\LB}\left[1-\frac{k}{2\omega}\left(1-\etai\right)\right]\right\rbrace-\frac{3k^2}{8}\frac{k^2}{4\omega^2}\frac{\Ln}{\LB}\etai\right)v^6\right]
\end{align*}

\begin{align*}
\frac{2}{\betai}\IphiA &= -1+\frac{k}{2\omega}\left(1-\etai\right)+1+\frac{\taue k}{2\omega}\left(1-\etae\right)+\frac{3k^2}{4}\left[1-\frac{k}{2\omega}\left(1-\etai\right)\right] \\
&\quad+2\frac{k}{2\omega}\left\lbrace\etai-\frac{\Ln}{\LB}\left[1-\frac{k}{2\omega}\left(1-\etai\right)\right]\right\rbrace+2\frac{\taue k}{2\omega}\left\lbrace\etae-\frac{\Ln}{\LB}\left[1+\frac{\taue k}{2\omega}\left(1-\etae\right)\right]\right\rbrace \\
&\quad-\frac{5k^4}{16}\left[1-\frac{k}{2\omega}\left(1-\etai\right)\right]-3\frac{3k^2}{4}\frac{k}{2\omega}\left\lbrace\etai-\frac{\Ln}{\LB}\left[1-\frac{k}{2\omega}\left(1-\etai\right)\right]\right\rbrace+6\frac{k^2}{4\omega^2}\frac{\Ln}{\LB}\etai-6\frac{\taue^2k^2}{4\omega^2}\frac{\Ln}{\LB}\etae \\
&\quad+\frac{35k^6}{384}\left[1-\frac{k}{2\omega}\left(1-\etai\right)\right]+\frac{5k^4}{4}\frac{k}{2\omega}\left\lbrace\etai-\frac{\Ln}{\LB}\left[1-\frac{k}{2\omega}\left(1-\etai\right)\right]\right\rbrace-9k^2\frac{k^2}{4\omega^2}\frac{\Ln}{\LB}\etai \\
\end{align*}
\begin{align*}
\frac{1}{\betai}\IphiA &= \frac{k}{4\omega}\left[1-\etai+\taue\left(1-\etae\right)\right]+\frac{3k^2}{8}\left[1-\frac{k}{2\omega}\left(1-\etai\right)\right] \\
&\quad+\frac{k}{2\omega}\left\lbrace\etai-\frac{\Ln}{\LB}\left[1-\frac{k}{2\omega}\left(1-\etai\right)\right]\right\rbrace+\frac{\taue k}{2\omega}\left\lbrace\etae-\frac{\Ln}{\LB}\left[1+\frac{\taue k}{2\omega}\left(1-\etae\right)\right]\right\rbrace \\
&\quad-\frac{5k^4}{32}\left[1-\frac{k}{2\omega}\left(1-\etai\right)\right]-3\frac{3k^2}{4}\frac{k}{4\omega}\left\lbrace\etai-\frac{\Ln}{\LB}\left[1-\frac{k}{2\omega}\left(1-\etai\right)\right]\right\rbrace+3\frac{k^2}{4\omega^2}\frac{\Ln}{\LB}\etai-3\frac{\taue^2k^2}{4\omega^2}\frac{\Ln}{\LB}\etae \\
&\quad+\frac{35k^6}{768}\left[1-\frac{k}{2\omega}\left(1-\etai\right)\right]+\frac{5k^4}{4}\frac{k}{4\omega}\left\lbrace\etai-\frac{\Ln}{\LB}\left[1-\frac{k}{2\omega}\left(1-\etai\right)\right]\right\rbrace-\frac{9k^2}{2}\frac{k^2}{4\omega^2}\frac{\Ln}{\LB}\etai \\
&= \frac{k}{4\omega}\left[1+\etai+\taue\left(1+\etae\right)\right]+\frac{3k^2}{8}\left[1-\frac{k}{2\omega}\left(1-\etai\right)-3\frac{k}{2\omega}\etai\right] \\
&\quad+\frac{\Ln}{\LB}\left(-\frac{k}{2\omega}\left\lbrace1-\frac{k}{2\omega}\left(1-\etai\right)+\taue\left[1+\frac{\taue k}{2\omega}\left(1-\etae\right)\right]-3\frac{k}{2\omega}\etai+3\taue\frac{\taue k}{2\omega}\etae\right\rbrace\right) \\
&\quad-\frac{5k^4}{32}\left[1-\frac{k}{2\omega}\left(1-\etai\right)\right]+3\frac{3k^2}{4}\frac{k}{4\omega}\frac{\Ln}{\LB}\left[1-\frac{k}{2\omega}\left(1-\etai\right)\right]-3\frac{3k^2}{4}\frac{k}{4\omega}\frac{\Ln}{\LB}\frac{k}{2\omega}4\etai \\
&\quad+\frac{35k^6}{768}\left[1-\frac{k}{2\omega}\left(1-\etai\right)\right]+\frac{5k^4}{4}\frac{k}{4\omega}\left\lbrace\etai-\frac{\Ln}{\LB}\left[1-\frac{k}{2\omega}\left(1-\etai\right)\right]\right\rbrace
\end{align*}
\begin{equation} \tag{26}
\begin{aligned}
\frac{1}{\betai}\IphiA &= \frac{k}{4\omega}\left[1+\etai+\taue\left(1+\etae\right)\right]+\frac{3k^2}{8}\left\lbrace1-\frac{k}{2\omega}\left(1+2\etai\right)+3\frac{k}{2\omega}\frac{\Ln}{\LB}\left[1-\frac{k}{2\omega}\left(1+3\etai\right)\right]\right\rbrace \\
&\quad+\frac{\Ln}{\LB}\left(-\frac{k}{2\omega}\left\lbrace1-\frac{k}{2\omega}\left(1+2\etai\right)+\taue\left[1+\frac{\taue k}{2\omega}\left(1+2\etae\right)\right]\right\rbrace\right) \\
&\quad-\frac{5k^4}{32}\left\lbrace1-\frac{k}{2\omega}\left(1+3\etai\right) 
-\frac{7k^2}{24}\left[1-\frac{k}{2\omega}\left(1-\etai\right)\right]-4\frac{k}{2\omega}\frac{\Ln}{\LB}\left[1-\frac{k}{2\omega}\left(1-\etai\right)\right]\right\rbrace
\end{aligned}
\end{equation}

\pagebreak
We briefly examine what happens if the assumption $1/\LB=0$ is used together with $p_0'\neq0$.
The product $\IphiQ\IBA$ is:
\begin{align*}
\IphiQ\IBA &= -\frac{k^2}{2}\left\lbrace 1-\frac{k}{2\omega}\left(1+\etai\right)-\frac{3k^2}{8}+\frac{3k^3}{16\omega}\left(1+2\etai\right) \right. \\
&\left.\quad+\frac{\Ln}{\LB}\left[\frac{1}{2\omega^2}\left(1+\etai+\taue+\taue\etae\right)+\frac{k}{\omega}-\frac{k^2}{2\omega^2}\left(1+2\etai\right)-\frac{9k^3}{16\omega}\right] \right. \\
&\left.\quad+\frac{1}{\omega^2}\left(\frac{\Ln}{\LB}\right)^2\left[-\left(1+\taue\right)+\frac{k}{2\omega}\left(1+2\etai-\taue^2-2\taue^2\etae\right)+\frac{3k^2}{2}-\frac{3k^3}{4\omega}\left(1+3\etai\right)\right]\right\rbrace \\
&\quad\left[1+\betai\left(1+\taue+\frac{k}{2\omega}\left[-\left(1+2\etai\right)+\taue^2\left(1+2\etae\right)\right]-3\frac{k^2}{4}\left\lbrace1-\frac{k}{2\omega}\left(1+3\etai\right) \right.\right.\right. \\
&\left.\left.\left.\quad-\frac{k^2}{3}\left[1-\frac{k}{2\omega}\left(1-\etai\right)\right]\right\rbrace+3\frac{\Ln}{\LB}\left\lbrace\frac{k}{2\omega}\left(1-\taue^2\right)-\left(\frac{k}{2\omega}\right)^2\left[\left(1+3\etai\right)+\taue^3 \left(1+3\etae\right)\right] \right.\right.\right.\\
&\left.\left.\left.\quad-k^2\frac{k}{2\omega}\left[1-\frac{k}{2\omega}\left(1-\etai\right)\right]\right\rbrace\right)\right]	
\end{align*}
keeping terms proportional to $k^2$ only:
\begin{align*}
\IphiQ\IBA &= -\frac{k^2}{2}\left[ 1+\frac{\Ln}{\LB}\frac{1}{2\omega^2}\left(1+\etai+\taue+\taue\etae\right)-\frac{1}{\omega^2}\left(\frac{\Ln}{\LB}\right)^2\left(1+\taue\right)\right]\left[1+\betai\left(1+\taue\right)\right]	\\
&= -\frac{k^2}{2}\left\lbrace 1+\betai\left(1+\taue\right)+\frac{\Ln}{\LB}\frac{1}{2\omega^2}\left(1+\etai+\taue+\taue\etae\right)\left[1+\betai\left(1+\taue\right)\right] \right.\\
&\left.\quad-\frac{1}{\omega^2}\left(\frac{\Ln}{\LB}\right)^2\left(1+\taue\right)\left[1+\betai\left(1+\taue\right)\right]	\right\rbrace	\\
&= -\frac{k^2}{2\omega^2}\left\lbrace \left[1+\betai\left(1+\taue\right)\right]\omega^2+\frac{\Ln}{\LB}\frac{\alpha_0}{2}\left[1+\betai\left(1+\taue\right)\right] \right.\\
&\left.\quad-\left(\frac{\Ln}{\LB}\right)^2\left(1+\taue\right)\left[1+\betai\left(1+\taue\right)\right]	\right\rbrace
\end{align*}
or retaining only $\mathcal{O}\left(\beta^2\right)$ terms:
\begin{equation} \tag{27a}
\IphiQ\IBA = -\frac{k^2}{2\omega^2}\left\lbrace \left[1+\betai\left(1+\taue\right)\right]\omega^2+\frac{\Ln}{\LB}\frac{\alpha_0}{2}\left[1+\betai\left(1+\taue\right)\right]-\left(\frac{\Ln}{\LB}\right)^2\left(1+\taue\right)	\right\rbrace
\end{equation}
The right side of the dispersion relation has, to order 2 in $k$:
\begin{align*}
\frac{2}{\betai}\IphiA^2 &= 2\betai\left(\frac{k}{4\omega}\left[1+\etai+\taue\left(1+\etae\right)\right]+\frac{3k^2}{8} \right.\\
&\left.\quad-\frac{\Ln}{\LB}\frac{k}{2\omega}\left\lbrace1-\frac{k}{2\omega}\left(1+2\etai\right)+\taue\left[1+\frac{\taue k}{2\omega}\left(1+2\etae\right)\right]\right\rbrace \right)^2 \\
&= \frac{\betai k^2}{2\omega^2}\left\lbrace\frac{1}{2}\left[1+\etai+\taue\left(1+\etae\right)\right]-\frac{\Ln}{\LB}\left(1+\taue\right) \right\rbrace^2 \\
&= \frac{\betai k^2}{2\omega^2}\left[\frac{\alpha_0^2}{4}-\LnOLB\alpha_0\left(1+\taue\right)+\left(\LnOLB\right)^2\left(1+\taue\right)^2 \right]
\end{align*}
and to $\mathcal{O}\left(\beta^2\right)$ this is simply
\begin{equation} \tag{27b}
\frac{2}{\betai}\IphiA^2 = \frac{\betai k^2}{2\omega^2}\left[\frac{\alpha_0^2}{4}-\LnOLB\alpha_0\left(1+\taue\right)\right]
\end{equation}
Such that the dispersion relation is
\begin{align*}
&-\frac{k^2}{2\omega^2}\left\lbrace \left[1+\betai\left(1+\taue\right)\right]\omega^2+\frac{\Ln}{\LB}\frac{\alpha_0}{2}\left[1+\betai\left(1+\taue\right)\right] \right.\\
&\left.\quad -\left(\frac{\Ln}{\LB}\right)^2\left(1+\taue\right)	\right\rbrace = \frac{\betai k^2}{2\omega^2}\left[\frac{\alpha_0^2}{4}-\LnOLB\alpha_0\left(1+\taue\right)\right] \\
&-\left[1+\betai\left(1+\taue\right)\right]\omega^2-\frac{\Ln}{\LB}\frac{\alpha_0}{2}\left[1+\betai\left(1+\taue\right)\right] \\
&\quad +\left(\frac{\Ln}{\LB}\right)^2\left(1+\taue\right)	 = \betai \frac{\alpha_0^2}{4}-\LnOLB\betai \alpha_0\left(1+\taue\right) \\
&\left[1+\betai\left(1+\taue\right)\right]\omega^2=-\frac{\Ln}{\LB}\frac{\alpha_0}{2}\left[1+\betai\left(1+\taue\right)\right] \\
&\quad +\left(\frac{\Ln}{\LB}\right)^2\left(1+\taue\right)-\betai \frac{\alpha_0^2}{4}+\LnOLB\betai \alpha_0\left(1+\taue\right) \\
&\omega^2=\left[1-\betai\left(1+\taue\right)\right]\left[-\betai \frac{\alpha_0^2}{4}-\frac{\Ln}{\LB}\frac{\alpha_0}{2}\left[1-\betai\left(1+\taue\right)\right]+\left(\frac{\Ln}{\LB}\right)^2\left(1+\taue\right)\right]
\end{align*}
Noting that $\betai\left(1+\taue\right)=\beta$ and keeping only $\mathcal{O}\left(\beta^2\right)$ terms this becomes
\begin{align*}
\omega^2 &= -\betai \frac{\alpha_0^2}{4}-\frac{\Ln}{\LB}\frac{\alpha_0}{2}\left[1-\betai\left(1+\taue\right)\right]+\left(\frac{\Ln}{\LB}\right)^2\left(1+\taue\right)+\beta\betai\frac{\alpha_0^2}{4}+\LnOLB\frac{\alpha_0}{2}\beta \\
&= -\betai \frac{\alpha_0^2}{4}-\frac{\Ln}{\LB}\frac{\alpha_0}{2}\left(1-\beta\right)+\left(\frac{\Ln}{\LB}\right)^2\left(1+\taue\right)+\frac{\beta\alpha_0}{4}\left(\betai\alpha_0+2\LnOLB\right)
\end{align*}
\begin{equation} \tag{28}
\omega^2 = -\betai \frac{\alpha_0^2}{4}-\frac{\Ln}{\LB}\frac{\alpha_0}{2}+\left(\frac{\Ln}{\LB}\right)^2\left(1+\taue\right)+\frac{\beta\alpha_0}{4}\left(\betai\alpha_0+4\LnOLB\right)
\end{equation}
Here if one takes $1/\LB=0$ one obtains, to lowest order, the GDC growth rate.

Let us step back from GDC and instead simplify the general dispersion relation by keeping only $\mathcal{O}\left(\beta\right)$ terms, assuming $\omega\propto k$ and using $\Ln/\LB=-\betai\alpha_0/2$, where $\alpha_0=1+\etai+\taue\left(1+\etae\right)$, $\alpha_1=1+\etai$, $\alpha_2=1+2\etai$, $\alpha_3=1+2\etai-\taue^2\left(1+2\etae\right)$, $u=2\omega/k$:
\begin{align*}
\IphiQ &\simeq -\frac{k^2}{2}\left\lbrace 1-\frac{k}{2\omega}\left(1+\etai\right)-\frac{3k^2}{8}+\frac{3k^3}{16\omega}\left(1+2\etai\right) \right. \\
&\left.\quad+\frac{\Ln}{\LB}\left[\frac{1}{2\omega^2}\left(1+\etai+\taue+\taue\etae\right)+\frac{k}{\omega}-\frac{k^2}{2\omega^2}\left(1+2\etai\right)-\frac{9k^3}{16\omega}\right]\right\rbrace
\end{align*}
\begin{equation}\tag{29a}
\IphiQ \simeq -\frac{k^2}{2}\left\lbrace 1-\frac{3k^2}{8}-\frac{\alpha_1}{u}-\frac{3k^2\alpha_2}{8u^2}-\frac{\betai\alpha_0}{2}\left[\frac{\alpha_0}{2\omega^2}+\frac{1}{u}\left(2-\frac{9k^2}{8}\right)-\frac{2\alpha_2}{u^2}\right]\right\rbrace
\end{equation}

\begin{align*}
\IBA &= 1+\betai\left(1+\taue+\frac{k}{2\omega}\left[-\left(1+2\etai\right)+\taue^2\left(1+2\etae\right)\right]-3\frac{k^2}{4}\left\lbrace1-\frac{k}{2\omega}\left(1+3\etai\right) \right.\right. \\
&\left.\left.\quad-\frac{k^2}{3}\left[1-\frac{k}{2\omega}\left(1-\etai\right)\right]\right\rbrace+3\frac{\Ln}{\LB}\left\lbrace\frac{k}{2\omega}\left(1-\taue^2\right)-\left(\frac{k}{2\omega}\right)^2\left[\left(1+3\etai\right)+\taue^3 \left(1+3\etae\right)\right] \right.\right.\\
&\left.\left.\quad-k^2\frac{k}{2\omega}\left[1-\frac{k}{2\omega}\left(1-\etai\right)\right]\right\rbrace\right) \\
&= 1+\betai\left\lbrace1+\taue-\frac{\alpha_3}{u}-\frac{3k^2}{4}\left[1-\frac{\alpha_2+\etai}{u}-\frac{k^2}{3}\left(1-\frac{\alpha_1+2\etai}{u}\right)\right]\right\rbrace \\
&= 1+\betai\left\lbrace1+\taue-\frac{3k^2}{4}+\frac{k^4}{4}-\frac{\alpha_3}{u}+\frac{3k^2}{4}\left[\frac{\alpha_2+\etai}{u}-\frac{k^2}{3}\frac{\alpha_1+2\etai}{u}\right]\right\rbrace
\end{align*}
\begin{equation} \tag{29b}
\IBA = 1+\betai\left(1+\taue-\frac{k^2}{4}\left(3-k^2\right)-\frac{1}{u}\left\lbrace\alpha_3-\frac{k^2}{4}\left[3\left(\alpha_2+\etai\right)-k^2\left(\alpha_1+2\etai\right)\right]\right\rbrace\right)
\end{equation}

\begin{align*}
\frac{1}{\betai}\IphiA &= \frac{k}{4\omega}\left[1+\etai+\taue\left(1+\etae\right)\right]+\frac{3k^2}{8}\left\lbrace1-\frac{k}{2\omega}\left(1+2\etai\right)+3\frac{k}{2\omega}\frac{\Ln}{\LB}\left[1-\frac{k}{2\omega}\left(1+3\etai\right)\right]\right\rbrace \\
&\quad+\frac{\Ln}{\LB}\left(-\frac{k}{2\omega}\left\lbrace1-\frac{k}{2\omega}\left(1+2\etai\right)+\taue\left[1+\frac{\taue k}{2\omega}\left(1+2\etae\right)\right]\right\rbrace\right) \\
&\quad-\frac{5k^4}{32}\left\lbrace1-\frac{k}{2\omega}\left(1+3\etai\right) 
-\frac{7k^2}{24}\left[1-\frac{k}{2\omega}\left(1-\etai\right)\right]-4\frac{k}{2\omega}\frac{\Ln}{\LB}\left[1-\frac{k}{2\omega}\left(1-\etai\right)\right]\right\rbrace \\
&= \frac{\alpha_0}{2u}+\frac{3k^2}{8}\left(1-\frac{\alpha_2}{u}\right)-\frac{5k^4}{32}\left\lbrace1-\frac{k}{2\omega}\left(1+3\etai\right) 
-\frac{7k^2}{24}\left[1-\frac{k}{2\omega}\left(1-\etai\right)\right]\right\rbrace \\
&= \frac{\alpha_0}{2u}+\frac{3k^2}{8}\left(1-\frac{\alpha_2}{u}\right)-\frac{5k^4}{32}\left\lbrace1-\frac{\alpha_2+\etai}{u}
-\frac{7k^2}{24}\left[1-\frac{\alpha_1-2\etai}{u}\right]\right\rbrace
\end{align*}
\begin{equation}\tag{29c}
\frac{1}{\betai}\IphiA = \frac{3k^2}{8}-\frac{5k^4}{32}+\frac{35k^6}{768}+\frac{1}{2u}\left\lbrace\alpha_0-\frac{3k^2\alpha_2}{4}+\frac{5k^4}{384}\left[24\left(\alpha_2+\etai\right)
-7k^2\left(\alpha_1-2\etai\right)\right]\right\rbrace
\end{equation}

The dispersion relation is then:
\begin{align*}
\IphiQ\IBA &= 2\betai\left(\frac{1}{\betai}\IphiA\right)^2 \\
\text{LHS} &=  -\frac{k^2}{2}\left\lbrace 1-\frac{3k^2}{8}-\frac{\alpha_1}{u}-\frac{3k^2\alpha_2}{8u^2}-\frac{\betai\alpha_0}{2}\left[\frac{\alpha_0}{2\omega^2}+\frac{1}{u}\left(2-\frac{9k^2}{8}\right)-\frac{2\alpha_2}{u^2}\right]\right\rbrace \\
&\quad\left[1+\betai\left(1+\taue-\frac{k^2}{4}\left(3-k^2\right)-\frac{1}{u}\left\lbrace\alpha_3-\frac{k^2}{4}\left[3\left(\alpha_2+\etai\right)-k^2\left(\alpha_1+2\etai\right)\right]\right\rbrace\right)\right] \\
&=  -\frac{k^2}{2}\left\lbrace 1-\frac{3k^2}{8}-\frac{\alpha_1}{u}-\frac{3k^2\alpha_2}{8u^2}-\frac{\betai\alpha_0}{2}\left[\frac{2\alpha_0}{k^2u^2}+\frac{1}{u}\left(2-\frac{9k^2}{8}\right)-\frac{2\alpha_2}{u^2}\right]\right. \\
&\left.\quad+\betai\left(1-\frac{3k^2}{8}\right)\left[1+\taue-\frac{k^2}{4}\left(3-k^2\right)\right]-\frac{\betai}{u}\left(\alpha_1\left[1+\taue-\frac{k^2}{4}\left(3-k^2\right)\right] \right.\right.\\
&\left.\left.\quad+\left(1-\frac{3k^2}{8}\right)\left\lbrace\alpha_3-\frac{k^2}{4}\left[3\left(\alpha_2+\etai\right)-k^2\left(\alpha_1+2\etai\right)\right]\right\rbrace\right)-\frac{\betai}{u^2}\left(\frac{3k^2\alpha_2}{8}\left[1+\taue-\frac{k^2}{4}\left(3-k^2\right)\right] \right.\right.\\
&\left.\left.\quad-\alpha_1\left\lbrace\alpha_3-\frac{k^2}{4}\left[3\left(\alpha_2+\etai\right)-k^2\left(\alpha_1+2\etai\right)\right]\right\rbrace\right)\right\rbrace \\
&=  -\frac{k^2}{2}\left\lbrace 1-\frac{3k^2}{8}+\betai\left(1-\frac{3k^2}{8}\right)\left[1+\taue-\frac{k^2}{4}\left(3-k^2\right)\right]-\frac{1}{u}\left[\alpha_1+\betai\left(\alpha_0\left(1-\frac{9k^2}{16}\right) \right.\right.\right.\\
&\left.\left.\left.\quad+\alpha_1\left[1+\taue-\frac{k^2}{4}\left(3-k^2\right)\right]+\left(1-\frac{3k^2}{8}\right)\left\lbrace\alpha_3-\frac{k^2}{4}\left[3\left(\alpha_2+\etai\right)-k^2\left(\alpha_1+2\etai\right)\right]\right\rbrace\right)\right] \right.\\
&\left.\quad-\frac{1}{u^2}\left[\frac{3k^2\alpha_2}{8}+\betai\left(\frac{\alpha_0^2}{k^2}-\alpha_0\alpha_2+\frac{3k^2\alpha_2}{8}\left[1+\taue-\frac{k^2}{4}\left(3-k^2\right)\right] \right.\right.\right.\\
&\left.\left.\left.\quad-\alpha_1\left\lbrace\alpha_3-\frac{k^2}{4}\left[3\left(\alpha_2+\etai\right)-k^2\left(\alpha_1+2\etai\right)\right]\right\rbrace\right)\right]\right\rbrace \\
\end{align*}

\begin{align*}
\left(\frac{1}{\betai}\IphiA\right)^2 &= \frac{k^2}{4}\left(\frac{3k}{4}-\frac{5k^3}{16}+\frac{35k^5}{384}+\frac{1}{u}\left\lbrace\frac{\alpha_0}{k}-\frac{3k\alpha_2}{4}+\frac{5k^3}{384}\left[24\left(\alpha_2+\etai\right)
-7k^2\left(\alpha_1-2\etai\right)\right]\right\rbrace\right)^2 \\
&= \frac{k^2}{4}\left(\left(\frac{3k}{4}-\frac{5k^3}{16}+\frac{35k^5}{384}\right)^2+\frac{2}{u}\left(\frac{3k}{4}-\frac{5k^3}{16}+\frac{35k^5}{384}\right)\left\lbrace\frac{\alpha_0}{k}-\frac{3k\alpha_2}{4}+\frac{5k^3}{384}\left[24\left(\alpha_2+\etai\right) \right.\right.\right.\\
&\left.\left.\left.\quad-7k^2\left(\alpha_1-2\etai\right)\right]\right\rbrace +\frac{1}{u^2}\left\lbrace\frac{\alpha_0}{k}-\frac{3k\alpha_2}{4}+\frac{5k^3}{384}\left[24\left(\alpha_2+\etai\right)
-7k^2\left(\alpha_1-2\etai\right)\right]\right\rbrace^2\right)
\end{align*}

Then the dispersion relation reads:
\begin{align*}
&1-\frac{3k^2}{8}+\betai\left(1-\frac{3k^2}{8}\right)\left[1+\taue-\frac{k^2}{4}\left(3-k^2\right)\right]-\frac{1}{u}\left[\alpha_1+\betai\left(\alpha_0\left(1-\frac{9k^2}{16}\right) \right.\right.\\
&\left.\left.\quad+\alpha_1\left[1+\taue-\frac{k^2}{4}\left(3-k^2\right)\right]+\left(1-\frac{3k^2}{8}\right)\left\lbrace\alpha_3-\frac{k^2}{4}\left[3\left(\alpha_2+\etai\right)-k^2\left(\alpha_1+2\etai\right)\right]\right\rbrace\right)\right] \\
&\quad-\frac{1}{u^2}\left[\frac{3k^2\alpha_2}{8}+\betai\left(\frac{\alpha_0^2}{k^2}-\alpha_0\alpha_2+\frac{3k^2\alpha_2}{8}\left[1+\taue-\frac{k^2}{4}\left(3-k^2\right)\right] \right.\right.\\
&\left.\left.\quad-\alpha_1\left\lbrace\alpha_3-\frac{k^2}{4}\left[3\left(\alpha_2+\etai\right)-k^2\left(\alpha_1+2\etai\right)\right]\right\rbrace\right)\right]+\betai\left(\left(\frac{3k}{4}-\frac{5k^3}{16}+\frac{35k^5}{384}\right)^2 \right. \\
&\left.\quad+\frac{2}{u}\left(\frac{3k}{4}-\frac{5k^3}{16}+\frac{35k^5}{384}\right)\left\lbrace\frac{\alpha_0}{k}-\frac{3k\alpha_2}{4}+\frac{5k^3}{384}\left[24\left(\alpha_2+\etai\right)
-7k^2\left(\alpha_1-2\etai\right)\right]\right\rbrace \right.\\
&\left.\quad+\frac{1}{u^2}\left\lbrace\frac{\alpha_0}{k}-\frac{3k\alpha_2}{4}+\frac{5k^3}{384}\left[24\left(\alpha_2+\etai\right)
-7k^2\left(\alpha_1-2\etai\right)\right]\right\rbrace^2\right) = 0
\end{align*}
\begin{align*}
&1-\frac{3k^2}{8}+\betai\left\lbrace\left(1-\frac{3k^2}{8}\right)\left[1+\taue-\frac{k^2}{4}\left(3-k^2\right)\right]+\frac{k^2}{16}\left(3-\frac{5k^2}{4}+\frac{35k^4}{96}\right)^2\right\rbrace-\frac{1}{u}\left[\alpha_1+\betai\left(\alpha_0\left(1-\frac{9k^2}{16}\right) \right.\right.\\
&\left.\left.\quad+\alpha_1\left[1+\taue-\frac{k^2}{4}\left(3-k^2\right)\right]+\left(1-\frac{3k^2}{8}\right)\left\lbrace\alpha_3-\frac{k^2}{4}\left[3\left(\alpha_2+\etai\right)-k^2\left(\alpha_1+2\etai\right)\right]\right\rbrace\right)\right] \\
&\quad-\frac{1}{u^2}\left[\frac{3k^2\alpha_2}{8}+\betai\left(\frac{\alpha_0^2}{k^2}-\alpha_0\alpha_2+\frac{3k^2\alpha_2}{8}\left[1+\taue-\frac{k^2}{4}\left(3-k^2\right)\right] \right.\right.\\
&\left.\left.\quad-\alpha_1\left\lbrace\alpha_3-\frac{k^2}{4}\left[3\left(\alpha_2+\etai\right)-k^2\left(\alpha_1+2\etai\right)\right]\right\rbrace\right)\right] \\
&\quad+\frac{\betai k}{2u}\left(3-\frac{5k^2}{4}+\frac{35k^4}{96}\right)\left\lbrace\frac{\alpha_0}{k}-\frac{3k\alpha_2}{4}+\frac{5k^3}{384}\left[24\left(\alpha_2+\etai\right)
-7k^2\left(\alpha_1-2\etai\right)\right]\right\rbrace \\
&\quad+\frac{\betai}{u^2}\left\lbrace\frac{\alpha_0}{k}-\frac{3k\alpha_2}{4}+\frac{5k^3}{384}\left[24\left(\alpha_2+\etai\right)
-7k^2\left(\alpha_1-2\etai\right)\right]\right\rbrace^2 = 0
\end{align*}
\begin{align*}
0 &= 1-\frac{3k^2}{8}+\betai\left\lbrace\left(1-\frac{3k^2}{8}\right)\left[1+\taue-\frac{k^2}{4}\left(3-k^2\right)\right]+\frac{k^2}{16}\left(3-\frac{5k^2}{4}+\frac{35k^4}{96}\right)^2\right\rbrace \\
&\quad-\frac{1}{u}\left[\alpha_1+\betai\left(\alpha_0\left(1-\frac{9k^2}{16}\right) 
+\alpha_1\left[1+\taue-\frac{k^2}{4}\left(3-k^2\right)\right] \right.\right. \\
&\left.\left.\quad-\frac{k}{2}\left(3-\frac{5k^2}{4}+\frac{35k^4}{96}\right)\left\lbrace\frac{\alpha_0}{k}-\frac{3k\alpha_2}{4}+\frac{5k^3}{384}\left[24\left(\alpha_2+\etai\right)
-7k^2\left(\alpha_1-2\etai\right)\right]\right\rbrace \right.\right.\\
&\left.\left.\quad+\left(1-\frac{3k^2}{8}\right)\left\lbrace\alpha_3-\frac{k^2}{4}\left[3\left(\alpha_2+\etai\right)-k^2\left(\alpha_1+2\etai\right)\right]\right\rbrace\right)\right] \\
&\quad-\frac{1}{u^2}\left[\frac{3k^2\alpha_2}{8}+\betai\left(\frac{\alpha_0^2}{k^2}-\alpha_0\alpha_2+\frac{3k^2\alpha_2}{8}\left[1+\taue-\frac{k^2}{4}\left(3-k^2\right)\right] \right.\right.\\
&\left.\left.\quad-\alpha_1\left\lbrace\alpha_3-\frac{k^2}{4}\left[3\left(\alpha_2+\etai\right)-k^2\left(\alpha_1+2\etai\right)\right]\right\rbrace-\left\lbrace\frac{\alpha_0}{k}-\frac{3k\alpha_2}{4}+\frac{5k^3}{384}\left[24\left(\alpha_2+\etai\right)
-7k^2\left(\alpha_1-2\etai\right)\right]\right\rbrace^2\right)\right]
\end{align*}
Multiply across by $u^2$
\begin{align*}
0 &= u^2\left(1-\frac{3k^2}{8}+\betai\left\lbrace\left(1-\frac{3k^2}{8}\right)\left[1+\taue-\frac{k^2}{4}\left(3-k^2\right)\right]+\frac{k^2}{16}\left(3-\frac{5k^2}{4}+\frac{35k^4}{96}\right)^2\right\rbrace\right) \\
&\quad-u\left[\alpha_1+\betai\left(\alpha_0\left(1-\frac{9k^2}{16}\right) 
+\alpha_1\left[1+\taue-\frac{k^2}{4}\left(3-k^2\right)\right] \right.\right.\\
&\left.\left.\quad-\frac{k}{2}\left(3-\frac{5k^2}{4}+\frac{35k^4}{96}\right)\left\lbrace\frac{\alpha_0}{k}-\frac{3k\alpha_2}{4}+\frac{5k^3}{384}\left[24\left(\alpha_2+\etai\right)
-7k^2\left(\alpha_1-2\etai\right)\right]\right\rbrace \right.\right.\\
&\left.\left.\quad+\left(1-\frac{3k^2}{8}\right)\left\lbrace\alpha_3-\frac{k^2}{4}\left[3\left(\alpha_2+\etai\right)-k^2\left(\alpha_1+2\etai\right)\right]\right\rbrace\right)\right] \\
&\quad-\frac{3k^2\alpha_2}{8}-\betai\left(\frac{\alpha_0^2}{k^2}-\alpha_0\alpha_2+\frac{3k^2\alpha_2}{8}\left[1+\taue-\frac{k^2}{4}\left(3-k^2\right)\right] \right.\\
&\left.\quad-\alpha_1\left\lbrace\alpha_3-\frac{k^2}{4}\left[3\left(\alpha_2+\etai\right)-k^2\left(\alpha_1+2\etai\right)\right]\right\rbrace-\left\lbrace\frac{\alpha_0}{k}-\frac{3k\alpha_2}{4}+\frac{5k^3}{384}\left[24\left(\alpha_2+\etai\right)
-7k^2\left(\alpha_1-2\etai\right)\right]\right\rbrace^2\right)
\end{align*}
To lowest order in $k$ this is
\begin{align*}
0 &= u^2\left[1+\betai\left(1+\taue\right)\right]-u\left\lbrace\alpha_1+\betai\left[\alpha_0+\alpha_1\left(1+\taue\right)-\frac{3\alpha_0}{2}+\alpha_3\right]\right\rbrace \\
&\qquad-\betai\left(\frac{\alpha_0^2}{k^2}-\alpha_0\alpha_2-\alpha_1\alpha_3-\frac{\alpha_0^2}{k^2}+\frac{3\alpha_0\alpha_2}{2}\right)
\end{align*}
\begin{align} \tag{32}
0 &= u^2\left[1+\betai\left(1+\taue\right)\right]-u\left\lbrace\alpha_1+\betai\left[-\frac{\alpha_0}{2}+\alpha_1\left(1+\taue\right)+\alpha_3\right]\right\rbrace+\betai\left(\alpha_1\alpha_3-\frac{\alpha_0\alpha_2}{2}\right)
\end{align}

\section*{High $\beta$ low $k\rho_i$ limit}

\begin{equation} \tag{39}
\begin{aligned}
\frac{1}{\omegabi}=\frac{1}{\omega-\frac{k}{2}\LnOLB v^2} &= -\frac{1}{\frac{k}{2}\LnOLB v^2} \frac{1}{1-\frac{\omega}{\frac{k}{2}\LnOLB v^2}} \\
&= -\frac{1}{\di}\frac{1}{1-\frac{\omega}{\di}}=-\frac{1}{\di}\left(1+\frac{\omega}{\di}+\frac{\omega^2}{\di^2}+\frac{\omega^3}{\di^3}+\dots\right)
\end{aligned}
\end{equation}

\begin{equation} \tag{40}
\begin{aligned}
\frac{1}{\omegabe}=\frac{1}{\omega+\taue\frac{k}{2}\LnOLB v^2} &= \frac{1}{\taue\frac{k}{2}\LnOLB v^2} \frac{1}{1+\frac{\omega}{\taue\frac{k}{2}\LnOLB v^2}} \\
&= \frac{1}{\de}\frac{1}{1+\frac{\omega}{\de}}=\frac{1}{\de}\left(1-\frac{\omega}{\de}+\frac{\omega^2}{\de^2}-\frac{\omega^3}{\de^3}+\dots\right)
\end{aligned}
\end{equation}

In the high $\beta$ limit the $\omega$ fractions are expanded as follows:
\begin{equation} \tag{41}
\begin{aligned}
\frac{\omegabari}{\omegabi} &= -\frac{\omega}{\di}\left[1-\frac{k}{2\omega}\left(1-\etai\right)-\frac{k}{2\omega}\etai v^2\right]\left(1+\frac{\omega}{\di}+\frac{\omega^2}{\di^2}+\frac{\omega^3}{\di^3}+\dots\right) \\
&= -\frac{\omega}{\di}\left\lbrace1-\frac{k}{2\omega}\left(1-\etai\right)-\frac{k}{2\di}\etai v^2-\frac{k}{2\omega}\etai v^2 \right.\\
&\left.\qquad\qquad+\frac{\omega}{\di}\left[1-\frac{k}{2\omega}\left(1-\etai\right)-\frac{k}{2\di}\etai v^2\right]+\frac{\omega^2}{\di^2}\left[1-\frac{k}{2\omega}\left(1-\etai\right)\right]\right\rbrace
\end{aligned}
\end{equation}

\begin{equation} \tag{42}
\begin{aligned}
\frac{\omegabare}{\omegabe} &= \frac{\omega}{\de}\left[1+\frac{\taue k}{2\omega}\left(1-\etae\right)+\frac{\taue k}{2\omega}\etae v^2\right]\left(1-\frac{\omega}{\de}+\frac{\omega^2}{\de^2}-\frac{\omega^3}{\de^3}+\dots\right) \\
&= \frac{\omega}{\de}\left\lbrace1+\frac{\taue k}{2\omega}\left(1-\etae\right)-\frac{\taue k}{2\de}\etae v^2+\frac{\taue k}{2\omega}\etae v^2 \right.\\
&\left.\qquad\qquad-\frac{\omega}{\de}\left[1+\frac{\taue k}{2\omega}\left(1-\etae\right)-\frac{\taue k}{2\de}\etae v^2\right]+\frac{\omega^2}{\de^2}\left[1+\frac{\taue k}{2\omega}\left(1-\etae\right)\right]\right\rbrace
\end{aligned}
\end{equation}

In terms of $\beta$ and $u$ these are:
\begin{align*}
\frac{\omegabari}{\omegabi} &= -\LBOLn\frac{u}{v^2}\left\lbrace1-\frac{1}{u}\left(1-\etai\right)-\LBOLn\etai-\frac{1}{u}\etai v^2+\LBOLn\frac{u}{v^2}\left[1-\frac{1}{u}\left(1-\etai\right)-\LBOLn\etai\right] \right.\\
&\left.\qquad\qquad\qquad+\left(\LBOLn\right)^2\frac{u^2}{v^4}\left[1-\frac{1}{u}\left(1-\etai\right)\right]\right\rbrace \\
&= \LBOLn\left[\etai-\frac{1}{v^2}\left(u-1+\etai\right)\right]+\left(\LBOLn\right)^2\frac{u}{v^2}\left[\etai-\frac{1}{v^2}\left(u-1+\etai\right)\right]+\left(\LBOLn\right)^3\frac{u^2}{v^4}\left[\etai-\frac{1}{v^2}\left(u-1+\etai\right)\right]
\end{align*}
\begin{align*}
\frac{\omegabare}{\omegabe} &= \LBOLn\frac{1}{\taue}\frac{u}{v^2}\left\lbrace1+\frac{\taue}{u}\left(1-\etae\right)-\LBOLn\etae+\frac{\taue}{u}\etae v^2 -\LBOLn\frac{u}{v^2}\frac{1}{\taue}\left[1+\frac{\taue}{u}\left(1-\etae\right)-\LBOLn\etae\right] \right.\\
&\left.\quad+\left(\LBOLn\right)^2\frac{1}{\taue^2}\frac{u^2}{v^4}\left[1+\frac{\taue}{u}\left(1-\etae\right)\right]\right\rbrace \\
&= \LBOLn\frac{1}{\taue}\left[\taue\etae+\frac{1}{v^2}\left(u+\taue-\taue\etae\right)\right]-\left(\LBOLn\right)^2\frac{1}{\taue^2}\frac{u}{v^2}\left[\taue\etae+\frac{1}{v^2}\left(u+\taue-\taue\etae\right)\right] \\
&\quad+\left(\LBOLn\right)^3\frac{1}{\taue^3}\frac{u^2}{v^4}\left[\taue\etae+\frac{1}{v^2}\left(u+\taue-\taue\etae\right)\right]
\end{align*}

\begin{align*}
\IphiQ &= 2\int dv e^{-v^2}v\left[ J_0^2\frac{\omegabari}{\omegabi}-1+
\taui\left(\frac{\omegabare}{\omegabe}-1\right)\right] \\
&=2\int dv e^{-v^2}v\left[\left(1-\frac{k^2}{4}v^2+\frac{k^4}{64}v^4\right)^2\frac{\omegabari}{\omegabi}-1+
\taui\left(\frac{\omegabare}{\omegabe}-1\right)\right] \\
&=2\int dv e^{-v^2}v\left(-\frac{\omega}{\di}\left(1-\frac{k^2}{2}v^2+\frac{3k^4}{32}v^4\right)\left\lbrace1-\frac{k}{2\omega}\left(1-\etai\right)-\frac{k}{2\di}\etai v^2 \right.\right.\\
&\left.\left.\quad -\frac{k}{2\omega}\etai v^2+\frac{\omega}{\di}\left[1-\frac{k}{2\omega}\left(1-\etai\right)-\frac{k}{2\di}\etai v^2\right]+\frac{\omega^2}{\di^2}\left[1-\frac{k}{2\omega}\left(1-\etai\right)\right]\right\rbrace-\left(1+\taui\right) \right.\\
&\left.\quad+\taui\frac{\omega}{\de}\left\lbrace1+\frac{\taue k}{2\omega}\left(1-\etae\right)-\frac{\taue k}{2\de}\etae v^2+\frac{\taue k}{2\omega}\etae v^2 -\frac{\omega}{\de}\left[1+\frac{\taue k}{2\omega}\left(1-\etae\right)-\frac{\taue k}{2\de}\etae v^2\right] \right.\right.\\
&\left.\left.\quad+\frac{\omega^2}{\de^2}\left[1+\frac{\taue k}{2\omega}\left(1-\etae\right)\right]\right\rbrace\right) \\
&=2\int dv e^{-v^2}v\left[-\frac{\omega}{\di}\left(\frac{\omega^2}{\di^2}\left[1-\frac{k}{2\omega}\left(1-\etai\right)\right]+\frac{\omega}{\di}\left\lbrace1-\frac{k}{2\omega}\left(1-\etai\right)-\frac{k}{2\di}\etai v^2 \right.\right.\right.\\
&\left.\left.\left.\quad-\frac{k^2}{2}\frac{\omega}{\di}v^2\left[1-\frac{k}{2\omega}\left(1-\etai\right)\right]\right\rbrace+1-\frac{k}{2\omega}\left(1-\etai\right)-\frac{k}{2\di}\etai v^2-\frac{k^2}{2}\frac{\omega}{\di}v^2\left[1-\frac{k}{2\omega}\left(1-\etai\right)-\frac{k}{2\di}\etai v^2\right] \right.\right.\\
&\left.\left.\quad+\frac{3k^4}{32}\frac{\omega^2}{\di^2}v^4\left[1-\frac{k}{2\omega}\left(1-\etai\right)\right]-v^2\left\lbrace\frac{k}{2\omega}\etai+\frac{k^2}{2}\left[1-\frac{k}{2\omega}\left(1-\etai\right)-\frac{k}{2\di}\etai v^2\right] \right.\right.\right.\\
&\left.\left.\left.\quad-\frac{3k^4}{32}\frac{\omega}{\di}v^2\left[1-\frac{k}{2\omega}\left(1-\etai\right)-\frac{k}{2\di}\etai v^2\right]\right\rbrace+v^4\left\lbrace\frac{3k^4}{32}\left[1-\frac{k}{2\omega}\left(1-\etai\right)-\frac{k}{2\di}\etai v^2\right]+\frac{k^3}{4\omega}\etai\right\rbrace\right) \right.\\
&\left.\quad-\left(1+\taui\right)+\taui\frac{\omega}{\de}\left\lbrace\frac{\omega^2}{\de^2}\left[1+\frac{\taue k}{2\omega}\left(1-\etae\right)\right]-\frac{\omega}{\de}\left[1+\frac{\taue k}{2\omega}\left(1-\etae\right)-\frac{\taue k}{2\de}\etae v^2\right] \right.\right.\\
&\left.\left.\quad+1+\frac{\taue k}{2\omega}\left(1-\etae\right)-\frac{\taue k}{2\de}\etae v^2+\frac{\taue k}{2\omega}\etae v^2 \right\rbrace\right]
\end{align*}

\begin{align*}
\IphiQ &= \frac{\omega v^2}{\di}\left\lbrace\frac{k}{2\omega}\etai+\frac{k^2}{2}\left[1-\frac{k}{2\omega}\left(1-\etai\right)-\frac{k}{2\di}\etai v^2\right]-\frac{3k^4}{32}\frac{\omega}{\di}v^2\left[1-\frac{k}{2\omega}\left(1-\etai\right)-\frac{k}{2\di}\etai v^2\right] \right. \\
&\left.\quad-\frac{3k^4}{32}\left[1-\frac{k}{2\omega}\left(1-\etai\right)-\frac{k}{2\di}\etai v^2\right]-\frac{k^3}{4\omega}\etai\right\rbrace-\left(1+\taui\right)+\frac{\taui}{\de}\frac{\taue k}{2}\etae v^2 \\
&\quad +2\int dv e^{-v^2}v\left[-\frac{\omega}{\di}\left(\frac{\omega^2}{\di^2}\left[1-\frac{k}{2\omega}\left(1-\etai\right)\right]+\frac{\omega}{\di}\left\lbrace1-\frac{k}{2\omega}\left(1-\etai\right)-\frac{k}{2\di}\etai v^2 \right.\right.\right.\\
&\left.\left.\left.\quad -\frac{k^2}{2}\frac{\omega}{\di}v^2\left[1-\frac{k}{2\omega}\left(1-\etai\right)\right]\right\rbrace+1-\frac{k}{2\omega}\left(1-\etai\right)-\frac{k}{2\di}\etai v^2-\frac{k^2}{2}\frac{\omega}{\di}v^2\left[1-\frac{k}{2\omega}\left(1-\etai\right)-\frac{k}{2\di}\etai v^2\right] \right.\right.\\
&\left.\left.\quad+\frac{3k^4}{32}\frac{\omega^2}{\di^2}v^4\left[1-\frac{k}{2\omega}\left(1-\etai\right)\right]\right)+\taui\frac{\omega}{\de}\left\lbrace\frac{\omega^2}{\de^2}\left[1+\frac{\taue k}{2\omega}\left(1-\etae\right)\right]-\frac{\omega}{\de}\left[1+\frac{\taue k}{2\omega}\left(1-\etae\right)-\frac{\taue k}{2\de}\etae v^2\right] \right.\right.\\
&\left.\left.\quad +1+\frac{\taue k}{2\omega}\left(1-\etae\right)-\frac{\taue k}{2\de}\etae v^2 \right\rbrace\right]
\end{align*}

To second order in $k$ the integrated part reads:
\begin{equation} \tag{42}
\begin{aligned}
\IphiQ &= \frac{\omega v^2}{\di}\left\lbrace\frac{k}{2\omega}\etai+\frac{k^2}{2}\left[1-\frac{k}{2\omega}\left(1-\etai\right)\right]-\frac{k^3}{4\omega}\etai\right\rbrace-\left(1+\taui\right)+\frac{\taui}{\de}\frac{\taue k}{2}\etae v^2 \\
&= \frac{2}{k}\LBOLn\left\lbrace\frac{k}{2}\etai+\frac{k^2}{2}\left[\omega-\frac{k}{2}\left(1-\etai\right)\right]-\frac{k^3}{4}\etai\right\rbrace-\left(1+\taui\right)+\frac{\taui}{\taue}\frac{2}{k}\LBOLn\frac{\taue k}{2}\etae \\
&= \frac{2}{k}\LBOLn\left(\frac{k}{2}\etai+\frac{k^2}{2}\omega-\frac{k^3}{4}\right)-\left(1+\taui\right)+\taui\LBOLn\etae \\
&= \frac{2}{k}\LBOLn\left(\frac{k}{2}\etai+\frac{k^3}{4}u-\frac{k^3}{4}\right)-\left(1+\taui\right)+\taui\LBOLn\etae \\
&= \frac{2}{k}\LBOLn\left(\frac{k}{2}\etai+\frac{k^3}{4}u-\frac{k^3}{4}\right)-\left(1+\taui\right)+\taui\LBOLn\etae \\
\IphiQ &= \LBOLn\frac{k^2}{2}\left(u-1\right)-\left(1+\taui\right) +\LBOLn\left(\etai+\taui\etae\right)
\end{aligned}
\end{equation}
Before moving on to the next integral, we note that to $\mathcal{O}\left(\beta^{-1}\right)$ the integral $\IphiQ$ including the non-integrated part is:
\begin{align*}
\IphiQ &= \LBOLn\frac{k^2}{2}\left(u-1\right)-\left(1+\taui\right) \\
&\quad +2\int dv e^{-v^2}v\left\lbrace-\frac{\omega}{\di}\left(1-\frac{k}{2\omega}\left(1-\etai\right)-\frac{k}{2\di}\etai v^2-\frac{k}{2\omega}\etai v^2\right) \right.\\
&\left.\quad+\taui\frac{\omega}{\de}\left[1+\frac{\taue k}{2\omega}\left(1-\etae\right)-\frac{\taue k}{2\de}\etae v^2 -\frac{\taue k}{2\omega}\etae v^2\right]\right\rbrace
\end{align*}
where the term proportional to $\left(\etai+\taui\etae\right)$ has been anti-integrated and will be removed later. This part of the integral comes from, to $\mathcal{O}\left(\beta^{-1}\right)$, the multiplication of the lowest order term in the expansion of $J_0^2$ and $\omegabari/\omegabi$ and the lowest order terms in $\omegabare/\omegabe$.
\begin{align*}
\IphiQ &= \LBOLn\frac{k^2}{2}\left(u-1\right)-\left(1+\taui\right)+2\int dv e^{-v^2}v\left(\frac{\omegabari}{\omegabi}+\taui\frac{\omegabare}{\omegabe}\right) = \LBOLn\frac{k^2}{2}\left(u-1\right)-\left(1+\taui\right)+R_{\phi Q}
\end{align*}
where this residue is
\begin{align*}
R_{\phi Q} &= 2\int dv e^{-v^2}v\left(\frac{\omega-\frac{k}{2}\left(1-\etai\right)-\frac{k}{2}\etai v^2}{\omega-\frac{k}{2}\LnOLB v^2}+\taui\frac{\omega+\frac{\taue k}{2}\left(1-\etae\right)+\frac{\taue k}{2}\etae v^2}{\omega+\taue\frac{k}{2}\LnOLB v^2}\right) \\
&= 2\int dv e^{-v^2}v\left(-\frac{\frac{k}{2}\etai v^2}{\omega-\frac{k}{2}\LnOLB v^2}+\taui\frac{\frac{\taue k}{2}\etae v^2}{\omega+\taue\frac{k}{2}\LnOLB v^2}+\frac{\omega-\frac{k}{2}\left(1-\etai\right)}{\omega-\frac{k}{2}\LnOLB v^2}+\taui\frac{\omega+\frac{\taue k}{2}\left(1-\etae\right)}{\omega+\taue\frac{k}{2}\LnOLB v^2}\right) \\
&= 2\int dv e^{-v^2}v\left(\frac{1}{\di}\frac{\frac{k}{2}\etai v^2}{1-\frac{\omega}{\di}}+\frac{\taui}{\de}\frac{\frac{\taue k}{2}\etae v^2}{1+\frac{\omega}{\de}}+\frac{\omega-\frac{k}{2}\left(1-\etai\right)}{\omega-\frac{k}{2}\LnOLB v^2}+\taui\frac{\omega+\frac{\taue k}{2}\left(1-\etae\right)}{\omega+\taue\frac{k}{2}\LnOLB v^2}\right) \\
&= 2\int dv e^{-v^2}v\left(\frac{1}{\di}\frac{k}{2}\etai v^2+\frac{\taui}{\de}\frac{\taue k}{2}\etae v^2+\frac{\omega-\frac{k}{2}\left(1-\etai\right)}{\omega-\frac{k}{2}\LnOLB v^2}+\taui\frac{\omega+\frac{\taue k}{2}\left(1-\etae\right)}{\omega+\taue\frac{k}{2}\LnOLB v^2}\right) \\
&= \frac{1}{\di}\frac{k}{2}\etai v^2+\frac{\taui}{\de}\frac{\taue k}{2}\etae v^2+2\int dv e^{-v^2}v\left(\frac{\omega-\frac{k}{2}\left(1-\etai\right)}{\omega-\frac{k}{2}\LnOLB v^2}+\taui\frac{\omega+\frac{\taue k}{2}\left(1-\etae\right)}{\omega+\taue\frac{k}{2}\LnOLB v^2}\right) \\
&= \LBOLn\left(\etai+\taui\etae\right)+2\int dv e^{-v^2}v\left(\frac{\omega-\frac{k}{2}\left(1-\etai\right)}{\omega-\frac{k}{2}\LnOLB v^2}+\taui\frac{\omega+\frac{\taue k}{2}\left(1-\etae\right)}{\omega+\taue\frac{k}{2}\LnOLB v^2}\right)
\end{align*}
Here we consider the integral
\begin{align*}
2\int^\infty_0 dv e^{-v^2} \frac{v}{c+v^2} &= \int^\infty_0 dz e^{-z} \frac{1}{c+z} \\
&= \int^{z_0}_0 dz \frac{1}{c+z} + \int^\infty_{z_0} dz e^{-z} \frac{1}{c+z}
\end{align*}
This second step considers a small $z_0$ and uses the fact that in the limit $z_0\rightarrow0$ one has $e^{-z_0}\rightarrow1$.
\begin{align*}
\int^\infty_0 dv e^{-v^2} \frac{v}{c+v^2} &= \int^{z_0}_c dx \frac{1}{x} + \int^\infty_{z_0} dz e^{-z} \frac{1}{\cancelto{\propto \beta^{-1}~\text{\scriptsize small}}{c}+z} \\
&= \ln\left(z_0\right) - \ln\left(c\right) + \Gamma\left(0,z_0\right),
\end{align*}
which in the limit $z_0\rightarrow0$ we can write as:
\begin{align*}
\int^\infty_0 dv e^{-v^2} \frac{v}{c+v^2} &= \ln\left(z_0\right) - \ln\left(c\right) - \gamma_{EM}-\ln\left(z_0\right) + \dots \\
&= - \ln\left(c\right) - \gamma_{EM}+ \dots.
\end{align*}
The the residue $R_{\phi Q}$ can be expressed as:
\begin{align*}
R_{\phi Q} &= \LBOLn\left(\etai+\taui\etae\right)-\frac{1}{\di v^2}\left[\omega-\frac{k}{2}\left(1-\etai\right)\right]\left[-\ln\left(-\frac{\omega}{\di v^2}\right)-\gamma_{EM}\right] \\
&\quad+\frac{\taui}{\de v^2}\left[\omega+\frac{\taue k}{2}\left(1-\etae\right)\right]\left[-\ln\left(\frac{\omega}{\de v^2}\right)-\gamma_{EM}\right].
\end{align*}
In terms of $u$ this is:
\begin{align*}
R_{\phi Q} &= \LBOLn\left(\etai+\taui\etae\right)+\LBOLn\left[u-\left(1-\etai\right)\right]\left[\ln\left(-\LBOLn u\right)+\gamma_{EM}\right] \\
&\quad-\frac{\taui}{\taue}\LBOLn\left[u+\taue\left(1-\etae\right)\right]\left[\ln\left(\taui\LBOLn u\right)+\gamma_{EM}\right].
\end{align*}
In terms of $\betai$ this is
\begin{equation} \tag{45}
\begin{aligned}
R_{\phi Q} &= -\frac{2}{\betai\alpha_0}\left\lbrace\left(\etai+\taui\etae\right)+\left[u-\left(1-\etai\right)\right]\left[\ln\left(\frac{2}{\betai\alpha_0} u\right)+\gamma_{EM}\right] \right.\\
&\left.\quad-\taui^2\left[u+\taue\left(1-\etae\right)\right]\left[\ln\left(-\frac{2\taui}{\betai\alpha_0} u\right)+\gamma_{EM}\right]\right\rbrace.
\end{aligned}
\end{equation}
Thought the term proportional to $\left(\etai+\taui\etae\right)$ is kept in some of the subsequent calculations in this supplement, it will later be dropped because it shares the same origin as the rest of $R_{\phi Q}$, and keeping this residue results in a transcendental equation.

Next, we compute the $\IBA$ integral.
\begin{align*}
\IBA &= 1+2\int dv e^{-v^2}v^3\left(J_1^2\frac{2\betai\omegabari}{k^2\omegabi}+v^2\frac{\betae\omegabare}{2\omegabe}\right) \\
&= 1+2\int dv e^{-v^2}v^3\left[\left(\frac{k}{2}v-\frac{k^3}{16}v^3+\frac{k^5}{384}v^5\right)^2\frac{2\betai\omegabari}{k^2\omegabi}+v^2\frac{\betae\omegabare}{2\omegabe}\right] \\
&= 1+2\int dv e^{-v^2}v^3\left[\left(\frac{k^2}{4}v^2-\frac{k^4}{16}v^4\right)\frac{2\betai\omegabari}{k^2\omegabi}+v^2\frac{\betae\omegabare}{2\omegabe}\right] \\
&= 1+\betai\int dv e^{-v^2}v^5\left(-\frac{\omega}{\di}\left(1-\frac{k^2}{4}v^2\right)\left\lbrace1-\frac{k}{2\omega}\left(1-\etai\right)-\frac{k}{2\di}\etai v^2-\frac{k}{2\omega}\etai v^2 \right.\right.\\
&\left.\left.\quad +\frac{\omega}{\di}\left[1-\frac{k}{2\omega}\left(1-\etai\right)-\frac{k}{2\di}\etai v^2\right]+\frac{\omega^2}{\di^2}\left[1-\frac{k}{2\omega}\left(1-\etai\right)\right]\right\rbrace+\taue\frac{\omega}{\de}\left\lbrace1+\frac{\taue k}{2\omega}\left(1-\etae\right) \right.\right.\\
&\left.\left.\quad-\frac{\taue k}{2\de}\etae v^2+\frac{\taue k}{2\omega}\etae v^2 -\frac{\omega}{\de}\left[1+\frac{\taue k}{2\omega}\left(1-\etae\right)-\frac{\taue k}{2\de}\etae v^2\right]+\frac{\omega^2}{\de^2}\left[1+\frac{\taue k}{2\omega}\left(1-\etae\right)\right]\right\rbrace\right) \\
&= 1+\betai\int dv e^{-v^2}v^5\left(-\frac{\omega}{\di}\left(\frac{\omega^2}{\di^2}\left[1-\frac{k}{2\omega}\left(1-\etai\right)\right]+\frac{\omega}{\di}\left\lbrace1-\frac{k}{2\omega}\left(1-\etai\right)-\frac{k}{2\di}\etai v^2 \right.\right.\right.\\
&\left.\left.\left.\quad-\frac{k^2}{4}\frac{\omega}{\di}v^2\left[1-\frac{k}{2\omega}\left(1-\etai\right)\right]\right\rbrace +1-\frac{k}{2\omega}\left(1-\etai\right)-\frac{k}{2\di}\etai v^2-\frac{k^2}{4}\frac{\omega}{\di}v^2\left[1-\frac{k}{2\omega}\left(1-\etai\right)-\frac{k}{2\di}\etai v^2\right] \right.\right.\\
&\left.\left.\quad-v^2\left\lbrace\frac{k}{2\omega}\etai+\frac{k^2}{4}\left[1-\frac{k}{2\omega}\left(1-\etai\right)-\frac{k}{2\di}\etai v^2\right]\right\rbrace+\frac{k^3}{8\omega}\etai v^4\right)+\taue\frac{\omega}{\de}\left\lbrace1+\frac{\taue k}{2\omega}\left(1-\etae\right)-\frac{\taue k}{2\de}\etae v^2 \right.\right.\\
&\left.\left.\quad +\frac{\taue k}{2\omega}\etae v^2-\frac{\omega}{\de}\left[1+\frac{\taue k}{2\omega}\left(1-\etae\right)-\frac{\taue k}{2\de}\etae v^2\right]+\frac{\omega^2}{\de^2}\left[1+\frac{\taue k}{2\omega}\left(1-\etae\right)\right]\right\rbrace\right) \\
&= 1-\frac{\betai}{2}\frac{\omega}{\di}v^2\left(\frac{\omega}{\di}v^2\left\lbrace1-\frac{k}{2\omega}\left(1-\etai\right)-\frac{k}{2\di}\etai v^2-\frac{k^2}{4}\frac{\omega}{\di}v^2\left[1-\frac{k}{2\omega}\left(1-\etai\right)\right]\right\rbrace+1-\frac{k}{2\omega}\left(1-\etai\right) \right.\\
&\left.\quad-\frac{k}{2\di}\etai v^2-\frac{k^2}{4}\frac{\omega}{\di}v^2\left[1-\frac{k}{2\omega}\left(1-\etai\right)-\frac{k}{2\di}\etai v^2\right]-2\left\lbrace\frac{k}{2\omega}\etai+\frac{k^2}{4}\left[1-\frac{k}{2\omega}\left(1-\etai\right)-\frac{k}{2\di}\etai v^2\right]\right\rbrace \right.\\
&\left.\quad+\frac{3k^3}{4\omega}\etai\right)-\frac{\betai}{2}\taue\frac{\omega}{\de}v^2\left\lbrace\frac{\omega}{\de}v^2\left[1+\frac{\taue k}{2\omega}\left(1-\etae\right)-\frac{\taue k}{2\de}\etae v^2\right]-1-\frac{\taue k}{2\omega}\left(1-\etae\right)+\frac{\taue k}{2\de}\etae v^2-2\frac{\taue k}{2\omega}\etae\right\rbrace \\
&\quad +\betai\int dv e^{-v^2}v^5\left(-\frac{\omega^3}{\di^3}\left[1-\frac{k}{2\omega}\left(1-\etai\right)\right]+\taue\frac{\omega^3}{\de^3}\left[1+\frac{\taue k}{2\omega}\left(1-\etae\right)\right]\right)
\end{align*}

Keeping only terms up to order 2 in $\LB/Ln$:
\begin{equation} \tag{43}
\begin{aligned}
\IBA &= 1-\frac{\betai}{2}\frac{\omega}{\di}v^2\left(\frac{\omega}{\di}v^2\left\lbrace1-\frac{k}{2\omega}\left(1-\etai\right)\right\rbrace+1-\frac{k}{2\omega}\left(1-\etai\right) \right.\\
&\left.\quad-\frac{k}{2\di}\etai v^2-\frac{k^2}{4}\frac{\omega}{\di}v^2\left[1-\frac{k}{2\omega}\left(1-\etai\right)\right]-2\left\lbrace\frac{k}{2\omega}\etai+\frac{k^2}{4}\left[1-\frac{k}{2\omega}\left(1-\etai\right)-\frac{k}{2\di}\etai v^2\right]\right\rbrace \right.\\
&\left.\quad+\frac{3k^3}{4\omega}\etai\right)-\frac{\betai}{2}\taue\frac{\omega}{\de}v^2\left\lbrace\frac{\omega}{\de}v^2\left[1+\frac{\taue k}{2\omega}\left(1-\etae\right)\right]-1-\frac{\taue k}{2\omega}\left(1-\etae\right)+\frac{\taue k}{2\de}\etae v^2-2\frac{\taue k}{2\omega}\etae\right\rbrace \\
&\quad +\betai\int dv e^{-v^2}v^5\left(-\frac{\omega^3}{\di^3}\left[1-\frac{k}{2\omega}\left(1-\etai\right)\right]+\taue\frac{\omega^3}{\de^3}\left[1+\frac{\taue k}{2\omega}\left(1-\etae\right)\right]\right) \\
&= 1-\frac{\betai}{2}\LBOLn u\left(\LBOLn u\left\lbrace1-\frac{1}{u}\left(1-\etai\right)\right\rbrace+1-\frac{1}{u}\left(1-\etai\right) \right.\\
&\left.\quad-\LBOLn\etai-\frac{k^2}{4}\LBOLn u\left[1-\frac{1}{u}\left(1-\etai\right)\right]-2\left\lbrace\frac{1}{u}\etai+\frac{k^2}{4}\left[1-\frac{1}{u}\left(1-\etai\right)-\LBOLn\etai\right]\right\rbrace+\frac{3k^2}{2u}\etai\right)  \\
&\quad-\frac{\betai}{2}\LBOLn u\left\lbrace\frac{1}{\taue}\LBOLn u\left[1+\frac{\taue}{u}\left(1-\etae\right)\right]-1-\frac{\taue}{u}\left(1-\etae\right)+\LBOLn\etae-2\frac{\taue}{u}\etae\right\rbrace \\
&\quad +\betai\int dv e^{-v^2}v^5\left(-\frac{\omega^3}{\di^3}\left[1-\frac{k}{2\omega}\left(1-\etai\right)\right]+\taue\frac{\omega^3}{\de^3}\left[1+\frac{\taue k}{2\omega}\left(1-\etae\right)\right]\right) \\
&= 1+\frac{\betai}{2}\LBOLn\left[-u+1-\etai+2\etai+\frac{k^2}{2}\left(u-1+\etai\right)-\frac{3k^2}{2}\etai+u+\taue-\taue\etae+2\taue\etae\right] \\
&\quad -\frac{\betai}{2}\left(\LBOLn\right)^2u\left[u-1+\etai-\etai-\frac{k^2}{4}\left(u-1+\etai\right)+\frac{k^2}{2}\etai+\taui u+1-\etae+\etae\right] \\
&\quad +\betai\int dv e^{-v^2}v^5\left(-\frac{\omega^3}{\di^3}\left[1-\frac{k}{2\omega}\left(1-\etai\right)\right]+\taue\frac{\omega^3}{\de^3}\left[1+\frac{\taue k}{2\omega}\left(1-\etae\right)\right]\right) \\
&= 1+\frac{\betai}{2}\LBOLn\left\lbrace1+\etai+\taue+\taue\etae+\frac{k^2}{2}\left[u-\left(1+2\etai\right)\right]\right\rbrace \\
&\quad -\frac{\betai}{2}\left(\LBOLn\right)^2u\left\lbrace\left(1+\taui\right)u-\frac{k^2}{4}\left[u-\left(1+\etai\right)\right]\right\rbrace \\
&\quad +\betai\int dv e^{-v^2}v^5\left(-\frac{\omega^3}{\di^3}\left[1-\frac{k}{2\omega}\left(1-\etai\right)\right]+\taue\frac{\omega^3}{\de^3}\left[1+\frac{\taue k}{2\omega}\left(1-\etae\right)\right]\right)
\end{aligned}
\end{equation}

One more integral!
\begin{align*}
\frac{1}{\betai}\IphiA &= \int dv e^{-v^2}v^2\left(-2J_0J_1\frac{\omegabari}{k\omegabi}+v\frac{\omegabare}{\omegabe}\right) \\
&= \int dv e^{-v^2}v^2\left[-2\left(1-\frac{k^2}{4}v^2+\frac{k^4}{64}v^4\right)\left(\frac{k}{2}v-\frac{k^3}{16}v^3+\frac{k^5}{384}v^5\right)\frac{\omegabari}{k\omegabi}+v\frac{\omegabare}{\omegabe}\right] \\
&= \int dv e^{-v^2}v^3\left[-\left(1-\frac{k^2}{4}v^2+\frac{k^4}{64}v^4\right)\left(1-\frac{k^2}{8}v^2+\frac{k^4}{192}v^4\right)\frac{\omegabari}{\omegabi}+\frac{\omegabare}{\omegabe}\right] \\
&= \int dv e^{-v^2}v^3\left[-\left(1-\frac{3k^2}{8}v^2+\frac{5k^4}{96}v^4\right)\frac{\omegabari}{\omegabi}+\frac{\omegabare}{\omegabe}\right] \\
&= \int dv e^{-v^2}v^3\left[\frac{\omega}{\di}\left(1-\frac{3k^2}{8}v^2+\frac{5k^4}{96}v^4\right) \left\lbrace1-\frac{k}{2\omega}\left(1-\etai\right)-\frac{k}{2\di}\etai v^2-\frac{k}{2\omega}\etai v^2 \right.\right.\\
&\left.\left.\quad+\frac{\omega}{\di}\left[1-\frac{k}{2\omega}\left(1-\etai\right)-\frac{k}{2\di}\etai v^2\right]+\frac{\omega^2}{\di^2}\left[1-\frac{k}{2\omega}\left(1-\etai\right)\right]\right\rbrace+\frac{\omega}{\de}\left\lbrace1+\frac{\taue k}{2\omega}\left(1-\etae\right)-\frac{\taue k}{2\de}\etae v^2 \right.\right.\\
&\left.\left.\quad+\frac{\taue k}{2\omega}\etae v^2 -\frac{\omega}{\de}\left[1+\frac{\taue k}{2\omega}\left(1-\etae\right)-\frac{\taue k}{2\de}\etae v^2\right]+\frac{\omega^2}{\de^2}\left[1+\frac{\taue k}{2\omega}\left(1-\etae\right)\right]\right\rbrace\right] \\
&= \int dv e^{-v^2}v^3\left[\frac{\omega}{\di}\left(\frac{\omega^2}{\di^2}\left[1-\frac{k}{2\omega}\left(1-\etai\right)\right]+\frac{\omega}{\di}\left\lbrace1-\frac{k}{2\omega}\left(1-\etai\right)-\frac{k}{2\di}\etai v^2 \right.\right.\right.\\
&\left.\left.\left.\quad -\frac{3k^2}{8}\frac{\omega}{\di}v^2\left[1-\frac{k}{2\omega}\left(1-\etai\right)\right]\right\rbrace+1-\frac{k}{2\omega}\left(1-\etai\right)-\frac{k}{2\di}\etai v^2-\frac{3k^2}{8}\frac{\omega}{\di}v^2\left[1-\frac{k}{2\omega}\left(1-\etai\right)-\frac{k}{2\di}\etai v^2\right] \right.\right.\\
&\left.\left.\quad +\frac{5k^4}{96}\frac{\omega^2}{\di^2}v^4\left[1-\frac{k}{2\omega}\left(1-\etai\right)\right]-v^2\left\lbrace\frac{k}{2\omega}\etai+\frac{3k^2}{8}\left[1-\frac{k}{2\omega}\left(1-\etai\right)-\frac{k}{2\di}\etai v^2\right] \right.\right.\right.\\
&\left.\left.\left.\quad -\frac{5k^4}{96}\frac{\omega}{\di}v^2\left[1-\frac{k}{2\omega}\left(1-\etai\right)-\frac{k}{2\di}\etai v^2\right]\right\rbrace+v^4\left\lbrace\frac{3k^3}{16\omega}\etai+\frac{5k^4}{96}\left[1-\frac{k}{2\omega}\left(1-\etai\right)-\frac{k}{2\di}\etai v^2\right]\right\rbrace \right.\right.\\
&\left.\left.\quad -\frac{5k^5}{192\omega}\etai v^6\right)+\frac{\omega}{\de}\left\lbrace1+\frac{\taue k}{2\omega}\left(1-\etae\right)-\frac{\taue k}{2\de}\etae v^2+\frac{\taue k}{2\omega}\etae v^2 -\frac{\omega}{\de}\left[1+\frac{\taue k}{2\omega}\left(1-\etae\right)-\frac{\taue k}{2\de}\etae v^2\right] \right.\right.\\
&\left.\left.\quad +\frac{\omega^2}{\de^2}\left[1+\frac{\taue k}{2\omega}\left(1-\etae\right)\right]\right\rbrace\right] \\
&= \frac{1}{2}\frac{\omega}{\di}v^2\left\lbrace1-\frac{k}{2\omega}\left(1-\etai\right)-\frac{k}{2\di}\etai v^2-\frac{3k^2}{8}\frac{\omega}{\di}v^2\left[1-\frac{k}{2\omega}\left(1-\etai\right)-\frac{k}{2\di}\etai v^2\right] \right.\\
&\left.\quad +\frac{5k^4}{96}\frac{\omega^2}{\di^2}v^4\left[1-\frac{k}{2\omega}\left(1-\etai\right)\right]-\frac{k}{2\omega}\etai-\frac{3k^2}{8}\left[1-\frac{k}{2\omega}\left(1-\etai\right)-\frac{k}{2\di}\etai v^2\right] \right.\\
&\left.\quad +\frac{5k^4}{96}\frac{\omega}{\di}v^2\left[1-\frac{k}{2\omega}\left(1-\etai\right)-\frac{k}{2\di}\etai v^2\right]+\frac{3k^3}{8\omega}\etai+\frac{5k^4}{48}\left[1-\frac{k}{2\omega}\left(1-\etai\right)-\frac{k}{2\di}\etai v^2\right]-\frac{5k^5}{32\omega}\etai\right\rbrace \\
&\quad +\frac{1}{2}\frac{\omega}{\de}v^2\left[1+\frac{\taue k}{2\omega}\left(1-\etae\right)-\frac{\taue k}{2\de}\etae v^2+\frac{\taue k}{2\omega}\etae\right] \\
&\quad+\int dv e^{-v^2}v^3\left[\frac{\omega}{\di}\left(\frac{\omega^2}{\di^2}\left[1-\frac{k}{2\omega}\left(1-\etai\right)\right]+\frac{\omega}{\di}\left\lbrace1-\frac{k}{2\omega}\left(1-\etai\right)-\frac{k}{2\di}\etai v^2 \right.\right.\right.\\
&\left.\left.\left.\quad -\frac{3k^2}{8}\frac{\omega}{\di}v^2\left[1-\frac{k}{2\omega}\left(1-\etai\right)\right]\right\rbrace\right)+\frac{\omega}{\de}\left\lbrace -\frac{\omega}{\de}\left[1+\frac{\taue k}{2\omega}\left(1-\etae\right)-\frac{\taue k}{2\de}\etae v^2\right] \right.\right.\\
&\left.\left.\quad+\frac{\omega^2}{\de^2}\left[1+\frac{\taue k}{2\omega}\left(1-\etae\right)\right]\right\rbrace\right]
\end{align*}
\begin{align*}
\frac{1}{\betai}\IphiA &= \frac{1}{2}\LBOLn u\left\lbrace1-\frac{1}{u}\left(1-\etai\right)-\LBOLn\etai-\frac{3k^2}{8}\LBOLn u\left[1-\frac{1}{u}\left(1-\etai\right)-\LBOLn\etai\right] \right.\\
&\left.\quad +\frac{5k^4}{96}\left(\LBOLn\right)^2 u^2\left[1-\frac{1}{u}\left(1-\etai\right)\right]-\frac{1}{u}\etai-\frac{3k^2}{8}\left[1-\frac{1}{u}\left(1-\etai\right)-\LBOLn\etai\right] \right.\\
&\left.\quad +\frac{5k^4}{96}\LBOLn u\left[1-\frac{1}{u}\left(1-\etai\right)-\LBOLn\etai\right]+\frac{3k^2}{4u}\etai+\frac{5k^4}{48}\left[1-\frac{1}{u}\left(1-\etai\right)-\LBOLn\etai\right]-\frac{5k^4}{16u}\etai\right\rbrace \\
&\quad +\frac{1}{2}\frac{1}{\taue}\LBOLn u\left[1+\frac{\taue}{u}\left(1-\etae\right)-\LBOLn\etae+\frac{\taue}{u}\etae\right] \\
&\quad+\int dv e^{-v^2}v^3\left[\frac{\omega}{\di}\left(\frac{\omega^2}{\di^2}\left[1-\frac{k}{2\omega}\left(1-\etai\right)\right]+\frac{\omega}{\di}\left\lbrace1-\frac{k}{2\omega}\left(1-\etai\right)-\frac{k}{2\di}\etai v^2 \right.\right.\right.\\
&\left.\left.\left.\quad -\frac{3k^2}{8}\frac{\omega}{\di}v^2\left[1-\frac{k}{2\omega}\left(1-\etai\right)\right]\right\rbrace\right)+\frac{\omega}{\de}\left\lbrace -\frac{\omega}{\de}\left[1+\frac{\taue k}{2\omega}\left(1-\etae\right)-\frac{\taue k}{2\de}\etae v^2\right] \right.\right.\\
&\left.\left.\quad+\frac{\omega^2}{\de^2}\left[1+\frac{\taue k}{2\omega}\left(1-\etae\right)\right]\right\rbrace\right]
\end{align*}
Keeping only terms of lowest order in $\beta$:
\begin{equation} \tag{44}
\begin{aligned}
\IphiA &= \frac{\betai}{2}\LBOLn u\left\lbrace1-\frac{1}{u}\left(1-\etai\right)-\frac{1}{u}\etai-\frac{3k^2}{8}\left[1-\frac{1}{u}\left(1-\etai\right)\right] +\frac{3k^2}{4u}\etai+\frac{5k^4}{48}\left[1-\frac{1}{u}\left(1-\etai\right)\right]\right.\\
&\left.\quad -\frac{5k^4}{16u}\etai+\frac{1}{\taue}\left[1+\frac{\taue}{u}\left(1-\etae\right)+\frac{\taue}{u}\etae\right]\right\rbrace \\
&= \frac{\betai}{2}\LBOLn \left\lbrace u-\left(1-\etai\right)-\etai-\frac{3k^2}{8}\left[u-\left(1-\etai\right)\right] +\frac{3k^2}{4}\etai+\frac{5k^4}{48}\left[u-\left(1-\etai\right)\right]\right.\\
&\left.\quad -\frac{5k^4}{16}\etai+\frac{1}{\taue}\left[u+\taue\left(1-\etae\right)+\taue\etae\right]\right\rbrace \\
&= \frac{\betai}{2}\LBOLn \left\lbrace u-1-\frac{3k^2}{8}\left[u-\left(1+\etai\right)\right] +\frac{5k^4}{48}\left[u-\left(1+2\etai\right)\right]+\frac{1}{\taue}u+1\right\rbrace \\
&= \frac{\betai}{2}\LBOLn \left\lbrace \left(1+\taui\right)u-\frac{3k^2}{8}\left[u-\left(1+\etai\right)\right] +\frac{5k^4}{48}\left[u-\left(1+2\etai\right)\right]\right\rbrace
\end{aligned}
\end{equation}

In this case the dispersion relation is:
\begin{align*}
\IphiQ\IBA &= 2\betai\left(\frac{1}{\betai}\IphiA\right)^2 \\
\text{LHS} &= \left\lbrace\LBOLn\left[\frac{k^2}{2}\left(u-1\right)+\left(\etai+\taui\etae\right)\right]-\left(1+\taui\right)\right\rbrace\left(1+\frac{\betai}{2}\LBOLn\left\lbrace1+\etai+\taue+\taue\etae+\frac{k^2}{2}\left[u-\left(1+2\etai\right)\right]\right\rbrace \right.\\
&\left.\quad -\frac{\betai}{2}\left(\LBOLn\right)^2u\left\lbrace\left(1+\taui\right)u-\frac{k^2}{4}\left[u-\left(1+\etai\right)\right]\right\rbrace\right) \\
&= -\left(1+\taui\right)+\LBOLn\left[\frac{k^2}{2}\left(u-1\right)+\left(\etai+\taui\etae\right)-\frac{\betai}{2}\left(1+\taui\right)\left\lbrace1+\etai+\taue+\taue\etae+\frac{k^2}{2}\left[u-\left(1+2\etai\right)\right]\right\rbrace \right.\\
&\left.\quad +\frac{\betai}{2}\LBOLn\left(\left[\frac{k^2}{2}\left(u-1\right)+\left(\etai+\taui\etae\right)\right]\left\lbrace1+\etai+\taue+\taue\etae+\frac{k^2}{2}\left[u-\left(1+2\etai\right)\right]\right\rbrace \right.\right.\\
&\left.\left.\quad +u\left(1+\taui\right)\left\lbrace\left(1+\taui\right)u-\frac{k^2}{4}\left[u-\left(1+\etai\right)\right]\right\rbrace\right)\right] \\
&= \frac{\betai}{2}\left(\LBOLn\right)^2\left[\frac{k^4}{4}+\left(1+\taui\right)^2-\frac{k^2}{4}\left(1+\taui\right)\right]u^2 \\
&\quad+\LBOLn\frac{k^2}{2}\left\lbrace1-\frac{\betai}{2}\left(1+\taui\right)+\frac{\betai}{2}\LBOLn\left[1+\etai+\taue+\taue\etae-\frac{k^2}{2}\left(1+2\etai\right)-\frac{k^2}{2}+\left(\etai+\taui\etae\right) \right.\right.\\
&\left.\left.\quad +\frac{1}{2}\left(1+\taui\right)\left(1+\etai\right)\right]\right\rbrace u+\LBOLn\left\lbrace-\frac{k^2}{2}+\left(\etai+\taui\etae\right)-\frac{\betai}{2}\left(1+\taui\right)\left[1+\etai+\taue+\taue\etae-\frac{k^2}{2}\left(1+2\etai\right)\right] \right.\\
&\left.\quad+\frac{\betai}{2}\LBOLn\left[-\frac{k^2}{2}+\left(\etai+\taui\etae\right)\right]\left[1+\etai+\taue+\taue\etae-\frac{k^2}{2}\left(1+2\etai\right)\right]\right\rbrace-\left(1+\taui\right)
\end{align*}

while the right side is 

\begin{align*}
\text{RHS} &= 2\betai \frac{1}{\betai^2} \frac{\betai^2}{4}\left(\LBOLn\right)^2 \left\lbrace \left(1+\taui\right)u-\frac{3k^2}{8}\left[u-\left(1+\etai\right)\right] +\frac{5k^4}{48}\left[u-\left(1+2\etai\right)\right]\right\rbrace^2 \\
&= \frac{\betai}{2}\left(\LBOLn\right)^2 \left\lbrace \left(1+\taui\right)^2u^2-\frac{3k^2}{4}u\left(1+\taui\right)\left[u-\left(1+\etai\right)\right]+\frac{9k^4}{64}\left[u-\left(1+\etai\right)\right]^2 \right.\\
&\left.\quad +\frac{5k^4}{24}u\left(1+\taui\right)\left[u-\left(1+2\etai\right)\right]-\frac{5k^4}{24}\frac{3k^2}{8}\left[u-\left(1+\etai\right)\right]\left[u-\left(1+2\etai\right)\right]+\frac{25k^8}{2304}\left[u-\left(1+2\etai\right)\right]^2\right\rbrace
\end{align*}

keeping only order 2 in $k$ terms, the dispersion relation is 
\begin{align*}
&\quad\frac{\betai}{2}\left(\LBOLn\right)^2\left[\frac{k^4}{4}+\left(1+\taui\right)^2-\frac{k^2}{4}\left(1+\taui\right)\right]u^2 \\
&\quad+\LBOLn\frac{k^2}{2}\left\lbrace1-\frac{\betai}{2}\left(1+\taui\right)+\frac{\betai}{2}\LBOLn\left[1+\etai+\taue+\taue\etae-\frac{k^2}{2}\left(1+2\etai\right)-\frac{k^2}{2}+\left(\etai+\taui\etae\right) \right.\right.\\
&\left.\left.\quad +\frac{1}{2}\left(1+\taui\right)\left(1+\etai\right)\right]\right\rbrace u-\left(1+\taui\right)+\LBOLn\left\lbrace-\frac{k^2}{2}+\left(\etai+\taui\etae\right)-\frac{\betai}{2}\left(1+\taui\right)\left[1+\etai+\taue+\taue\etae \right.\right.\\
&\left.\left.\quad-\frac{k^2}{2}\left(1+2\etai\right)\right]+\frac{\betai}{2}\LBOLn\left[-\frac{k^2}{2}+\left(\etai+\taui\etae\right)\right]\left[1+\etai+\taue+\taue\etae-\frac{k^2}{2}\left(1+2\etai\right)\right]\right\rbrace \\
&\quad= \frac{\betai}{2}\left(\LBOLn\right)^2 \left\lbrace \left(1+\taui\right)^2u^2-\frac{3k^2}{4}u\left(1+\taui\right)\left[u-\left(1+\etai\right)\right]+\frac{9k^4}{64}\left[u-\left(1+\etai\right)\right]^2 \right.\\
&\left.\quad +\frac{5k^4}{24}u\left(1+\taui\right)\left[u-\left(1+2\etai\right)\right]-\frac{5k^4}{24}\frac{3k^2}{8}\left[u-\left(1+\etai\right)\right]\left[u-\left(1+2\etai\right)\right]+\frac{25k^8}{2304}\left[u-\left(1+2\etai\right)\right]^2\right\rbrace
\end{align*}
canceling some terms, dropping $k^6$ and higher, and using $\alpha$ notation:
\begin{align*}
&-\frac{\betai}{2}\left(\LBOLn\right)^2\frac{k^2}{4}\taui u^2 +\LBOLn\frac{k^2}{2}\left\lbrace1-\frac{\betai}{2}\tauit+\frac{\betai}{2}\LBOLn\left[\alpha_0-\frac{k^2}{2}\alpha_2-\frac{k^2}{2}+\left(\etai+\taui\etae\right)+\frac{1}{2}\tauit\alpha_1\right]\right\rbrace u \\
&\quad-\tauit+\LBOLn\left\lbrace-\frac{k^2}{2}+\left(\etai+\taui\etae\right)-\frac{\betai}{2}\tauit\left[\alpha_0-\frac{k^2}{2}\alpha_2\right]+\frac{\betai}{2}\LBOLn\left[-\frac{k^2}{2}+\left(\etai+\taui\etae\right)\right]\left[\alpha_0-\frac{k^2}{2}\alpha_2\right]\right\rbrace \\
&= \frac{\betai}{2}\left(\LBOLn\right)^2 \left\lbrace -\frac{3k^2}{4}u\tauit\left(u-\alpha_1\right)+\frac{9k^4}{64}\left(u-\alpha_1\right)^2+\frac{5k^4}{24}u\tauit\left(u-\alpha_2\right)\right\rbrace \\
0&=\frac{\betai}{2}\left(\LBOLn\right)^2\left[\frac{k^2}{4}\taui-\frac{3k^2}{4}\tauit+\frac{9k^4}{64}+\frac{5k^4}{24}\tauit\right]u^2-\LBOLn\frac{k^2}{2}\left\lbrace1-\frac{\betai}{2}\tauit \right.\\
&\left.\quad+\frac{\betai}{2}\LBOLn\left[\alpha_0-\frac{k^2}{2}\alpha_2-\frac{k^2}{2}+\left(\etai+\taui\etae\right)-\frac{3}{2}\tauit\alpha_1+\frac{9k^2}{16}\alpha_1+\frac{5k^2}{12}\tauit\alpha_2+\frac{1}{2}\tauit\alpha_1\right]\right\rbrace u \\
&\quad+\tauit+\LBOLn\left(\frac{k^2}{2}-\left(\etai+\taui\etae\right)+\frac{\betai}{2}\tauit\left[\alpha_0-\frac{k^2}{2}\alpha_2\right]-\frac{\betai}{2}\LBOLn\left\lbrace-\frac{k^2}{2}\left[\alpha_0+\alpha_2\left(\etai+\taui\etae\right)\right]\right.\right.\\
&\left.\left.\quad +\alpha_0\left(\etai+\taui\etae\right)+\frac{k^4}{4}\alpha_2 -\frac{9k^4}{64}\alpha_1^2\right\rbrace\right)
\end{align*}
At this point we employ the useful relation
\begin{align*}
\frac{1}{\Lp} &= \frac{1}{p}\d{p}{x} =\frac{1}{p}\left(\d{p_e}{x}+\d{p_i}{x}\right) =\frac{1}{p}\left(n\d{T_e}{x}+T_e\d{n}{x}+n\d{T_i}{x}+T_i\d{n}{x}\right) \\
&=\frac{nT_i}{p}\left(\taue\LTe^{-1}+\taue\Ln^{-1}+\LTi^{-1}+\Ln^{-1}\right) \\
&=\frac{p_i}{p\Ln}\left(\taue\etae+\taue+\etai+1\right) = \frac{p_i}{p\Ln}\alpha_0
\end{align*}
such that using the equilibrium relation $1/\LB=-\beta/(2\Lp)$ we can write
\begin{align*}
\LBOLn &= -\frac{2\Lp}{\beta\Ln} = -\frac{2}{\beta\Ln}\frac{p\Ln}{p_i\alpha_0} = -\frac{2}{\betai\alpha_0}
\end{align*}
We can therefore write the dispersion relation as:
\begin{align*}
0 &= \frac{\betai}{2}\frac{4}{\betai^2\alpha_0^2}\left[\frac{k^2}{4}\taui-\frac{3k^2}{4}\tauit+\frac{9k^4}{64}+\frac{5k^4}{24}\tauit\right]u^2+\frac{2}{\betai\alpha_0}\frac{k^2}{2}\left\lbrace1-\frac{\betai}{2}\tauit \right.\\
&\left.\quad-\frac{\betai}{2}\frac{2}{\betai\alpha_0}\left[\alpha_0-\frac{k^2}{2}\alpha_2-\frac{k^2}{2}+\left(\etai+\taui\etae\right)-\frac{3}{2}\tauit\alpha_1+\frac{9k^2}{16}\alpha_1+\frac{5k^2}{12}\tauit\alpha_2+\frac{1}{2}\tauit\alpha_1\right]\right\rbrace u \\
&\quad+\tauit-\frac{2}{\betai\alpha_0}\left(\frac{k^2}{2}-\left(\etai+\taui\etae\right)+\frac{\betai}{2}\tauit\left[\alpha_0-\frac{k^2}{2}\alpha_2\right]+\frac{\betai}{2}\frac{2}{\betai\alpha_0}\left\lbrace-\frac{k^2}{2}\left[\alpha_0+\alpha_2\left(\etai+\taui\etae\right)\right] \right.\right.\\
&\left.\left.\quad +\alpha_0\left(\etai+\taui\etae\right)+\frac{k^4}{4}\alpha_2-\frac{9k^4}{64}\alpha_1^2\right\rbrace\right)
\end{align*}
or multiplying by $\betai\alpha_0^2$ and keeping order 2 in $k$ terms only:
\begin{align*}
&\quad\frac{2}{\betai\alpha_0^2}\left(\frac{k^2}{4}\taui-\frac{3k^2}{4}\tauit\right)u^2+\frac{2}{\betai\alpha_0}\frac{k^2}{2}\left\lbrace1-\frac{\betai}{2}\tauit-\frac{1}{\alpha_0}\left[\alpha_0-\frac{k^2}{2}\alpha_2-\frac{k^2}{2}+\left(\etai+\taui\etae\right)-\tauit\alpha_1\right]\right\rbrace u \\
&\quad+\tauit-\frac{2}{\betai\alpha_0}\left(\frac{k^2}{2}-\left(\etai+\taui\etae\right)+\frac{\betai}{2}\tauit\left[\alpha_0-\frac{k^2}{2}\alpha_2\right] \right.\\
&\left.\qquad\qquad\qquad\quad +\frac{1}{\alpha_0}\left\lbrace-\frac{k^2}{2}\left[\alpha_0+\alpha_2\left(\etai+\taui\etae\right)\right]+\alpha_0\left(\etai+\taui\etae\right)\right\rbrace\right) = 0
\end{align*}
\begin{align*}
&-2\frac{k^2}{2}\tauit u^2+2\alpha_0\frac{k^2}{2}\left\lbrace1-\frac{\betai}{2}\tauit-\frac{1}{\alpha_0}\left[\alpha_0-\frac{k^2}{2}\alpha_2-\frac{k^2}{2}+\left(\etai+\taui\etae\right)-\tauit\alpha_1\right]\right\rbrace u \\
&\quad+\betai\alpha_0^2\tauit-2\alpha_0\left(\frac{k^2}{2}-\left(\etai+\taui\etae\right)+\frac{\betai}{2}\tauit\left[\alpha_0-\frac{k^2}{2}\alpha_2\right] \right.\\
&\left.\qquad\qquad\qquad\qquad +\frac{1}{\alpha_0}\left\lbrace-\frac{k^2}{2}\left[\alpha_0+\alpha_2\left(\etai+\taui\etae\right)\right]+\alpha_0\left(\etai+\taui\etae\right)\right\rbrace\right) = 0
\end{align*}
Drop the $\left(\etai+\taui\etae\right)$ terms since they belong to $R_{\phi Q}$ (see discussion above):
\begin{align*}
&-2\frac{k^2}{2}\tauit u^2+2\alpha_0\frac{k^2}{2}\left\lbrace1-\frac{\betai}{2}\tauit-\frac{1}{\alpha_0}\left[\alpha_0-\frac{k^2}{2}\alpha_2-\frac{k^2}{2}-\tauit\alpha_1\right]\right\rbrace u \\
&\quad+\betai\alpha_0^2\tauit-2\alpha_0\left[\frac{k^2}{2}+\frac{\betai}{2}\tauit\left(\alpha_0-\frac{k^2}{2}\alpha_2\right)-\frac{1}{\alpha_0}\frac{k^2}{2}\alpha_0\right] = 0 \\
&-2\frac{k^2}{2}\tauit u^2+2\alpha_0\frac{k^2}{2}\left\lbrace-\frac{\betai}{2}\tauit+\frac{1}{\alpha_0}\frac{k^2}{2}\left(\alpha_2+1\right)+\frac{\tauit\alpha_1}{\alpha_0}\right\rbrace u \\
&\quad+\betai\alpha_0^2\tauit-2\alpha_0\frac{\betai}{2}\tauit\left(\alpha_0-\frac{k^2}{2}\alpha_2\right) = 0 \\
&-2\frac{k^2}{2}\tauit u^2+2\alpha_0\frac{k^2}{2}\left\lbrace-\frac{\betai}{2}\tauit+\frac{1}{\alpha_0}\frac{k^2}{2}\left(\alpha_2+1\right)+\frac{\tauit\alpha_1}{\alpha_0}\right\rbrace u+\alpha_0\betai\tauit\frac{k^2}{2}\alpha_2 = 0 \\
\end{align*}
Divide by $-\tauit k^2/2$:
\begin{align*}
&2 u^2-2\alpha_0\left[-\frac{\betai}{2}+\frac{1}{\tauit\alpha_0}\frac{k^2}{2}\left(\alpha_2+1\right)+\frac{\alpha_1}{\alpha_0}\right] u-\alpha_0\betai\alpha_2 = 0 \\
\end{align*}
Only keep lower order $k$ terms:
\begin{align} \tag{46}
&2 u^2+\left(\betai\alpha_0-2\alpha_1\right) u-\alpha_0\betai\alpha_2 = 0
\end{align}

\pagebreak
\section*{Specious origin of the GDC instability}

Now we examine how the GDC comes about in this formulism. We do not assume a magnitude for $\beta$, but take $1/\LB=0$, such that $\omegabi=\omegabe=\omega$. The integrals in the dispersion relation are manipulated as follows:
\begin{equation*}
\begin{aligned}
\IphiQ &= 2\int dv e^{-v^2}v\left[ J_0^2\frac{\omegabari}{\omegabi}-1+
\taui\left(\frac{\omegabare}{\omegabe}-1\right)\right] \\
&\simeq 2\int dv e^{-v^2}v\left[ \left(1-\frac{k^2}{4}v^2+\frac{k^4}{64}v^4\right)^2\frac{\omegabari}{\omegabi}-1+
\taui\left(\frac{\omegabare}{\omegabe}-1\right)\right] \\
&\simeq 2\int dv e^{-v^2}v\left\lbrace \left(1-\frac{k^2}{2}v^2+\frac{3k^4}{32}v^4\right)\left[1-\frac{k}{2\omega}\left(1-\etai\right)-\frac{k}{2\omega}\etai v^2\right]-\left(1+\taui\right) \right.\\
&\left.\quad+\taui\left[1+\frac{\taue k}{2\omega}\left(1-\etae\right)+\frac{\taue k}{2\omega}\etae v^2\right]\right\rbrace \\
&\simeq 2\int dv e^{-v^2}v\left( -\left(1+\taui\right)+1-\frac{k}{2\omega}\left(1-\etai\right)- v^2\left\lbrace\frac{k}{2\omega}\etai+\frac{k^2}{2}\left[1-\frac{k}{2\omega}\left(1-\etai\right)\right]\right\rbrace \right.\\
&\left.\quad +v^4\left\lbrace\frac{k^2}{2}\frac{k}{2\omega}\etai+\frac{3k^4}{32}\left[1-\frac{k}{2\omega}\left(1-\etai\right)\right]\right\rbrace+\taui\left[1+\frac{\taue k}{2\omega}\left(1-\etae\right)+\frac{\taue k}{2\omega}\etae v^2\right]\right) \\
&\simeq -\left(1+\taui\right)+1-\frac{k}{2\omega}\left(1-\etai\right)-\frac{k}{2\omega}\etai-\frac{k^2}{2}\left[1-\frac{k}{2\omega}\left(1-\etai\right)\right] \\
&\quad +2\frac{k^2}{2}\frac{k}{2\omega}\etai+\frac{3k^4}{16}\left[1-\frac{k}{2\omega}\left(1-\etai\right)\right]+\taui\left[1+\frac{\taue k}{2\omega}\left(1-\etae\right)+\frac{\taue k}{2\omega}\etae\right]
\end{aligned}
\end{equation*}
\begin{align} \tag{50a}
\IphiQ &\simeq -\frac{k^2}{2}\left\lbrace1-\frac{k}{2\omega}\left(1+\etai\right)-\frac{3k^2}{8}\left[1-\frac{k}{2\omega}\left(1-\etai\right)\right]\right\rbrace
\end{align}

\begin{align*}
\IBA &= 1+2\int dv e^{-v^2}v^3\left(J_1^2\frac{2\betai\omegabari}{k^2\omegabi}+v^2\frac{\betae\omegabare}{2\omegabe}\right) \\
&= 1+2\int dv e^{-v^2}v^3\left[\left(\frac{k}{2}v-\frac{k^3}{16}v^3+\frac{k^5}{384}v^5\right)^2\frac{2\betai\omegabari}{k^2\omegabi}+v^2\frac{\betae\omegabare}{2\omegabe}\right] \\
&= 1+2\int dv e^{-v^2}v^3\left[\frac{2\betai}{k^2}\left(\frac{k^2}{4}v^2-\frac{k^4}{16}v^4\right)\left(1-\frac{k}{2\omega}\left(1-\etai\right)-\frac{k}{2\omega}\etai v^2\right) \right. \\
&\left.\quad+v^2\frac{\betae}{2}\left(1+\frac{\taue k}{2\omega}\left(1-\etae\right)+\frac{\taue k}{2\omega}\etae v^2\right)\right] \\
&= 1+2\frac{\betai}{2}\int dv e^{-v^2}v^5\left[\left(1-\frac{k^2}{4}v^2\right)\left(1-\frac{k}{2\omega}\left(1-\etai\right)-\frac{k}{2\omega}\etai v^2\right) \right. \\
&\left.\quad+\taue\left(1+\frac{\taue k}{2\omega}\left(1-\etae\right)+\frac{\taue k}{2\omega}\etae v^2\right)\right] \\
&= 1+2\frac{\betai}{2}\int dv e^{-v^2}v^5\left[1-\frac{k}{2\omega}\left(1-\etai\right)-v^2\left\lbrace\frac{k}{2\omega}\etai+\frac{k^2}{4}\left[1-\frac{k}{2\omega}\left(1-\etai\right)\right]\right\rbrace+\frac{k^3}{8\omega}\etai v^4 \right. \\
&\left.\quad+\taue\left(1+\frac{\taue k}{2\omega}\left(1-\etae\right)+\frac{\taue k}{2\omega}\etae v^2\right)\right] \\
&= 1+\frac{\betai}{2}\left(2\left[1-\frac{k}{2\omega}\left(1-\etai\right)\right]-6\left\lbrace\frac{k}{2\omega}\etai+\frac{k^2}{4}\left[1-\frac{k}{2\omega}\left(1-\etai\right)\right]\right\rbrace+\frac{3k^3}{\omega}\etai \right. \\
&\left.\quad+\taue\left\lbrace2\left[1+\frac{\taue k}{2\omega}\left(1-\etae\right)\right]+\frac{3\taue k}{\omega}\etae\right\rbrace\right] \\
&= 1+\betai\left\lbrace1-\frac{k}{2\omega}\left(1-\etai\right)-3\frac{k}{2\omega}\etai-3\frac{k^2}{4}\left[1-\frac{k}{2\omega}\left(1-\etai\right)\right]+\frac{3k^3}{2\omega}\etai \right. \\
&\left.\quad+\taue\left[1+\frac{\taue k}{2\omega}\left(1-\etae\right)+\frac{3\taue k}{2\omega}\etae\right]\right\rbrace \\
&= 1+\betai\left\lbrace1-\frac{k}{2\omega}\left(1+2\etai\right)-3\frac{k^2}{4}\left[1-\frac{k}{2\omega}\left(1-\etai\right)\right]+\frac{3k^3}{2\omega}\etai+\taue\left[1+\frac{\taue k}{2\omega}\left(1+2\etae\right)\right]\right\rbrace
\end{align*}
\begin{align} \tag{50b}
\IBA &= 1+\betai\left\lbrace1+\taue-\frac{k}{2\omega}\left[\left(1+2\etai\right)-\taue^2\left(1+2\etae\right)\right]-3\frac{k^2}{4}\left[1-\frac{k}{2\omega}\left(1+3\etai\right)\right]\right\rbrace
\end{align}

\begin{align*}
\frac{1}{\betai}\IphiA &= \int dv e^{-v^2}v^2\left(-2J_0J_1\frac{\omegabari}{k\omegabi}+v\frac{\omegabare}{\omegabe}\right) \\
&= \int dv e^{-v^2}v^2\left[-2\left(1-\frac{k^2}{4}v^2+\frac{k^4}{64}v^4\right)\left(\frac{k}{2}v-\frac{k^3}{16}v^3+\frac{k^5}{384}v^5\right)\frac{\omegabari}{k\omegabi}+v\frac{\omegabare}{\omegabe}\right] \\
&= \int dv e^{-v^2}v^2\left[-\frac{2}{k}\left(\frac{k}{2}v-\frac{3k^3}{16}v^3+\frac{5k^5}{192}v^5\right)\left(1-\frac{k}{2\omega}\left(1-\etai\right) \right.\right.\\
&\left.\left.\quad-\frac{k}{2\omega}\etai v^2\right)+v\left(1+\frac{\taue k}{2\omega}\left(1-\etae\right)+\frac{\taue k}{2\omega}\etae v^2\right)\right] \\
&= \int dv e^{-v^2}v^3\left[-\left(1-\frac{3k^2}{8}v^2+\frac{5k^4}{96}v^4\right)\left(1-\frac{k}{2\omega}\left(1-\etai\right) \right.\right.\\
&\left.\left.\quad-\frac{k}{2\omega}\etai v^2\right)+1+\frac{\taue k}{2\omega}\left(1-\etae\right)+\frac{\taue k}{2\omega}\etae v^2\right] \\
&= \int dv e^{-v^2}v^3\left[-1+\frac{k}{2\omega}\left(1-\etai\right)+v^2\left\lbrace\frac{3k^2}{8}\left[1-\frac{k}{2\omega}\left(1-\etai\right)\right]+\frac{k}{2\omega}\etai\right\rbrace \right.\\
&\left.\quad -v^4\left\lbrace\frac{5k^4}{96}\left[1-\frac{k}{2\omega}\left(1-\etai\right)\right]+\frac{3k^2}{8}\frac{k}{2\omega}\etai\right\rbrace+1+\frac{\taue k}{2\omega}\left(1-\etae\right)+\frac{\taue k}{2\omega}\etae v^2\right] \\
\frac{2}{\betai}\IphiA &= -1+\frac{k}{2\omega}\left(1-\etai\right)+2\left\lbrace\frac{3k^2}{8}\left[1-\frac{k}{2\omega}\left(1-\etai\right)\right]+\frac{k}{2\omega}\etai\right\rbrace \\
&\quad -6\left\lbrace\frac{5k^4}{96}\left[1-\frac{k}{2\omega}\left(1-\etai\right)\right]+\frac{3k^2}{8}\frac{k}{2\omega}\etai\right\rbrace+1+\frac{\taue k}{2\omega}\left(1-\etae\right)+2\frac{\taue k}{2\omega}\etae \\
\frac{2}{\betai}\IphiA &= \frac{k}{2\omega}\left(1-\etai\right)+\frac{3k^2}{4}\left[1-\frac{k}{2\omega}\left(1-\etai\right)\right]+2\frac{k}{2\omega}\etai \\
&\quad -\frac{5k^4}{16}\left[1-\frac{k}{2\omega}\left(1-\etai\right)\right]-\frac{9k^2}{4}\frac{k}{2\omega}\etai+\frac{\taue k}{2\omega}\left(1-\etae\right)+2\frac{\taue k}{2\omega}\etae \\
\frac{2}{\betai}\IphiA &= \frac{k}{2\omega}\left(1+\etai\right)+\frac{3k^2}{4}\left[1-\frac{k}{2\omega}\left(1-\etai\right)\right] \\
&\quad -\frac{5k^4}{16}\left[1-\frac{k}{2\omega}\left(1-\etai\right)\right]-\frac{9k^2}{4}\frac{k}{2\omega}\etai+\frac{\taue k}{2\omega}\left(1+\etae\right)
\end{align*}
\begin{align} \tag{50c}
\frac{1}{\betai}\IphiA &= \frac{k}{4\omega}\left[\left(1+\etai\right)+\taue\left(1+\etae\right)\right]+\frac{3k^2}{8}\left[1-\frac{k}{2\omega}\left(1+2\etai\right)\right] \\
&\quad -\frac{5k^4}{32}\left[1-\frac{k}{2\omega}\left(1-\etai\right)\right]
\end{align}

To third order in $k$, and in terms of $\alpha$'s and $u$, these integrals are:

\begin{equation} \tag{50}
\begin{aligned}
\IphiQ &= -\frac{k^2}{2}\left(1-\frac{\alpha_1}{u}\right) \\
\IBA &= 1+\betai\left(1+\taue-\frac{\alpha_3}{u}\right) \\
\frac{1}{\betai}\IphiA &= \frac{\alpha_0}{2u}+\frac{3k^2}{8}\left(1-\frac{\alpha_2}{u}\right)
\end{aligned}
\end{equation} 

The dispersion relation then leads to the following quadratic equation:

\begin{align*}
\IphiQ\IBA &= 2\betai\left(\frac{1}{\betai}\IphiA\right)^2 \\
-&\frac{k^2}{2}\left(1-\frac{\alpha_1}{u}\right)\left[1+\betai\left(1+\taue-\frac{\alpha_3}{u}\right)\right] = 2\betai \left[\frac{\alpha_0}{2u}+\frac{3k^2}{8}\left(1-\frac{\alpha_2}{u}\right)\right]^2 \\
-&\frac{k^2}{2}\left(1+\betai\left(1+\taue\right)-\frac{1}{u}\left\lbrace\betai\alpha_3+\alpha_1\left[1+\betai\left(1+\taue\right)\right]\right\rbrace+\betai\alpha_1\alpha_3\frac{1}{u^2}\right) \\
&\quad= 2\betai \left[\frac{\alpha_0^2}{4}\frac{1}{u^2}+\alpha_0\frac{3k^2}{8}\left(1-\frac{\alpha_2}{u}\right)\frac{1}{u}+\frac{9k^4}{64}\left(1-\frac{\alpha_2}{u}\right)^2\right] \\
-&\frac{k^2}{2}\left(1+\betai\left(1+\taue\right)-\frac{1}{u}\left\lbrace\betai\alpha_3+\alpha_1\left[1+\betai\left(1+\taue\right)\right]\right\rbrace+\betai\alpha_1\alpha_3\frac{1}{u^2}\right) \\
&\quad= 2\betai \left[\left(\frac{\alpha_0^2}{4}-\alpha_0\alpha_2\frac{3k^2}{8}+\alpha_2^2\frac{9k^4}{64}\right)\frac{1}{u^2}+\left(\alpha_0\frac{3k^2}{8}-\alpha_2\frac{9k^4}{32}\right)\frac{1}{u}+\frac{9k^4}{64}\right] \\
&\betai \left[\left(\alpha_0^2-\alpha_0\alpha_2\frac{3k^2}{2}+\alpha_2^2\frac{9k^4}{16}\right)\frac{1}{u^2}+\left(\alpha_0\frac{3k^2}{2}-\alpha_2\frac{9k^4}{8}\right)\frac{1}{u}+\frac{9k^4}{16}\right] \\
&\quad +k^2\left(1+\betai\left(1+\taue\right)-\frac{1}{u}\left\lbrace\betai\alpha_3+\alpha_1\left[1+\betai\left(1+\taue\right)\right]\right\rbrace+\betai\alpha_1\alpha_3\frac{1}{u^2}\right) = 0\\
&k^2\left[1+\betai\left(1+\taue\right)+\betai\frac{9k^2}{16}\right]u^2-k^2\left\lbrace\betai\alpha_3+\alpha_1\left[1+\betai\left(1+\taue\right)\right]-\betai\left(\frac{3}{2}\alpha_0-\alpha_2\frac{9k^2}{8}\right)\right\rbrace u \\
&\quad+k^2\betai\alpha_1\alpha_3+\betai\left(\alpha_0^2-\alpha_0\alpha_2\frac{3k^2}{2}+\alpha_2^2\frac{9k^4}{16}\right) = 0
\end{align*}
and noting that $\betai\left(1+\taue\right)=\beta$ we get
\begin{equation} \tag{51}
\begin{aligned}
&k^2\left(1+\beta+\betai\frac{9k^2}{16}\right)u^2-k^2\left[\betai\alpha_3+\alpha_1\left(1+\beta\right)-\betai\left(\frac{3}{2}\alpha_0-\alpha_2\frac{9k^2}{8}\right)\right] u \\
&\quad+k^2\betai\alpha_1\alpha_3+\betai\left(\alpha_0^2-\alpha_0\alpha_2\frac{3k^2}{2}+\alpha_2^2\frac{9k^4}{16}\right) = 0.
\end{aligned}
\end{equation}

The instability condition is given in terms of
\begin{align*}
\DeltaGDC &= \left\lbrace-k^2\left[\betai\alpha_3+\alpha_1\left(1+\beta\right)-\betai\left(\frac{3}{2}\alpha_0-\alpha_2\frac{9k^2}{8}\right)\right]\right\rbrace^2 \\
&\quad -4k^2\left(1+\beta+\betai\frac{9k^2}{16}\right)\left[k^2\betai\alpha_1\alpha_3+\betai\left(\alpha_0^2-\alpha_0\alpha_2\frac{3k^2}{2}+\alpha_2^2\frac{9k^4}{16}\right)\right] \\
&= k^4\left[\betai\alpha_3+\alpha_1\left(1+\beta\right)-\betai\frac{3}{2}\alpha_0\right]^2-4k^2\left(1+\beta\right)\betai\alpha_0^2 \\
&\quad -4k^4\left(1+\beta\right)\left(\betai\alpha_1\alpha_3-\frac{3}{2}\betai\alpha_0\alpha_2\right)-4k^2\betai\frac{9k^2}{16}\betai\alpha_0^2+\mathcal{O}\left(k^6\right) \\
&= -4\left(1+\beta\right)\betai\alpha_0^2k^2+k^4\left[\betai^2\alpha_3^2+2\betai\alpha_1\alpha_3\left(1+\beta\right)+\alpha_1^2\left(1+\beta\right)^2-3\betai^2\alpha_0\alpha_3-3\betai\alpha_0\alpha_1\left(1+\beta\right) \right. \\
&\left.\quad +\frac{9}{4}\betai^2\alpha_0^2\right]-4k^4\left(1+\beta\right)\betai\left(\alpha_1\alpha_3-\frac{3}{2}\alpha_0\alpha_2\right)-4k^2\betai\frac{9k^2}{16}\betai\alpha_0^2+\mathcal{O}\left(k^6\right) \\
&= -4\left(1+\beta\right)\betai\alpha_0^2k^2+k^4\left[\betai^2\alpha_3^2-2\betai\alpha_1\alpha_3\left(1+\beta\right)+\alpha_1^2\left(1+\beta\right)^2-3\betai^2\alpha_0\alpha_3-3\betai\alpha_0\alpha_1\left(1+\beta\right) \right. \\
&\left.\quad +6\left(1+\beta\right)\betai\alpha_0\alpha_2\right]+\mathcal{O}\left(k^6\right)
\end{align*}
\begin{equation} \tag{54}
\begin{aligned}
\DeltaGDC &= -4\left(1+\beta\right)\betai\alpha_0^2k^2 \\
&\quad+k^4\left\lbrace\left[\betai\alpha_3-\alpha_1\left(1+\beta\right)\right]^2+\betai\alpha_0\left[6\left(1+\beta\right)\alpha_2-3\betai\alpha_3-3\alpha_1\left(1+\beta\right)\right]\right\rbrace+\mathcal{O}\left(k^6\right).
\end{aligned}
\end{equation}

So that to lowest order the GDC growth rate is given by:
\begin{align*}
\gamma_{\text{GDC}} &= \frac{\lvert\Delta_{\text{GDC}}\rvert^{1/2}k}{4k^2\left(1+\beta+\betai\frac{9k^2}{16}\right)} \simeq \frac{\left[4\left(1+\beta\right)\betai\alpha_0^2k^2\right]^{1/2}}{4k\left(1+\beta\right)} \\
&\simeq \left[\frac{\betai\alpha_0^2}{4\left(1+\beta\right)}\right]^{1/2} \\
&\simeq \left[\frac{\betai}{4\left(1+\beta\right)}\left(\frac{\beta}{\betai}\frac{\Ln}{\Lp}\right)^2\right]^{1/2} \\
&\simeq \left[\frac{\beta}{4\left(1+\beta\right)}\right]^{1/2}\left(\frac{\beta}{\betai}\right)^{1/2}\left\lvert\frac{\Ln}{\Lp}\right\rvert \\
&\simeq \left[\frac{\beta}{2\left(1+\beta\right)}\right]^{1/2}\left(\frac{1+\taue}{2}\right)^{1/2}\left\lvert\frac{\Ln}{\Lp}\right\rvert.
\end{align*}
In physical units this is
\begin{align*}
\gamma_{\text{GDC}} &= \frac{\vti}{\left\lvert\Ln\right\rvert}\left[\frac{\beta}{2\left(1+\beta\right)}\right]^{1/2}\left(\frac{1+\taue}{2}\right)^{1/2}\left\lvert\frac{\Ln}{\Lp}\right\rvert = \left[\frac{\beta}{2\left(1+\beta\right)}\right]^{1/2}\left[\frac{\left(1+\taue\right)2 T_i}{2m_i}\right]^{1/2}\frac{1}{\left\lvert\Lp\right\rvert} \\
&= \left[\frac{\beta}{2\left(1+\beta\right)}\right]^{1/2}\left[\frac{p_0}{n_0m_i}\right]^{1/2}\frac{1}{\left\lvert\Lp\right\rvert}
\end{align*}
\begin{equation} \tag{9}
\gamma_{\text{GDC}} = \left[\frac{\beta}{2\left(1+\beta\right)}\right]^{1/2}\frac{\cs}{\left\lvert\Lp\right\rvert}.
\end{equation}

\end{document}